\documentclass[11pt, a4paper]{article}
\usepackage{url}
\usepackage{a4wide}
\usepackage{graphicx}
\usepackage{natbib}
\usepackage{enumitem}
\usepackage{hyperref}
\usepackage{setspace}
\usepackage{multirow, booktabs}
\usepackage{amsmath, amsthm, amssymb, amsfonts}
\usepackage{bm, bbm}
\usepackage[T1]{fontenc}
\usepackage[ruled, vlined]{algorithm2e}
\usepackage{color}

\usepackage{xr}

\allowdisplaybreaks

\DeclareMathAlphabet\mathbfcal{OMS}{cmsy}{b}{n}

\newcommand{\mbf}{\mathbf}
\newcommand{\mc}{\mathcal}
\newcommand{\bmx}{\begin{bmatrix}}
\newcommand{\emx}{\end{bmatrix}}

\newcommand{\vep}{\varepsilon}

\renewcommand{\l}{\left}
\renewcommand{\r}{\right}

\def\wh{\widehat}
\def\wt{\widetilde}

\newcommand{\E}[0]{\mathsf{E}}
\newcommand{\Var}[0]{\mathsf{Var}}
\newcommand{\Cov}[0]{\mathsf{Cov}}

\newcommand{\p}{\mathsf{P}}
\newcommand{\R}{\mathbb{R}}
\newcommand{\Z}{\mathbb{Z}}

\newcommand{\iid}{\text{\upshape iid}}
\newcommand{\nn}{\nonumber}
\newcommand{\lft}{\text{\tiny \upshape L}}
\newcommand{\rgt}{\text{\tiny \upshape R}}
\newcommand{\vertiii}[1]{{\left\vert\kern-0.25ex\left\vert\kern-0.25ex\left\vert #1 
    \right\vert\kern-0.25ex\right\vert\kern-0.25ex\right\vert}}
    
\newcommand{\crsc}{C_{\text{\tiny \upshape RE}}}
\newcommand{\cdev}{C_{\text{\tiny \upshape dev}}}
\newcommand{\cp}{\theta}
\newcommand{\Cp}{\Theta}

\theoremstyle{definition}
\newtheorem{thm}{Theorem}
\newtheorem{cor}[thm]{Corollary}
\newtheorem{prop}[thm]{Proposition}
\newtheorem{lem}[thm]{Lemma}
\newtheorem{assum}{Assumption}
\newtheorem{rem}{Remark}

\newtheorem{cond}{Condition}

\newtheorem{massuminner}{Assumption}
\newenvironment{massum}[1]{%
  \renewcommand\themassuminner{#1}%
  \massuminner
}{\endmassuminner}

\title{High-dimensional data segmentation in regression settings
permitting temporal dependence and non-Gaussianity}
\author{Haeran Cho$^1$ \and Dom Owens$^2$}

\begin{document}

\maketitle

\footnotetext[1]{School of Mathematics, University of Bristol.
Email: \url{haeran.cho@bristol.ac.uk}. Supported by the Leverhulme Trust (RPG-2019-390).}

\footnotetext[2]{School of Mathematics, University of Bristol.
Email: \url{domowens1@gmail.com}. Supported by EPSRC Centre for Doctoral Training in Computational Statistics and Data Science (EP/S023569/1).}


\begin{abstract}
We propose a data segmentation methodology for the high-dimensional linear regression problem where regression parameters are allowed to undergo multiple changes. 
The proposed methodology, MOSEG, proceeds in two stages: first, the data are scanned for multiple change points using a moving window-based procedure, which is followed by a location refinement stage. 
MOSEG enjoys computational efficiency thanks to the adoption of a coarse grid in the first stage, and achieves theoretical consistency in estimating both the total number and the locations of the change points, under general conditions permitting serial dependence and non-Gaussianity.
We also propose MOSEG.MS, a multiscale extension of MOSEG which, while comparable to MOSEG in terms of computational complexity, achieves theoretical consistency for a broader parameter space where large parameter shifts over short intervals and small changes over long stretches of stationarity are simultaneously allowed. 
We demonstrate good performance of the proposed methods in comparative simulation studies and in an application 
to predicting the equity premium. 
\end{abstract}

\section{Introduction}

Regression modelling in high dimensions 
has received great attention with the development of data collection and storage technologies, and numerous applications are found in natural and social sciences, economics, finance and genomics, to name a few.
There is a mature literature on high-dimensional linear regression modelling under the sparsity assumption, 
see \cite{buhlmann2011statistics} and \cite{tibshirani2011regression} for an overview.
When observations are collected over time in highly nonstationary environments,
it is natural to allow for shifts in the regression parameters.
Permitting the parameters to vary over time in a piecewise constant manner,
data segmentation, a.k.a.\ multiple change point detection, provides a conceptually simple framework for handling nonstationarity in the data.

In this paper, we consider the problem of multiple change point detection under the following model:
We observe $(Y_t, \mbf x_t), \, t = 1, \ldots, n$, 
with $\mbf x_t = (X_{1t}, \ldots, X_{pt})^\top \in \R^p$ where
\begin{align} 
\label{eq:model}
Y_t = \l\{\begin{array}{ll}
\mbf{x}_t^\top\bm{\beta}_0 + \vep_t & \text{for } \cp_0 = 0 < t \le \cp_1, \\
\mbf{x}_t^\top\bm{\beta}_1 + \vep_t & \text{for } \cp_1 < t \le \cp_2, \\
\vdots \\
\mbf{x}_t^\top\bm{\beta}_q + \vep_t & \text{for } \cp_q < t \le n = \cp_{q+1}.
\end{array} \r.	 
\end{align}
Here, $\{\vep_t\}_{t = 1}^n$ denotes a sequence of errors satisfying $\E(\vep_t) = 0$ and $\Var(\vep_t) = \sigma_\vep^2 \in (0, \infty)$ for all $t$, which may be serially correlated.
At each change point $\cp_j$, the vector of parameters undergoes a change such that $\bm\beta_{j  - 1} \ne \bm\beta_j$ for all $j = 1, \ldots, q$.
Then, our aim is to estimate the set of change points $\Cp = \{\cp_j, \, 1 \le j \le q\}$ by estimating both the total number $q$ and the locations $\cp_j$ of the change points.

The data segmentation problem under~\eqref{eq:model} is considered by \cite{bai1998estimating}, \cite{qu2007estimating}, \cite{zhao2022segmenting} and \cite{kirch2022}, among others, when the dimension $p$ is fixed.
In high-dimensional settings, when there exists at most one change point ($q = 1$),
\cite{lee2016lasso} and \cite{kaul2019efficient} consider the problem of detecting and locating the change point, respectively.
For the general case with unknown $q$, several data segmentation methods exist which adopt dynamic programming \citep{leonardi2016computationally, rinaldo2020localizing, xu2022change},
fused Lasso \citep{wang2021denoising, bai2022unified}
or wild binary segmentation \citep{wang2021statistically} algorithms
for the detection of multiple change points,
and Bayesian approaches also exist \citep{datta2019bayesian}.
A related yet distinct problem of testing for the presence of a single change point under the regression model has been considered in \cite{wang2022optimal} and \cite{liu2022change}.
\cite{gao2022sparse} consider the case where $\bm{\beta}_j - \bm{\beta}_{j - 1}$ is sparse without requiring the sparsity of $\bm\beta_j, \, j = 0, \ldots, q$, under the conditions that $p < n$ and $X_{it} \sim_{\iid} \mc N(0, 1)$ for all $i$ and $t$.

\begin{figure}[h!t!]
\centering
\begin{tabular}{cc}
	\includegraphics[width=.475\textwidth]{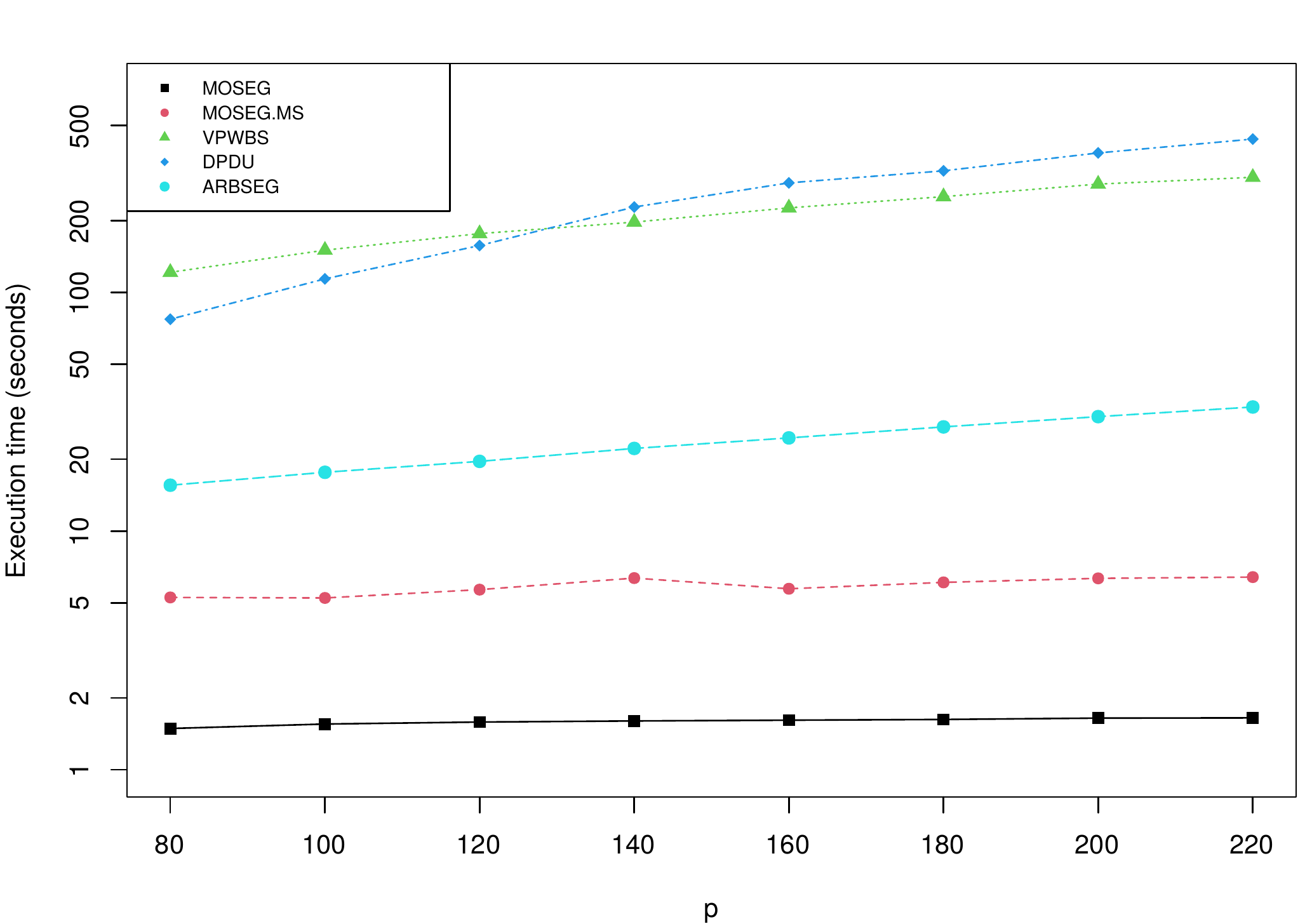} &
	\includegraphics[width=.475\textwidth]{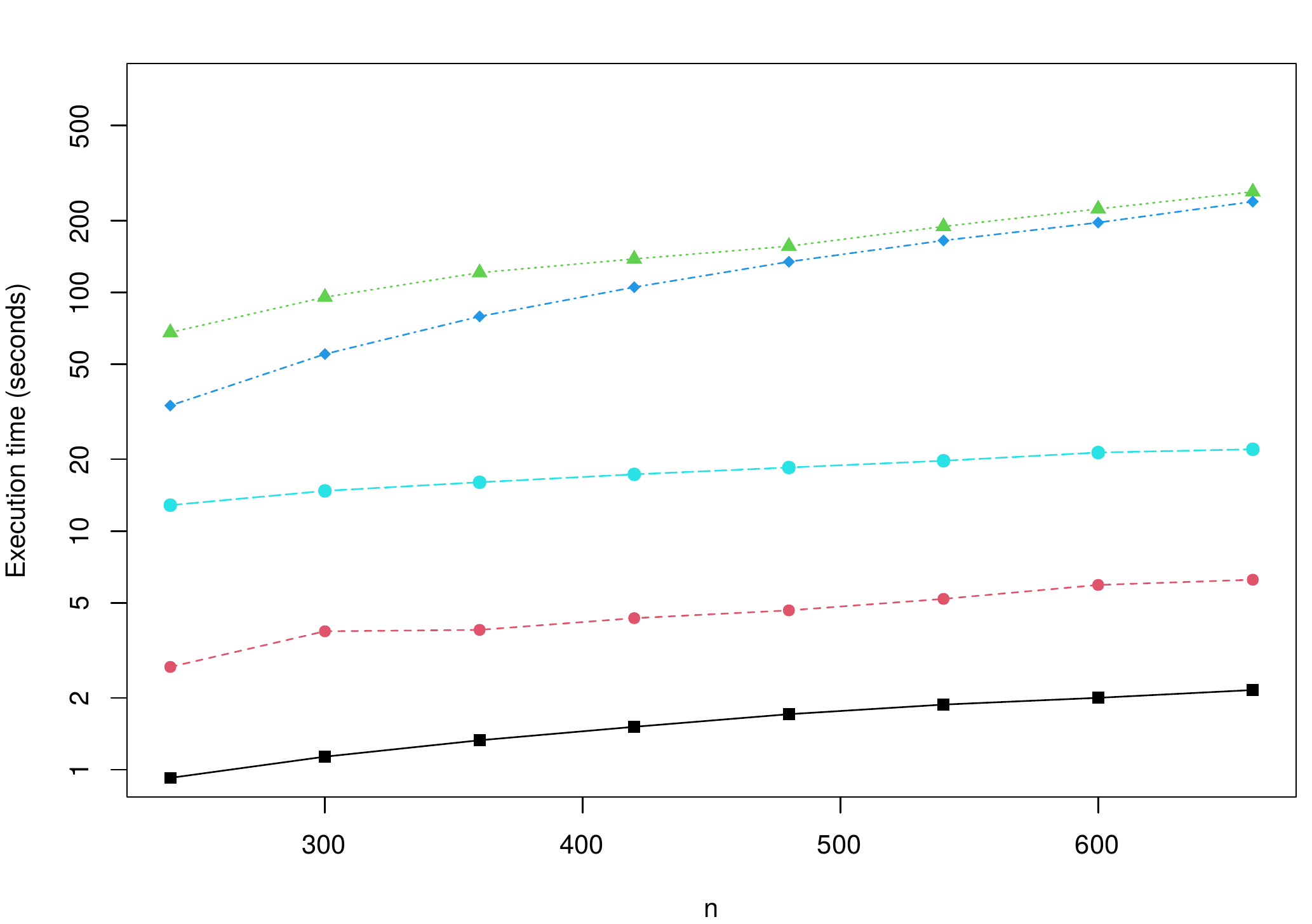}
\end{tabular}
\caption{Execution time in seconds of MOSEG and MOSEG.MS and competing methodologies on simulated datasets ($y$-axis is in log scale for ease of comparison).
Left: $p$ varies while $n = 450$ is fixed.
Right: $n$ varies while $p = 100$ is fixed.
For each setting, $100$ realisations are generated and the average execution time is reported.
See Section~\ref{sec:runtime} for full details.}
\label{fig:runtime}
\end{figure}

Against the above literature background, we list the contributions made in this paper
by proposing computationally and statistically efficient data segmentation methods.
\begin{enumerate}[label = (\roman*)]
\item {\bf Computational efficiency.} For the data segmentation problem under~\eqref{eq:model}, often the computational bottleneck is the local estimation of the regression parameters via penalised $M$-estimation such as Lasso.
We propose MOSEG, a moving window-based two-stage methodology, and its multiscale extension, which are both highly efficient computationally.
In the first stage, MOSEG scans the data for multiple change points using a moving window of length $G$ on a {\it coarse} grid of size $O(nG^{-1})$, 
which is followed by a simple location refinement step minimising the local residual sum of squares.
The adoption of a coarse grid in the first stage contributes greatly to the reduction of Lasso estimation steps while losing little detection power.
Figure~\ref{fig:runtime} demonstrates the computational competitiveness of the proposed MOSEG and MOSEG.MS where they greatly outperform the existing methodologies in their execution time for a range of $n$ and $p$.

\item {\bf Multiscale change point detection.}  
We propose a multiscale extension of the single-bandwidth methodology MOSEG. 
Referred to as MOSEG.MS, it is fully adaptive to the difficult scenarios with {\it multiscale} change points, where large frequent parameter shifts and small changes over long stretches of stationarity are simultaneously present, while still enjoying computational competitiveness.
To the best of our knowledge, MOSEG.MS is the only data segmentation methodology under the model~\eqref{eq:model} for which the detection and localisation consistency is derived explicitly for the broad parameter space that permits multiscale change points.
Also, while there exist several data segmentation methods that propose to apply moving window-based procedures with multiple bandwidths, MOSEG.MS is the first extension in high dimensions with a guaranteed rate of localisation. 

\item {\bf Theoretical consistency in general settings.} 
We show the consistency of MOSEG and MOSEG.MS in estimating the total number and the locations of multiple change points.
Under Gaussianity, their separation and localisation rates nearly match the minimax lower bounds up to a logarithmic factor.
Moreover, in our theoretical investigation, we permit temporal dependence as well as tail behaviour heavier than sub-Gaussianity.
This, compared to the existing literature where independence and (sub-)Gaussianity assumptions are commonly made, shows that the proposed methods work well in situations that are more realistic for empirical applications.
\end{enumerate}

The rest of the paper is organised as follows.
Section~\ref{sec:single} introduces MOSEG, the single-bandwidth methodology, and establishes its theoretical consistency.
Then in Section~\ref{sec:multiscale}, we propose its multiscale extension, MOSEG.MS, and show that it achieves theoretical consistency in a broader parameter space.
Numerical experiments in Section~\ref{sec:sim} demonstrate the competitiveness of the proposed methods in comparison with the existing data segmentation algorithms
and Section~\ref{sec:real} provides a real data application to equity premium data.
In the Appendix, we provide a comprehensive comparison between the existing methods and MOSEG and MOSEG.MS both on their theoretical and computational properties, and present all the proofs and additional numerical results.
The R software implementing MOSEG and MOSEG.MS is available from \url{https://github.com/Dom-Owens-UoB/moseg}.

\paragraph{Notation.} 

For a random variable $X$, we write $\Vert X \Vert_\nu = [\E(\vert X \vert^\nu)]^{1/\nu}$ for $\nu > 0$.
For $\mbf a = (a_1, \ldots, a_p)^\top \in \R^p$, we write $\text{supp}(\mbf a) = \{i, \, 1 \le i \le p: \, a_i \ne 0\}$, $\l\vert \mbf a \r\vert_0 = \sum_{i = 1}^p \mathbb{I}_{\{ a_i \ne 0 \}}$,  $\l\vert \mbf a\r\vert_1 = \sum_{i = 1}^p \l\vert a_i \r\vert$, $\l\vert \mbf a \r\vert_2 = (\sum_{i = 1}^p a_i^2)^{1/2}$ and $\l\vert \mbf a \r\vert_\infty = \max_{1 \le i \le p} \l\vert a_i \r\vert$.
For a square matrix $\mbf A$, let $\Lambda_{\max}(\mbf A)$ and $\Lambda_{\min}\l(\mbf A\r)$ denote its maximum and minimum eigenvalues, respectively.
For a set $\mc A$, we denote its cardinality by $\l\vert \mc A\r\vert$.
For sequences of positive numbers $\{a_n\}$ and $\{b_n\}$, we write $a_n \lesssim b_n$ if there exists some constant $C > 0$ such that $a_n/b_n \le C$ as $n \to \infty$. 
Finally, we write $a \vee b = \max(a, b)$ and $a \wedge b = \min(a, b)$.

\section{Single-bandwidth methodology}
\label{sec:single}

We introduce MOSEG, a single-bandwidth two-stage methodology for data segmentation in regression settings. 
We first describe its two stages in Section~\ref{sec:method}, establish its theoretical consistency in Section~\ref{sec:theory} and verify meta-assumptions made for the theoretical analysis in Section~\ref{sec:linear} for a class of linear processes with serial dependence and heavier tails than that permitted under sub-Gaussianity. 

\subsection{MOSEG}
\label{sec:method}

\subsubsection{Stage~1: Moving window procedure on a coarse grid}  
\label{sec:mosum}  

Single-bandwidth moving window procedures have successfully been adopted for univariate \citep{preuss2015, yau2016, eichinger2018mosum}, multivariate \citep{kirch2022} and high-dimensional \citep{cho2022} time series segmentation.
When applying a moving window-based procedure to a data segmentation problem, the key challenge is to carefully design a detector statistic which, when adopted for scanning the data for changes, has good detection power against the type of changes which is of interest to detect.

For a given bandwidth $G \in \mathbb{N}$ satisfying $G \le n/2$, our proposed detector statistic is
\begin{align}
\label{eq:wald}  
T_k(G) = \sqrt{\frac{G}{2}} \l\vert
\wh{\bm{\beta}}_{k, k+G} - \wh{\bm{\beta}}_{k-G, k} \r\vert_2, \quad 
G \le k \le n - G.
\end{align}
Here, $\wh{\bm\beta}_{s, e}$ denotes an estimator of the vector of parameters 
obtained from $(Y_t, \mbf x_t), \, s + 1 \le t \le e$,
for any $0 \le s < e \le n$.
The statistic $T_k(G)$ contrasts the local parameter estimators from two adjacent data sections over $\{k - G + 1, \ldots, k\}$ and $\{k + 1, \ldots, k + G\}$.
Then, $T_k(G)$ is expected to form local maxima near the change points where the local parameter estimators differ the most, 
and thus it is well-suited for detecting and locating the change points under the model~\eqref{eq:model}.

We propose to obtain the local estimator $\wh{\bm\beta}_{s, e}$ via Lasso, as
\begin{align} 
\label{eq:lasso}
\wh{\bm{\beta}}_{s, e}(\lambda) = {\arg\min}_{\bm\beta \in \R^p}
\sum_{t = s + 1}^e (Y_t - \mbf x_t^\top \bm\beta)^2 +
\lambda \sqrt{e - s} \vert \bm\beta \vert_1
\end{align} 
for some tuning parameter $\lambda > 0$.
In what follows, we suppress the dependence of this estimator on $\lambda$ when there is no confusion.
The estimand of $\wh{\bm{\beta}}_{k - G, k}$ is 
\begin{align} 
\label{eq:mixture}
\bm\beta^*_{k - G, k} = \frac{1}{G} \sum_{j = L(k - G + 1)}^{L(k)} 
\l\{(\cp_{j + 1} \wedge k) - ((k - G) \vee \cp_j)\r\} \bm\beta_j,
\end{align}
where $L(t) = \{j, \, 0 \le j \le q: \, \cp_j + 1 \le t \}$ denotes the index of a change point $\cp_j$ that is the closest to~$t$ while lying strictly to its left.
In short, $\bm\beta^*_{k - G, k}$ is a weighted sum of $\bm\beta_j$ with the weights corresponding to the proportion of the intervals $\{k - G + 1, \ldots, k\}$ overlapping with $\{\cp_j + 1, \ldots, \cp_{j + 1 }\}$.

Scanning the detector statistic $T_k(G)$ over all $k \in \{G, \ldots, n - G\}$ requires the computation of the Lasso estimator $O(n)$ times.
This is far fewer than $O(n^2)$ times required by dynamic programming algorithms for $\ell_0$-penalised cost minimisation \citep{rinaldo2020localizing, xu2022change}, but it may still pose a computational bottleneck when the data sequence is very long or its dimensionality ultra high.
Instead, we propose to evaluate $T_k(G)$ on a coarser grid only for generating {\it pre-estimators} of the change points.
Let $\mc T$ denote the grid over which we evaluate $T_k(G)$, which is given by
\begin{align}
\label{eq:grid} 
\mc T = \mc T(r, G) = \l\{ G + \lfloor rG \rfloor m, \,
0 \le m \le \l\lfloor \frac{n - 2G}{rG} \r\rfloor \r\} 
\end{align}
with some constant $r \in [G^{-1}, 1)$ that controls the coarseness of the grid.
When $r = G^{-1}$, we have the finest grid $\mc T = \{G, \ldots, n - G\}$ and the grid becomes coarser with increasing $r$. 

Motivated by \cite{eichinger2018mosum}, who considered the problem of detecting multiple shifts in the mean of univariate time series using a moving window procedure, we propose to accept all significant local maximisers of $T_k(G)$ over $k \in \mc T$ as the pre-estimators of the change points.
That is, for some threshold $D > 0$ and a tuning parameter $\eta \in (0, 1]$, we accept all $\wt\cp \in \mc T$ that simultaneously satisfy
\begin{align} 
\label{eq:local:max}
T_{\wt\cp}(G) > D \quad \text{and} \quad
\wt\cp = {\arg\max}_{k \in \mc T: \, \vert k - \wt\cp \vert \le \eta G} \, T_k(G).
\end{align}
That is, at such $\wt\cp$, the detector $T_{\wt\cp}(G)$ exceeds the threshold and attains a local maximum over the grid within the interval of length $\eta G$.
We denote the set collecting all pre-estimators fulfilling~\eqref{eq:local:max}, by
$\wt{\Cp} = \{\wt\cp_j, \, 1 \le j \le \wh q: \, \wt\cp_1 < \ldots < \wt\cp_{\wh q} \}$
with $\wh q = \vert \wt{\Cp} \vert$ as the estimator of the number of change points.
This grid-based approach substantially reduces the computational complexity by requiring the Lasso estimators to be computed only 
$O(n/\lfloor rG \rfloor)$ times.
Even so, it is sufficient for detecting the presence of all $q$ change points, provided that $r$ is chosen not too large (see Theorem~\ref{thm:one}~\ref{thm:one:one} below). 
We remark that the idea of utilising only a sub-sample of the data for detecting the presence of change points, has been proposed for handling long data sequences in the context of univariate mean change point detection \citep{lu2017intelligent}.
The next section describes the location refinement step applied to the pre-estimators of change point locations. 

\subsubsection{Stage~2: Location refinement}
\label{sec:refine}

Once the set of pre-estimators $\wt\Cp$ is generated by the first-stage moving window procedure on a coarse grid, we further refine the location estimators.
It involves the local evaluation and minimisation 
of the following objective function
\begin{align}
\label{eq:q} 
Q\l( k; a, b, \wh{\bm\gamma}^{\lft}, \wh{\bm\gamma}^{\rgt} \r)
= \sum_{t = a + 1}^k (Y_t - \mbf x_t^\top \wh{\bm\gamma}^{\lft})^2
+ 
\sum_{t = k + 1}^b (Y_t - \mbf x_t^\top \wh{\bm\gamma}^{\rgt})^2
\text{ for } k = a + 1, \ldots, b,
\end{align}
for suitably chosen $a$, $b$, $\wh{\bm\gamma}^{\lft}$ and $\wh{\bm\gamma}^{\rgt}$.

For each $j = 1, \ldots, \wh q$, let 
$\wt\cp_j^{\lft} = \wt\cp_j - \lfloor G/2 \rfloor$ and
$\wt\cp_j^{\rgt} = \wt\cp_j + \lfloor G/2 \rfloor$,
and consider the following local parameter estimators
\begin{align} \label{eq:beta:lr}
\wh{\bm\beta}^{\lft}_j = \wh{\bm{\beta}}_{0 \vee (\wt\cp_j^{\lft} - G), \wt\cp_j^{\lft}}
\quad \text{and} \quad 
\wh{\bm\beta}^{\rgt}_j = \wh{\bm{\beta}}_{\wt\cp_j^{\rgt}, (\wt\cp_j^{\rgt} +G) \wedge n},
\end{align} 
which serve as the estimators of $\bm\beta_{j - 1}$ and $\bm\beta_j$, respectively.
Then in Stage~2, we propose to obtain a refined location estimator of $\cp_j$ 
from its pre-estimator $\wt\cp_j$, as
\begin{align} 
\label{eq:refined}
\wh \cp_j = {\arg\min}_{\wt\cp_j - G + 1 \le k \le \wt\cp_j + G} \,
Q\l(k; \wt\cp_j  - G, \wt\cp_j + G, \wh{\bm\beta}^{\lft}_j, \wh{\bm\beta}^{\rgt}_j\r),
\end{align}
for all $j = 1, \ldots, \wh q$.
{A similar approach has commonly been taken in the change point literature, see e.g.\ \cite{kaul2019efficient} and \cite{xu2022change} for the data segmentation problem under the model~\eqref{eq:model}.
Our proposal differs from theirs only in that the interval over which the search is performed in~\eqref{eq:refined}, is chosen to contain exactly one change point with high probability under Assumption~\ref{assum:bandwidth}~\ref{assum:bandwidth:a} below.}
Referring to the methodology combining the two stages as MOSEG, we provide its algorithmic description in Algorithm~\ref{algo:mosumsr} of Appendix~\ref{app:alg}.

\subsection{Consistency of MOSEG}
\label{sec:theory}

We make the following assumptions on $(\mbf x_t, \vep_t), \, 1 \le t \le n$.
\begin{assum}
\label{assum:xe}
We assume that $\E(\mbf x_t) = \mbf 0$, $\E(\vep_t) = 0$ and $\Var(\vep_t) = \sigma_\vep^2$ for all $t = 1, \ldots, n$,
and that $\Cov(\mbf x_t) = \bm\Sigma_x$ has its eigenvalues bounded,
i.e.\ there exist $0 \le \omega \le \bar{\omega} < \infty$ such that
\begin{align*}
\omega \le \Lambda_{\min}(\bm\Sigma_x) \le \Lambda_{\max}(\bm\Sigma_x) \le \bar{\omega}.
\end{align*}
\end{assum}
{The condition on the eigenvalues of $\bm\Sigma_x$ can also be found e.g.\ in \cite{rinaldo2020localizing} and \cite{wang2019statistically}. 
Where relevant, we explicitly specify the roles played by $\omega$ and $\bar{\omega}$ in presenting the theoretical results, which indicates how this condition may be relaxed.}
Assumptions~\ref{assum:dev} and~\ref{assum:rsc} below extend the deviation bound and {restricted eigenvalue (RE)} conditions required for high-dimensional $M$-estimation \citep{van2009conditions, loh2012high, negahban2012unified}, to change point settings.
{We later verify the assumptions under general conditions accommodating serial dependence as well as non-Gaussian tail behaviour.}
Here, we explicitly state these meta-assumptions to highlight the full generality of the consistency of MOSEG derived in this section. 

\begin{assum}[Deviation bound]
\label{assum:dev}
There exist fixed constants $C_0, \cdev > 0$ and some $\rho_{n, p}~\to~\infty$
as $n, p \to \infty$,
such that $\p(\mc D^{(1)} \cap \mc D^{(2)}) \to 1$, where
\begin{align*}
\mc D^{(1)} &= \l\{ 
\max_{0 \le s < e \le n, \, e - s \ge C_0 \rho^2_{n, p}}
\l\vert \frac{1}{\sqrt{e - s}} \sum_{t = s + 1}^e \vep_t \mbf x_t \r\vert_\infty 
\le \cdev \rho_{n, p} \r\},
\\
\mc D^{(2)} &= \l\{ 
\max_{\substack{0 \le s < e \le n, \,  e - s \ge C_0 \rho^2_{n, p} \\ \vert \{s + 1, \ldots, e\} \cap \Cp \vert \le 1}}
\l\vert \frac{1}{\sqrt{e - s}} \sum_{t = s + 1}^e (Y_t - \mbf x_t^\top \bm\beta^*_{s, e}) \mbf x_t \r\vert_\infty 
\le \cdev \rho_{n, p} \r\}.
\end{align*}
\end{assum}

\begin{assum}[{Restricted eigenvalue}]
\label{assum:rsc}
There exist fixed constants $\crsc > 0$ and $\tau \in [0, 1)$ 
such that $\p(\mc R^{(1)} \cap \mc R^{(2)}) \to~1$, where
\begin{align*}
\mc R^{(1)} &= \l\{ \sum_{t = s + 1}^e \mbf a^\top \mbf x_t \mbf x_t^\top \mbf a 
\ge (e - s) \omega \vert \mbf a \vert_2^2 - \crsc \log(p) (e - s)^\tau \vert \mbf a \vert_1^2 \text{ for all } \r.
\\
& \qquad \l.0 \le s < e \le n \text{ satisfying } e - s \ge C_0 \rho^2_{n, p} \text{ and } \mbf a \in \R^p \r\},
\\
\mc R^{(2)} &= \l\{ \sum_{t = s + 1}^e \mbf a^\top \mbf x_t \mbf x_t^\top \mbf a 
\le (e - s) \bar{\omega} \vert \mbf a \vert_2^2 + \crsc \log(p) (e - s)^\tau \vert \mbf a \vert_1^2 \text{ for all } \r.
\\
& \qquad \l.0 \le s < e \le n \text{ satisfying } e - s \ge C_0 \rho^2_{n, p} \text{ and } \mbf a \in \R^p \r\}.
\end{align*}
\end{assum}

For each $j = 0, \ldots, q$, we denote by $\mc S_j = \text{supp}(\bm\beta_j)$ the support of $\bm\beta_j$, and by $\mathfrak{s} = \max_{0 \le j \le q} \vert \mc S_j \vert$ the maximum segment-wise sparsity of the regression parameters.
We make the following assumptions on the size of change $\delta_j = \vert \bm{\beta}_j - \bm{\beta}_{j - 1} \vert_2$ and the spacing between the neighbouring change points.

\begin{assum}
\label{assum:bounded}  
There exists some constant $C_{\delta} > 0$
such that $\max_{1 \le j \le q} \delta_j \le C_{\delta}$.
\end{assum}

Assumption~\ref{assum:bounded} is a technical condition under which we focus on the more challenging regime where the size of change is allowed to tend to zero; 
an analogous condition is found in \cite{lee2016lasso}, \cite{kaul2019efficient}, \cite{wang2021statistically} and \cite{xu2022change}.
{It follows immediately if we assume that $\Var(Y_t) < \infty$, since
$\Var(Y_t) \ge \omega \sum_{j = 0}^q \vert \bm\beta_j \vert_2^2 \cdot \mathbb{I}_{\{\cp_j + 1 \le t \le \cp_j\}} + \sigma_\vep^2$.
Without Assumption~\ref{assum:bounded}, it incurs an extra multiplicative factor $\mathfrak{s}$ in the detection boundary (see Assumption~\ref{assum:bandwidth}~\ref{assum:bandwidth:b} below) and the rate of localisation, see also \cite{rinaldo2020localizing}.}

\begin{assum}
\label{assum:bandwidth}
The bandwidth $G$ fulfils the following conditions with $\tau$, $\rho_{n, p}$ and $\omega$ introduced in Assumptions~\ref{assum:xe}, \ref{assum:dev} and~\ref{assum:rsc}.
\begin{enumerate}[label = (\alph*)]
\item \label{assum:bandwidth:a} $2 G \le \min_{1 \le j \le q + 1} (\cp_j - \cp_{j - 1})$.
\item \label{assum:bandwidth:b} There exists a fixed constant $C_1 > 0$ such that
\begin{align*}
\min_{1 \le j \le q} \delta_j^2 G \ge C_1 
\max\l\{ \omega^{-2} \mathfrak{s} \rho_{n, p}^2, \l(\omega^{-1} \mathfrak{s} \log(p)\r)^{1/(1 - \tau)} \r\}.
\end{align*}
\end{enumerate} 
\end{assum} 

Assumption~\ref{assum:bandwidth}~\ref{assum:bandwidth:a} relates the choice of bandwidth $G$ to the minimum spacing between the change points.
Together, \ref{assum:bandwidth:a} and~\ref{assum:bandwidth:b}
specify the {\it separation rate} imposing a lower bound on
\begin{align}
\label{eq:sep:rate:one}
\Delta^{(1)} = \min_{1 \le j \le q} \delta_j^2 \cdot \min_{0 \le j \le q} (\cp_{j + 1} - \cp_j),
\end{align}
for all the $q$ change points to be detectable by MOSEG.
Later in Section~\ref{sec:multiscale}, we propose a multiscale extension of MOSEG which achieves consistency under a more relaxed condition than Assumption~\ref{assum:bandwidth}.

\begin{thm} 
\label{thm:one}
Suppose that Assumptions~\ref{assum:xe}, \ref{assum:dev}, \ref{assum:rsc}, \ref{assum:bounded} and~\ref{assum:bandwidth} hold.
Let the tuning parameters satisfy $\lambda \ge 4 \cdev \rho_{n, p}$, $r \in [1/G, 1/4)$, $\eta \in (4r, 1]$ and
\begin{align} \label{eq:threshold}
\frac{48 \sqrt{\mathfrak{s}} \lambda}{\omega} < D 
< \frac{\eta}{4\sqrt{2}} \min_{1 \le j \le q} \delta_j \sqrt{G}.
\end{align} 
Then conditional on $\mc D^{(1)} \cap \mc D^{(2)} \cap \mc R^{(1)} \cap \mc R^{(2)}$,
the following holds.
\begin{enumerate}[label = (\roman*)]
\item \label{thm:one:one}
Stage~1 of MOSEG returns $\wt{\Cp} = \{\wt\cp_j, \, 1 \le j \le \wh q: \, \wt\cp_1 < \ldots < \wt\cp_{\wh q}\}$ which satisfies
\begin{align*} 
\wh{q} = q  \quad \text{and} \quad 
\vert \wt\cp_j - \cp_j \vert \le \frac{48\sqrt{2\mathfrak{s} G}\lambda}{\omega \delta_j}
+ \lfloor rG \rfloor < \l\lfloor \frac{G}{2} \r\rfloor
\text{ for each $j = 1, \ldots, q$.}
\end{align*}
\item \label{thm:one:two} 
There exists a large enough constant $c_0 > 0$ such that  Stage~2 of MOSEG returns $\wh{\Cp} = \{\wh \cp_j, \, 1 \le j \le \wh q: \, \wh \cp_1 < \ldots < \wh \cp_{\wh q}\}$ which satisfies 
\begin{align*}
\max_{1 \le j \le q} \delta_j^2 \vert \wh\cp_j  - \cp_j \vert
\le c_0 \max\l( \mathfrak{s} \rho_{n, p}^2, 
\l(\mathfrak{s} \log(p)\r)^{\frac{1}{1 - \tau}} \r).
\end{align*}
\end{enumerate}
\end{thm}

Theorem~\ref{thm:one}~\ref{thm:one:one} establishes that Stage~1 of MOSEG correctly 
estimates the number of change points as well as identifying their locations by the pre-estimators 
with some accuracy.
There is a trade-off between computational efficiency and theoretical consistency with respect to the choice of $r$.
On one hand, increasing $r$ leads to a coarser grid $\mc T$ with its cardinality $\vert \mc T \vert = O(n/(rG))$, and thus reduces the computational cost.
On the other, the pre-estimators lie in the grid such that the best approximation to each change point $\cp_j$ can be as far from $\cp_j$ as $\lfloor rG \rfloor / 2$, 
which is reflected on the localisation property of the pre-estimators. 

Theorem~\ref{thm:one}~\ref{thm:one:two} derives the rate of estimation for the second-stage estimators $\wh\cp_j$ which shows that the location estimation is more challenging when the size of change $\delta_j$ is small. 
Note that we always have $\max_{1 \le j \le q} \delta_j^{-2} \max( \mathfrak{s} \rho_{n, p}^2, (\mathfrak{s} \log(p))^{1/(1 - \tau)} ) \lesssim G \lesssim \min_{1 \le j \le q + 1} (\cp_j - \cp_{j - 1})$ under Assumption~\ref{assum:bounded}.
{In Section~\ref{sec:linear}, we consider two scenarios permitting temporal dependence and non-Gaussianity on $(\mbf x_t, \vep_t)$, and give concrete rates in place of $\rho_{n, p}$ and $\tau$. In particular, as shown in Corollary~\ref{cor:one}~\ref{cor:one:two}, the rate derived in Theorem~\ref{thm:one}~\ref{thm:one:two} is near-minimax optimal under Gaussianity.}

\subsection{Verification of Assumptions~\ref{assum:dev} and~\ref{assum:rsc}}
\label{sec:linear}

Assumptions~\ref{assum:dev} and~\ref{assum:rsc} generalise the deviation bound and restricted eigenvalue conditions which are often found in the high-dimensional $M$-estimation literature, to accommodate change points, serial dependence and heavy-tailedness.
Condition~\ref{assum:lin} gives instances of $\{(\mbf x_t, \vep_t)\}_{t = 1}^n$ that fulfil Assumptions~\ref{assum:dev} and~\ref{assum:rsc} and specify the corresponding $\rho_{n, p}$ and $\tau$.

\begin{cond}
\label{assum:lin} 
Suppose that for i.i.d.\ random vectors 
$\bm\xi_t = (\xi_{1t}, \ldots, \xi_{p + 1, t})^\top \in \R^{p + 1}, \, t \in \Z$, 
with $\E(\bm\xi_t) = \mbf 0$ and $\Cov(\bm\xi_t) = \mbf I$,
we have
\begin{align}
\label{eq:wold}
\bmx \mbf x_t \\ \vep_t \emx &= \sum_{\ell = 0}^\infty \mbf D_\ell \bm\xi_{t - \ell} 
\quad \text{with} \quad
\mbf D_\ell = [D_{\ell, ik}, \, 1 \le i, k \le p + 1] \in \R^{(p + 1) \times (p + 1)}
\end{align}
subject to $\E(\mbf x_t \vep_t) = \mbf 0$.
Further, there exist constants $\Xi > 0$ and $\varsigma > 2$ such that 
\begin{align}
\label{eq:polynomial}
& \vert D_{\ell, ik} \vert \le C_{ik} (1 + \ell)^{-\varsigma} \quad \text{with} \quad
\max\l\{ \max_{1 \le k \le p + 1} \sum_{i = 1}^{p + 1} C_{ik}, \, 
\max_{1 \le i \le p + 1} \sum_{k = 1}^{p + 1} C_{ik}
\r\} \le \Xi
\end{align}
for all $\ell \ge 0$.
Finally, we impose {\it either} of the two conditions on $\xi_{it}$.
\begin{enumerate}[label = (\alph*)]
\item \label{cond:exp} There exist some constants $C_\xi > 0$ and $\gamma \in (0, 2]$ such that 
$(\E(\vert \xi_{it} \vert^\nu))^{1/\nu} = \Vert \xi_{it} \Vert_\nu \le C_\xi \nu^\gamma$ for all $\nu \ge 1$.
In other words, $\Vert \xi_{it} \Vert_{\psi_\nu} := \sup_{\nu \ge 1} \nu^{-1/\gamma} \Vert \xi_{it} \Vert_\nu \le C_\xi$.

\item \label{cond:gauss} $\xi_{it} \sim_{\iid} \mc N(0, 1)$.
\end{enumerate}
\end{cond}

\begin{prop}
\label{prop:xe}
Suppose that Assumptions~\ref{assum:xe} and~\ref{assum:bounded} and Condition~\ref{assum:lin} hold. 
Then, there exist some constants $c_1, c_2 > 0$ such that
$\p(\mc D^{(1)} \cap \mc D^{(2)} \cap \mc R^{(1)} \cap \mc R^{(2)}) 
\ge 1 - c_1 (p \vee n)^{-c_2}$,
with $\omega = \Lambda_{\min}(\bm\Sigma_x)/2$, 
$\bar{\omega} = 3\Lambda_{\max}(\bm\Sigma_x)/2$,
and $\tau$ and $\rho_{n, p}$ chosen as below.
\begin{enumerate}[label = (\roman*)]
\item \label{prop:xe:one} Under Condition~\ref{assum:lin}~\ref{cond:exp},
we set $\tau = (4\gamma + 2) / (4\gamma + 3)$
and $\rho_{n, p} = \log^{2\gamma + 3/2}(p \vee n)$.

\item \label{prop:xe:two} Under Condition~\ref{assum:lin}~\ref{cond:gauss},
we set $\tau = 0$ and $\rho_{n, p} = \sqrt{\log(p \vee n)}$.
\end{enumerate}
\end{prop}

Under Condition~\ref{assum:lin}, $\{(\mbf x_t, \vep_t)\}_{t = 1}^n$ is a linear process with algebraically decaying serial dependence.
Also, Condition~\ref{assum:lin}~\ref{cond:exp} permits heavier tail behaviour than that allowed under sub-Gaussianity or sub-exponential distributions when $\gamma > 1/2$ and $\gamma > 1$, respectively.
\begin{rem}
\label{rem:lit}
In the change point literature, \cite{wang2022optimal} propose a change point test and investigate its properties under $\beta$-mixing, and \cite{xu2022change} analyse the $\ell_0$-penalised least squares estimation approach when the functional dependence of $\{\mbf x_t\}_{t = 1}^n$ and $\{\vep_t\}_{t = 1}^n$ decays exponentially.
Relaxing the Gaussianity, it is typically required that $\mbf x_t$ is a sub-Weibull random vector, i.e.\ $\sup_{\mbf a \in \mathbb{B}_2(1)} \Vert \mbf a^\top \mbf x_t \Vert_{\psi_\nu} < \infty$ for some $\gamma > 0$ (where $\mathbb{B}_d(r) = \{\mbf a: \, \vert \mbf a \vert_d \le r\}$) and similarly, $\Vert \vep_t \Vert_{\psi_\gamma} < \infty$.
Under these assumptions, the common approach is to verify the deviation bound and restricted eigenvalue conditions analogous to those made in Assumptions~\ref{assum:dev}--\ref{assum:rsc}, with which the uniform consistency of the local Lasso estimators is derived.
Instead, we explicitly state the meta-assumptions to highlight that the proposed MOSEG achieves consistency 
whenever Assumptions~\ref{assum:dev}--\ref{assum:rsc} are met.
These in turn can be shown to hold in settings beyond those considered in Condition~\ref{assum:xe} by using the arguments adopted in the literature establishing the consistency of Lasso-type estimator (when $q = 0$) under a variety of characterisations of serial dependence and non-Gaussianity \citep{wu2016performance, adamek2020lasso, han2020high, wong2020lasso}. 

\end{rem}

Corollary~\ref{cor:one} follows immediately from
Theorem~\ref{thm:one} and Proposition~\ref{prop:xe}.

\begin{cor}
\label{cor:one}
Suppose that Assumptions~\ref{assum:xe}, \ref{assum:bounded} and~\ref{assum:bandwidth} and Condition~\ref{assum:lin} hold, and $\lambda$, $r$, $\eta$ and $D$ are chosen as in Theorem~\ref{thm:one}.
Then, there exist constants $c_i > 0, \, i = 0, 1, 2$, such that $\wh{\Cp} = \{\wh \cp_j, \, 1 \le j \le \wh q: \, \wh \cp_1 < \ldots < \wh \cp_{\wh q}\}$
returned by MOSEG satisfies the following.
\begin{enumerate}[label = (\roman*)]
\item \label{cor:one:one} Under Condition~\ref{assum:lin}~\ref{cond:exp},
we have
\begin{align*}
\p\l(\wh q = q \text{ \ and \ } \max_{1 \le j \le q} \delta_j^2 \vert \wh \cp_j - \cp_j \vert 
\le c_0 \l(\mathfrak{s} \log(p \vee n) \r)^{4\gamma + 3} \r) \ge 1 - c_1 (p \vee n)^{-c_2}.
\end{align*}
\item \label{cor:one:two} Under Condition~\ref{assum:lin}~\ref{cond:gauss},
we have
\begin{align*}
\p\l(\wh q = q \text{ \ and \ } \max_{1 \le j \le q} \delta_j^2 \vert \wh \cp_j - \cp_j \vert 
\le c_0 \mathfrak{s} \log(p \vee n) \r) \ge 1 - c_1 (p \vee n)^{-c_2}.
\end{align*}
\end{enumerate}
\end{cor}

Corollary~\ref{cor:one}~\ref{cor:one:two} shows that under Gaussianity, the rate of localisation attained by MOSEG matches the minimax lower bound up to $\log(p \vee n)$, see Lemma~4 of \cite{rinaldo2020localizing}.
At the same time, Assumption~\ref{assum:bandwidth}~\ref{assum:bandwidth:b} translates to $\Delta^{(1)} \gtrsim \mathfrak{s}\log(p \vee n)$ in this setting,
nearly matching the minimax lower bound on the separation rate derived in Lemma~3 of \cite{rinaldo2020localizing} up to the logarithmic term.
{We refer to Appendix~\ref{sec:comparison} for comprehensive comparison between MOSEG, its multiscale extension to be introduced in Section~\ref{sec:multiscale}, and the existing methods for the change point detection problem under~\eqref{eq:model}, on their theoretical and computational properties.}


\section{Multiscale methodology}
\label{sec:multiscale}

The single-bandwidth methodology proposed in Section~\ref{sec:single} enjoys theoretical consistency as well as computational efficiency, but faces the difficulty arising from identifying a bandwidth that satisfies Assumption~\ref{assum:bandwidth}~\ref{assum:bandwidth:a}--\ref{assum:bandwidth:b} simultaneously.
In this section, we propose MOSEG.MS, a multiscale extension of MOSEG, and show that it achieves consistency in a parameter space broader than that allowed by Assumption~\ref{assum:bandwidth}.

\subsection{MOSEG.MS: Multiscale extension of MOSEG}
\label{sec:multiscale:alg}

Similarly to MOSEG, MOSEG.MS consists of moving window-based data scanning and location refinement steps but it takes a set of bandwidths as an input.
The key innovation lies in that for each change point, MOSEG.MS learns the bandwidth best-suited for its detection and localisation from the given set of bandwidths. 
While there exist multiscale extensions of moving sum procedures, they are mostly developed for univariate time series segmentation \citep{messer2014multiple, cho2021two} and to the best of our knowledge, this is a first attempt at rigorously studying such an extension in a high-dimensional setting.

Below we describe MOSEG.MS step-by-step. An algorithmic description of MOSEG.MS is given in Algorithm~\ref{algo:multiscale} of Appendix~\ref{app:alg}.

\paragraph{Step~1: Pre-estimator generation.}
Given a set of bandwidths $\mc G = \{G_h, \, 1 \le h \le H: \, G_1 < \ldots < G_H\}$, we generate the coarse grid associated with each $G_h$ and the parameter $r$ by $\mc T_h = \mc T(r, G_h)$, see~\eqref{eq:grid}.
As in Stage~1 of MOSEG, the sets of pre-estimators $\wt{\Cp}(G_h)$ are generated for $h = 1, \ldots, H$, and we denote by $\wt{\Cp}(\mc G) = \cup_{h = 1}^H \wt{\Cp}(G_h)$ the pooled set of all such pre-estimators.
By~\eqref{eq:local:max}, at each $\wt\cp \in \wt{\Cp}(G_h)$, we have $T_{\wt\cp}(G_h) > D$ and $\wt\cp = \arg\max_{k \in \mc I_\eta(\wt\cp) \cap \mc T_h} T_k(G_h)$, where $\mc I_\eta(\wt\cp) = \{\wt\cp - \lfloor \eta G_h \rfloor + 1, \ldots, \wt\cp + \lfloor \eta G_h \rfloor\}$ denotes the detection interval associated with $\wt\cp$.
For simplicity, we write $\mc I_1(\wt\cp) = \mc I(\wt\cp)$.
Below, we sometimes write $\wt\cp(G) \in \wt{\Cp}(G)$ to highlight that the pre-estimator is obtained with the bandwidth $G$, and denote by $G(\wt\cp)$ the bandwidth involved in the detection of a pre-estimator~$\wt\cp$.
If some $\wt\cp$ is detected with more than one bandwidths, we distinguish between them.

\paragraph{Step~2: Anchor estimator identification.}
Next, we identify {\it anchor} change point estimators $\wt\cp^A(G) \in \wt{\Cp}(\mc G)$ detected at some $G \in \mc G$ which satisfy
\begin{align}
\label{eq:alg:multiscale:anchor}
\bigcup_{h: \, G_h < G} \, \bigcup_{k \in \wt{\Cp}(G_h)} \l\{ \mc I(k) \cap \mc I(\wt\cp^A(G)) \r\} = \emptyset.
\end{align}
That is, each anchor change point estimator does not have its detection interval
overlap with the detection interval of any pre-estimator that is detected with a finer bandwidth.
Denote the set of all such anchor change point estimators by
$\wt{\Cp}^A = \{\wt\cp^A_j, \, 1 \le j \le \wh q: \, 
\wt\cp^A_1 < \ldots < \wt\cp^A_{\wh q}\}$,
with $\wh q = \vert \wt{\Cp}^A \vert$ 
as an estimator of the number of change points $q$.

\paragraph{Step~3: Pre-estimator clustering.}
We find subsets of the pre-estimators in $\wt{\Cp}(\mc G)$ denoted by $\mc C_j, \, j = 1, \ldots, \wh q$, as described below.
Initialised as $\mc C_j = \emptyset$, for each $j$, we add to $\mc C_j$ the $j$th anchor estimator $\wt\cp^A_j$ as well as all $\wt\cp \in \wt{\Cp}(\mc G)$ which simultaneously fulfil
\begin{align}
& \mc I(\wt\cp) \cap \mc I(\wt\cp^A_j) \ne \emptyset, \quad \text{and}
\nn \\
& \{\wt\cp - G(\wt\cp) - \lfloor G(\wt\cp)/2 \rfloor + 1, \ldots, \wt\cp + G(\wt\cp) + \lfloor G(\wt\cp)/2 \rfloor\} \cap \mc I(\wt\cp^A_{j^\prime}) = \emptyset
\text{ for all } j^\prime \ne j.
\label{eq:alg:multiscale:clustering}
\end{align}
{It is possible that some pre-estimators do not belong to any of $\mc C_j, \, 1 \le j \le \wh q$, but each cluster contains at least one estimator by construction.}

\paragraph{Step~4: Location refinement.} 
For each $\mc C_j, \, j = 1, \ldots, \wh q$, we denote the smallest and the largest bandwidths 
associated with the detection of the pre-estimators in $\mc C_j$, by $G^m_j$ and $G^M_j$, respectively, and 
the corresponding pre-estimators by $\wt\cp^m_j$ and $\wt\cp^M_j$
(when $\vert \mc C_j \vert = 1$, we have $\wt\cp^m_j = \wt\cp^M_j = \wh\cp^A_j$ and $G^m_j = G^M_j$).
Setting $G^*_j = \lfloor 3 G^m_j/4 + G^M_j/4 \rfloor$, 
we identify the local minimiser of the objective function
defined in~\eqref{eq:q}, as
\begin{align}
& \check \cp_j = {\arg\min}_{\wt\cp^m_j - G^*_j + 1 \le k \le \wt\cp^m_j + G^*_j}
Q\l(k; 
\wt\cp^m_j - G^*_j, \wt\cp^m_j + G^*_j, \wh{\bm\beta}^{\lft}_j, \wh{\bm\beta}^{\rgt}_j \r),
\label{eq:alg:multiscale:final}
\\
& \text{with} \quad
\wh{\bm\beta}^{\lft}_j = \wh{\bm\beta}_{(\wt\cp^m_j - G^m_j - G^*_j) \vee 0, \wt\cp^m_j - G^m_j}
\quad \text{and} \quad
\wh{\bm\beta}^{\rgt}_j = \wh{\bm\beta}_{\wt\cp^m_j + G^m_j, (\wt\cp^m_j + G^m_j + G^*_j) \wedge n}. \nn
\end{align}
Repeatedly performing~\eqref{eq:alg:multiscale:final} for $j = 1, \ldots, \wh q$, we obtain $\check{\Cp} = \{\check \cp_j, \, 1 \le j \le \wh q\}$.
\medskip

{
The steps of MOSEG.MS algorithm have been devised to (i)~group the change point estimators across multiple bandwidths, into those detecting the identical change points, and (ii)~adaptively learn the bandwidth best suited to locate each change point from the bandwidths associated with the estimators in each group. 
For~(i), we adopt the anchor estimators in Step~2 which closely resemble the final estimators produced by the bottom-up merging proposed in \cite{messer2014multiple} in the context of univariate data segmentation. 
While the merging procedure is known to achieve detection consistency, the resultant estimators do not come with a guaranteed rate of localisation. 
Nonetheless, they serve as an adequate `anchor' for clustering the pre-estimators in Step~3. 
Moreover, the restriction imposed in~\eqref{eq:alg:multiscale:clustering} ensures that the bandwidths associated with the detection of pre-estimators clustered in $\mc C_j$, inform us a good choice of bandwidth for the detection of the $j$-th change point, with which we perform location refinement in Step~4.}

\begin{rem}[Bandwidth generation]
\label{rem:bandwidths}
\cite{cho2021two} propose to use $\mc G$ generated as a sequence of Fibonacci numbers, for a multiscale extension of the moving sum procedure for univariate mean change point detection \citep{eichinger2018mosum}.
For some finest bandwidth $G_0 = G_1$, we iteratively produce $G_h, \, h \ge 2$, as $G_h = G_{h - 1} + G_{h - 2}$. 
Equivalently, we set $G_h = F_h G_0$ where $F_h = F_{h - 1} + F_{h - 2}$ with $F_0 = F_1 = 1$.
This is repeated until for some $H$, it holds that $G_H < \lfloor n/2 \rfloor$ while $G_{H + 1} \ge \lfloor n/2 \rfloor$. 
By induction, $F_h = O(((1 + \sqrt{2})/2)^h)$ such that the thus-generated bandwidth set $\mc G$ satisfies $\vert \mc G \vert = O(\log(n))$. 
\end{rem}

\subsection{Consistency of MOSEG.MS}
\label{sec:multiscale:theory}

{We make the following assumption on the size of changes by placing a condition on
\begin{align}
\label{eq:sep:rate:two}
\Delta^{(2)} = \min_{1 \le j \le q} \delta_j^2 \cdot \min(\cp_{j + 1} - \cp_j, \cp_j - \cp_{j - 1}).
\end{align}
\begin{massum}{\ref{assum:bandwidth}$^\prime$}
\label{assum:multiscale}
Let $\mc G$ denote the set of bandwidths generated as in Remark~\ref{rem:bandwidths}
with $G_1 \ge C_0 \max\{\rho_{n, p}^2, (\omega^{-1} \mathfrak{s} \log(p))^{1/(1 - \tau)} \}$.
Then, 
we assume that
\begin{align*}
\Delta^{(2)} \ge 32 C_1 \max\l\{
\omega^{-2} \mathfrak{s} \rho_{n, p}^2,
\l(\omega^{-1} \mathfrak{s} \log(p)\r)^{1/(1 - \tau)} \r\},   \end{align*}
with $C_1$ from Assumption~\ref{assum:bandwidth}.
\end{massum}
In essence, Assumption~\ref{assum:multiscale} relaxes Assumption~\ref{assum:bandwidth} by requiring that for each $\cp_j$, there exists one bandwidth $G_{(j)} \in \mc G$ fulfilling the requirements imposed on a single bandwidth in the latter for all $j = 1, \ldots, q$, see~\ref{assum:multiscale:a}--\ref{assum:multiscale:b} in Appendix~\ref{sec:pf:thm:multiscale} for further details.}
Compared to $\Delta^{(1)}$ defined in~\eqref{eq:sep:rate:one}, we always have $\Delta^{(1)} \le \Delta^{(2)}$ and, if frequent large changes and small changes over long stretches of stationarity are simultaneously present, the former can be considerably smaller than the latter \citep{cho2021data}.
To the best of our knowledge, Theorem~\ref{thm:multiscale} below provides a first result obtained under the larger parameter space defined with $\Delta^{(2)}$, in establishing the consistency of a data segmentation methodology for the problem in~\eqref{eq:model}.
We refer to Appendix~\ref{sec:comparison} for further discussions and comprehensive comparison between MOSEG, MOSEG.MS and competing methodologies on their theoretical properties.

\begin{thm} 
\label{thm:multiscale}
Suppose that Assumptions~\ref{assum:xe}, \ref{assum:dev}, \ref{assum:rsc}, \ref{assum:bounded} and~\ref{assum:multiscale} hold. 
Let the tuning parameters satisfy $\lambda \ge 4 \cdev \rho_{n, p}$, $r \in [G_1^{-1}, 1/4)$, $\eta \in (4r, 1]$ and
\begin{align}
\label{eq:multiscale:threshold:one}
\frac{48 \sqrt{\mathfrak{s}} \lambda}{\omega} < D 
< {\frac{\eta}{2} \sqrt{\Delta^{(2)}}.}
\end{align}
Then, there exists a constant $c_0 > 0$ such that conditional on $\mc D^{(1)} \cap \mc D^{(2)} \cap \mc R^{(1)} \cap \mc R^{(2)}$, 
MOSEG.MS returns
$\check{\Cp} = \{\check \cp_j, \, 1 \le j \le \wh q: \, \check \cp_1 < \ldots < \check \cp_{\wh q}\}$
which satisfies 
\begin{align*}
\wh q = q \quad \text{and} \quad
\max_{1 \le j \le q} \delta_j^2 \vert \check\cp_j  - \cp_j \vert
\le c_0 \max\l(\mathfrak{s} \rho^2_{n, p}, \l(\mathfrak{s}\log(p)\r)^{\frac{1}{1 - \tau}} \r).
\end{align*}
\end{thm}

\begin{cor}
\label{cor:multiscale}
Suppose that Assumptions~\ref{assum:xe}, \ref{assum:bounded} and~\ref{assum:multiscale} and Condition~\ref{assum:lin} hold, and $\lambda$, $r$ and $D$ are chosen
as in Theorem~\ref{thm:multiscale}.
Then, there exist constants $c_i > 0, \, i = 0, 1, 2$, such that
$\check{\Cp} = \{\check \cp_j, \, 1 \le j \le \wh q: \, \check \cp_1 < \ldots < \check \cp_{\wh q}\}$ returned by MOSEG.MS satisfies the following.
\begin{enumerate}[label = (\roman*)]
\item \label{cor:multiscale:one} Under Condition~\ref{assum:lin}~\ref{cond:exp},
we have
\begin{align*}
\p\l(\wh q = q \text{ \ and \ } \max_{1 \le j \le q} \delta_j^2 \vert \check \cp_j - \cp_j \vert 
\le c_0 \l(\mathfrak{s} \log(p \vee n) \r)^{4 + 3\gamma} \r) \ge 1 - c_1 (p \vee n)^{-c_2}.
\end{align*}
\item \label{cor:multiscale:two} Under Condition~\ref{assum:lin}~\ref{cond:gauss},
we have
\begin{align*}
\p\l(\wh q = q \text{ \ and \ } \max_{1 \le j \le q} \delta_j^2 \vert \check \cp_j - \cp_j \vert 
\le c_0 \mathfrak{s}\log(p \vee n) \r) \ge 1 - c_1 (p \vee n)^{-c_2}.
\end{align*}
\end{enumerate}
\end{cor}

\section{Numerical experiments}
\label{sec:sim}

\subsection{Choice of tuning parameters}
\label{sec:tuning}


We discuss the selection of tuning parameters involved in MOSEG and MOSEG.MS, namely the set of bandwidths $\mc G$, the grid $\mc T(r, G)$ in~\eqref{eq:grid}, $\eta \in (0, 1]$ involved in the pre-estimation of the change points (see~\eqref{eq:local:max}), the penalty parameter~$\lambda$ and the threshold~$D$.

\paragraph{Selection of $\mc G$.} {As described in Remark~\ref{rem:bandwidths}}, the set of bandwidths $\mc G$ is determined once the finest bandwidth $G_1$ is chosen.
{To gain insights about the minimum bandwidth required for the reasonable performance of the local Lasso estimators, we conducted numerical experiments by simulating datasets under~\eqref{eq:model} with $q = 0$, $\mbf x_t \sim_{\iid} \mc N_p(\mbf 0, \mbf I_p)$, $\vep_t \sim_{\iid} \mc N(0, 1)$ and $\bm\beta_0 = (\beta_{0, 1}, \ldots, \beta_{0, p})^\top$ where $\beta_{0, i}, \, 1 \le i \le \mathfrak{s}$, are sampled uniformly from $[-1, 1]$ while $\beta_{0, i} = 0, \, \mathfrak{s} + 1 \le i \le p$.
Generating $100$ realisations for each setting with varying $(n, p, \mathfrak{s}, G)$, we record the relative $\ell_2$-error $\max_{0 \le k \le n - G} \vert \bm\beta_0 \vert_2^{-1} \vert \wh{\bm\beta}_{k, k + G} - \bm\beta_0 \vert_2$ for each realisation. 
Then, we obtain a simple rule to determine the finest bandwidth as $G_1 = G_1(n, p) = \lfloor c^*_0 \exp(c^*_1 \log\log(n) + c^*_2 \log\log(p)) \rfloor$ with pre-determined constants $c^*_i, \, i = 0, 1, 2$, which are chosen as transforms of the estimated regression coefficients from regressing the $90\%$-percentile of the logarithm of the estimation errors over $100$ realisations, onto the corresponding $\log(G)$, $\log\log(p)$ and $\log\log(n)$ (with $R^2 = 0.8945$).} 
Adopting the Fibonacci sequence in Remark~\ref{rem:bandwidths} sometimes gives a sequence of bandwidths that grows too quickly when the sample size $n$ is small.
Therefore, with the finest bandwidth $G_1$ chosen as above, we recommend generating bandwidths as $G_h = \lfloor (h + 2)G_1/3 \rfloor$ for $h \ge 2$. 
Throughout the simulation studies and real data applications, we set $H = 3$.

\paragraph{Selection of $D$ and $\lambda$.} 
Theorems~\ref{thm:one} and~\ref{thm:multiscale} provide ranges of values for $\lambda$ and $D$ for theoretical consistency, but they involve unknown parameters as is typically the case in the change point literature. 
For their simultaneous selection, we adopt a cross validation (CV) method motivated by \cite{zou2020consistent}.
Let $\Lambda = \Lambda(G)$ denote the grid of possible values for~$\lambda$ which, dependent on the bandwidth $G$, is chosen as an exponentially increasing sequence from $10^{-3} \lambda_{\max}$ up to $\lambda_{\max}$ with $\lambda_{\max} = \max_{0 \leq k \leq n - G} \vert \sum_{t = k + 1}^{k + G} \mbf x_t Y_t \vert_\infty / \sqrt{G}$ the smallest value with which we obtain $\wh{\bm\beta}_{k, k + G} = \mbf 0$ for all $0 \le k \le n - G$.
For given $G \in \mc G$ and $\lambda \in \Lambda$, 
we generate $\wt{\Cp}(G, \lambda) = \{\wt\cp_j(G, \lambda), \, 1 \le j \le \wt q_0(G, \lambda)\}$, the set of pre-estimators with $D = 0$, i.e.\ we take all local maximisers of the MOSUM statistics according to~\eqref{eq:local:max};
due to the detection rule, we always have $\wt q_0(G, \lambda) \le n/(2 \eta G)$.
Sorting the elements of $\wt{\Cp}(G, \lambda)$ in the decreasing order of the associated MOSUM detector values,
we generate a sequence of nested change point models
\begin{align}
\emptyset = \wt{\Cp}_{[0]}(G, \lambda) \subset \wt{\Cp}_{[1]}(G, \lambda) \subset \ldots \subset \wt{\Cp}_{[\wt q_0(G, \lambda)]}(G, \lambda) = \wt{\Cp}(G, \lambda). \nn
\end{align}
Then, using the odd-indexed observations $(Y_t, \mbf x_t), \, t \in \mc J_1 = \{2t + 1, \, t = 0, \ldots, \lfloor (n - 1)/2 \rfloor\}$, we produce local estimators of the regression parameters and the even-indexed observations $(Y_t, \mbf x_t)$, $t \in \mc J_0 = \{1, \ldots, n\} \setminus \mc J_1$, is used for validation.
Specifically, we evaluate 
$\text{CV}(G, \lambda, m) = \text{RSS}_0(\wt{\Cp}_{[m]}(G, \lambda), \lambda)$, where for any $\mc L = \{\ell_j, \, 1 \le \ell_j \le L: \, 0 = \ell_0 < \ell_1 < \ldots < \ell_L < \ell_{L + 1} = n \}$,
\begin{align*}
\text{RSS}_0(\mc L, \lambda) &= \sum_{j = 0}^{L} \sum_{\substack{t \in \mc J_0 \cap \\ \{\ell_j + 1, \ldots, \ell_{j + 1}\}}} \l(Y_t - \mbf x_t^\top \wh{\bm\beta}^{(1)}_j(\mc L, \lambda) \r)^2.
\end{align*}
Here, $\wh{\bm\beta}^{(1)}_j(\mc L, \lambda)$ denotes the Lasso estimator obtained using $(Y_t, \mbf x_t), \, t \in \mc J_1 \cap \{\ell_j, \ldots, \ell_{j + 1}\}$ with the penalty parameter $\lambda$.
Then for each $G_h \in \mc G$, we find
\begin{align}
(\lambda^*, m^*) = {\arg\min}_{\substack{(\lambda, m): \, \lambda \in \Lambda, \\ 0 \le m \le \wt q_0(G_h, \lambda)}} \text{CV}(G_h, \lambda, m)
\nn
\end{align}
and obtain the set of pre-estimators $\wt{\Cp}(G_h) = \wt{\Cp}_{[m^*]}(G_h, \lambda^*)$ using $\lambda^*$ and $m^*$. 
This amounts to selecting the bandwidth-dependent threshold $D$ at a value just below the $m^*$th largest MOSUM detector value.
Such $\wt{\Cp}(G_h), \, G_h \in \mc G$, serve as an input to Steps~2--4 of MOSEG.MS.
In all numerical experiments reported in this paper, we set $\vert \Lambda \vert = 5$.

\paragraph{Selection of other tuning parameters.} 
For change point estimation, we recommend to use $\eta = 0.5$ in~\eqref{eq:local:max} based on extensive simulations, which show that the performance of MOSEG and MOSEG.MS is not too sensitive to its choice.
As noted in Section~\ref{sec:runtime}, MOSEG.MS is highly competitive computationally against the existing methods even without adopting a coarse grid. 
Therefore, we report the results obtained with $r = G^{-1}$ (i.e.\ $\mc T =\{G, \ldots, n - G\}$ in~\eqref{eq:grid}) in the main text and provide the results obtained with {coarser grids in Appendix~\ref{sec:sim:grid}, where we observe that adopting a coarse grid does not undermine the performance of MOSEG provided that $r$ is sufficiently small, say $r \le 1/5$.}

\subsection{Computational complexity and run time} 
\label{sec:runtime}

Let $\text{Lasso}(p)$ denote the cost of solving a Lasso problem with $p$ variables.
For the coordinate descent algorithm \citep{friedman2009glmnet}, each complete iteration of the coordinate descent has the cost $O(p^2)$.
Then, the combined computational cost of Stages~1 and~2 of MOSEG is $O(n (rG)^{-1} \text{Lasso}(p))$, and the memory cost is $O(np)$.
Similarly, with the set of bandwidths generated as described in Remark~\ref{rem:bandwidths}, the complexity of the multiscale extension MOSEG.MS is $O(n (rG_1)^{-1} \text{Lasso}(p))$ with $G_1$ denoting the finest scale, which follows from that $\sum_{h = 1}^H n/(r G_h) \le n / (r G_1) \sum_{h = 1}^\infty F_h^{-1} = O(n (r G_1)^{-1})$ (see Remark~\ref{rem:bandwidths} for the notations).
For the CV outlined in Section~\ref{sec:tuning}, we generate pre-estimators and evaluate the CV objective function on a sequence of nested models for each $\lambda \in \Lambda$, which brings the computational complexity of the complete MOSEG.MS methodology to $O( \vert \Lambda \vert n (rG_1)^{-1} \text{Lasso}(p))$.

We investigate the run time of change point detection methodologies for the problem in~\eqref{eq:model}. 
MOSEG (with $G = \lfloor n / 6 \rfloor$) and MOSEG.MS are applied with the tuning parameters chosen as in Section~\ref{sec:tuning} and the finest grid (i.e.\ $\mc T = \{G, \ldots, n - G\}$); we include the CV procedure in run time.
For comparison, we consider VPWBS \citep{wang2021statistically}, DPDU \citep{xu2022change} and ARBSEG \citep{kaul2019detection} applied with the recommended tuning parameters.
In particular, the default implementations of VPWBS and DPDU adopt a grid of size $3$ and $4$, respectively, for the Lasso tuning parameter while MOSEG and MOSEG.MS are applied with the grid $\Lambda$ of size $\vert \Lambda \vert = 5$.
We generate the data under the model~\ref{sim:three} described in Section~\ref{sec:sim:model} below, with $\delta = 1.6$ and varying $(n, p)$.

Figure~\ref{fig:runtime} reports the average execution time (in seconds) over $100$ realisations for each setting, for the five methods in consideration. 
Both MOSEG and MOSEG.MS take only a fraction of time taken by the competing methodologies in their computation even when we do not use the coarser grid for Stage~1, and their run time does not vary much with increasing $n$ or $p$ in the ranges considered.
As expected, MOSEG is faster than MOSEG.MS but the difference in execution time is much smaller than that between MOSEG.MS and other competitors.

\subsection{Simulation settings}
\label{sec:sim:model}

We apply MOSEG.MS to datasets simulated with varying $(n, p, \mathfrak{s})$ and change point configurations. 
In each setting, we generate $\mbf x_t$ as i.i.d.\ Gaussian random vectors with mean $\mbf 0$ and the covariance matrix $\bm\Sigma_x$ which are specified below, and $\vep_t \sim_{\iid} \mc N(0, \sigma_\vep^2)$; unless specified otherwise, we use $\sigma_\vep = 1$.
We report the results from non-Gaussian and serially dependent data in Appendix~\ref{sec:sim:heavy} where overall, the results are not sensitive to tail behaviour or temporal dependence.
Additionally, we report the results when $p = 1000$ in Appendix~\ref{sec:sim:hd} which, together with the experiments reported in Section~\ref{sec:runtime}, demonstrate the scalability of MOSEG.MS.

The models~\ref{sim:one}--\ref{sim:three} below are taken from \cite{wang2021statistically}; in~\ref{sim:two} where non-diagonal $\bm\Sigma_x$ is considered, we adapt their model by randomly generating the set $\mc S$ on each realisation and in~\ref{sim:three}, we consider a broader range of values for $\delta$.
In what follows, we assume that for given $\mc S \subset \{1, \ldots, p\}$ with $\vert \mc S \vert = \mathfrak{s}$, the parameter vector $\bm\beta_0 = (\beta_{0, 1}, \ldots, \beta_{0, p})^\top$ has $\beta_{0, i} \ne 0$ for $i \in \mc S$ and $\beta_{0, i} = 0$ otherwise, i.e.\ $\mc S$ is the support of $\bm\beta_0$.
For each setting, we generate $100$ realisations.

\begin{enumerate}[label = (M\arabic*)]
\item \label{sim:one} 
Setting $p = 100$, $q = 3$ and $\bm\Sigma_x = \mbf I$, we vary 
$n \in \{480, 560, 640, 720, 800\}$ and the change points are at $\cp_j = jn/4$, $j = 1, 2, 3$. 
Fixing $\mc S = \{1, \ldots, \mathfrak{s}\}$ with $\mathfrak{s}=4$, we set $\beta_{0, i} = 0.4 \cdot (-1)^{i - 1}$ for $i \in \mc S$ and $\bm\beta_j = (-1)^j \cdot \bm\beta_0$.

\item \label{sim:two} We set $n = 300$, $p = 100$ and $q=2$, and $\bm\Sigma_x = [0.6^{\vert i - i' \vert}]_{i, i' = 1}^p$.
The change points are at $\cp_j = jn/3$, $j = 1, 2$, and we vary $\mathfrak{s} \in \{10, 20, 30\}$.
For each realisation, we randomly draw $\mc S \subset \{1, \ldots, p\}$ of size $\mathfrak{s}$, and set $\beta_{0, i} = 1/\sqrt{4\mathfrak{s}}$ for $i \in \mc S$ and $\bm\beta_j = (- 1)^j \cdot \bm\beta_0$.

\item \label{sim:three} 
We have $n = 300$, $p = 100$, $q = 2$, $\mathfrak{s} = 10$ and $\bm\Sigma_x = [0.6^{\vert i - i' \vert}]_{i, i' = 1}^p$.
The change points are at $\cp_j= jn/3$ and fixing $\mc S = \{1, \ldots, \mathfrak{s}\}$, we set $\beta_{0, i} = \delta \cdot (-1)^{i - 1}$ for $i \in \mc S$ with varying $\delta \in \{0.2, 0.4, 0.8, 1.6\}/\sqrt{\mathfrak{s}}$,
and $\bm\beta_j = (-1)^j \cdot \bm\beta_0$. 

\item \label{sim:four}
We set $n = 840$, $p = 50$, $q = 5$, $\mathfrak{s} = 10$ and $\bm\Sigma_x = \mbf I$. 
The change points are at $\cp_1 = 60$, $\cp_2 = 120$, $\cp_3 = 240$, $\cp_4 = 360$ and $\cp_5 = 600$ and fixing $\mc S = \{1, \ldots, \mathfrak{s}\}$, we set $\beta_{0, i} = \delta \cdot (-1)^{i - 1}$ for $i \in \mc S$ with varying $\delta \in \{0.2, 0.4, 0.8, 1.6\}/\sqrt{\mathfrak{s}}$, 
and $\bm\beta_1 = - \bm\beta_2 =  - 2 \bm\beta_0$, $\bm\beta_3 = -\bm\beta_4 = - \sqrt{2} \bm\beta_0$ and $\bm\beta_5 = - \bm\beta_6 = - \bm\beta_0$.

\item \label{sim:zero}
The data are generated as in~\ref{sim:three} except for that $q = 0$, $\bm\Sigma_x = [10^2 \cdot 0.6^{\vert i - i' \vert}]_{i, i' = 1}^p$ and $\sigma_\vep = 10$, and we use $\delta \in \{1, 1.2, 1.4, 1.6\}$. 
\end{enumerate} 

In setting~\ref{sim:four}, the change points are multiscale in the sense that 
the size of change and spacing between the change points vary, but $\delta_j^2 \cdot \min(\cp_{j + 1} - \cp_j, \cp_j - \cp_{j - 1})$ is kept constant for $j = 1, 3, 5$ and for $j = 2, 4$, respectively.
This results in $\Delta^{(1)}$ being much smaller than $\Delta^{(2)}$, see Equations~\eqref{eq:sep:rate:one} and~\eqref{eq:sep:rate:two} for their definitions.
The setting~\ref{sim:zero} is designed to test the performance of data segmentation methods when $q = 0$, where we scale the data to examine the sensitivity of the tuning parameter choices discussed in Section~\ref{sec:tuning}.

\subsection{Simulation results}
\label{sec:sim:res}

We apply MOSEG.MS with the tuning parameters selected as described in Section~\ref{sec:tuning}. 
For the purpose of illustration only, we also apply MOSEG with the bandwidth chosen with the knowledge of the minimum spacing between the change points; for~\ref{sim:one}--\ref{sim:three} where change points are evenly spaced, we set $G = 3/4 \cdot \min_{0 \le j \le q} (\cp_{j + 1} - \cp_j)$. 
For~\ref{sim:four} with multiscale change points, there does not exist a single bandwidth that works well in detecting all change points so we simply set $G = 125$.
For~\ref{sim:zero} with $q = 0$, we set $G = G_1$ selected as described in Section~\ref{sec:tuning}.
{For comparison, we apply the methods proposed by \cite{wang2021statistically} (referred to as VPWBS) and \cite{xu2022change} (DPDU).
The VPWBS method learns the projections well-suited for the detection of the change points and applies the wild binary segmentation algorithm to the projected univariate series, and has been shown to outperform the methods proposed in \cite{leonardi2016computationally} and \cite{lee2016lasso}.
Based on dynamic programming, the DPDU algorithm minimises the $\ell_0$-penalised cost function for multiple change point detection. 
Both methods have been applied with the default tuning parameters recommended by the authors.
We also considered the method proposed by \cite{kaul2019detection} but omit the results due to its poor performance on the simulation models considered in this paper. 
}

In Tables~\ref{tab:sim:one}--\ref{tab:sim:four}, we report the distribution of the bias in change point number estimation ($\wh q - q$) for each method over the $100$ realisations generated under each setting.
Additionally, we report the scaled Hausdorff distance between the sets of estimated ($\wh\Cp$) and true ($\Cp)$ change points, i.e.\
\begin{align}
\label{eq:hausdorff}
d_H(\wh{\Cp}, \Cp) = 
\frac{1}{n} \max\l\{\max_{\wh{\cp} \in \wh{\Cp}} \min_{\cp \in \Cp} \vert \wh{\cp}-\cp \vert, 
\max_{\cp \in \Cp} \min _{\wh{\cp} \in \wh{\Cp}} \vert \wh{\cp} - \cp \vert \r\},
\end{align}
averaged over $100$ realisations; by convention, we set $d_H(\emptyset, \Cp) = 1$.
We remark that the Hausdorff distance tends to favour the cases when the change points are over-detected, than when they are under-detected. 
In Table~\ref{tab:sim:zero} (considering the case $q = 0$), we report the proportion of realisations where any false positive is returned.

Generally, as expected, we observe better performance from all methods with increasing sample size in~\ref{sim:one} or increasing change size with $\delta$ in~\ref{sim:three}--\ref{sim:four}
while varying the sparsity level $\mathfrak{s}$ brings in less clear patterns in the performance. 
In the presence of homogeneous change points under~\ref{sim:one}--\ref{sim:three}, MOSEG performs as well as MOSEG.MS in terms of correctly estimating the number of change points, but it suffers from the lack of adaptivity in the presence of multiscale change points under~\ref{sim:four} where both large frequent shifts and small changes over long intervals are present.
Here, we observe the benefit of the multiscale approach taken by MOSEG.MS particularly as $\delta$ grows, where it achieves better accuracy in detection and localisation against MOSEG. 
Comparing the performance of MOSEG.MS and VPWBS, we note that the former generally attains better detection power while the latter exhibits better localisation properties under~\ref{sim:two} and~\ref{sim:three} (when $\delta$ is large).
{DPDU tends to show good detection power in the more challenging scenarios, such as when $n$ is small (under~\ref{sim:one}), $\mathfrak{s}$ is large (under~\ref{sim:two}) or the change size is small (see Table~\ref{tab:heavy-dep}). At the same time, it is observed to over-estimate the number of change points across all scenarios.}

Under~\ref{sim:zero}, where no changes are present, our methods are shown to control the number of false positives well.
Here, we do not include VPWBS or DPDU in Table~\ref{tab:sim:zero} as they tend to detect false positives in most cases. 

\begin{table}[h!t!b!]
\caption{\ref{sim:one} Performance of MOSEG, MOSEG.MS, VPWBS and {DPDU} over $100$ realisations.
The best performer in each setting is denoted in bold.}
\label{tab:sim:one}
\centering
{\footnotesize
\begin{tabular}{cc ccccccc c}
\toprule
& & \multicolumn{7}{c}{$\wh q - q$} &  \\
$n$ & Method & $-3$ & $-2$ & $-1$ & 0 & 1 & 2 & $\geq$ 3 & $d_H$ \\
\cmidrule(lr){1-2}\cmidrule(lr){3-9}\cmidrule(lr){10-10} 
$480$ & MOSEG & 3 & 3 & 6 & 81 & 7 & 0 & 0 & 0.0852 \\
 & MOSEG.MS & 1 & 6 & 7 & \bf{84} & 2 & 0 & 0 & 0.0710 \\
 & VPWBS & 1 & 3 & 14 & 58 & 16 & 5 & 3 & 0.0795 \\
  & DPDU & 0 & 0 & 3 & 80 & 13 & 4 & 0 & \bf{0.0405}  \\
\cmidrule(lr){1-2}\cmidrule(lr){3-9}\cmidrule(lr){10-10} 
560 & MOSEG & 2 & 3 & 5 & 72 & 17 & 1 & 0 & 0.0742 \\
 & MOSEG.MS & 0 & 1 & 5 & \bf{93} & 1 & 0 & 0 & \bf{0.0299} \\
 & VPWBS & 1 & 0 & 10 & 73 & 5 & 8 & 3 & 0.0579 \\
 & DPDU & 3 & 0 & 1 & 79 & 15 & 2 & 1 & 0.0660 \\
\cmidrule(lr){1-2}\cmidrule(lr){3-9}\cmidrule(lr){10-10} 
640 & MOSEG & 1 & 3 & 5 & 64 & 23 & 4 & 0 & 0.0652 \\
 & MOSEG.MS & 0 & 1 & 2 & \bf{91} & 6 & 0 & 0 & \bf{0.0203} \\
 & VPWBS & 0 & 1 & 3 & 89 & 3 & 2 & 2 & 0.0291 \\
 & DPDU & 0 & 0 & 0 & 77 & 18 & 5 & 1 & 0.0344 \\
\cmidrule(lr){1-2}\cmidrule(lr){3-9}\cmidrule(lr){10-10} 
720 & MOSEG & 1 & 3 & 1 & 76 & 18 & 1 & 0 & 0.0433 \\
 & MOSEG.MS & 0 & 0 & 0 & \bf{97} & 3 & 0 & 0 & \bf{0.0104} \\
 & VPWBS & 0 & 0 & 1 & 92 & 3 & 3 & 1 & 0.0190 \\
 & DPDU & 1 & 0 & 0 & 75 & 22 & 2 & 0 & 0.0390 \\
\cmidrule(lr){1-2}\cmidrule(lr){3-9}\cmidrule(lr){10-10} 
800 & MOSEG & 2 & 3 & 7 & 61 & 25 & 2 & 0 & 0.0753 \\
 & MOSEG.MS & 0 & 0 & 0 & \bf{100} & 0 & 0 & 0 & \bf{0.0073} \\
 & VPWBS & 0 & 0 & 2 & 92 & 3 & 2 & 1 & 0.0202 \\
 & DPDU & 0 & 0 & 0 & 68 & 25 & 7 & 0 & 0.0385 \\
 \bottomrule
 \end{tabular}
}
\end{table}
 
\begin{table}[h!t!b!]
\caption{\ref{sim:two} Performance of MOSEG, MOSEG.MS, VPWBS and {DPDU} over $100$ realisations.
The best performer in each setting is denoted in bold.}
\label{tab:sim:two}
\centering
{\footnotesize
\begin{tabular}{cc cccccc c}
\toprule
& & \multicolumn{6}{c}{$\wh q - q$} &  \\
$\mathfrak{s}$ & Method & $-2$ & $-1$ & 0 & 1 & 2 & $\geq$ 3 & $d_H$ \\
\cmidrule(lr){1-2}\cmidrule(lr){3-8}\cmidrule(lr){9-9}
10 & MOSEG & 27 & 28 & 35 & 10 & 0 & 0 & 0.4204 \\
 & MOSEG.MS & 11 & 32 & \bf{44} & 13 & 0 & 0 & 0.3117 \\
 & VPWBS & 45 & 17 & 11 & 9 & 15 & 3 & \bf{0.2465} \\
 & DPDU & 55 & 5 & 33 & 7 & 0 & 0 & 0.5981 \\
\cmidrule(lr){1-2}\cmidrule(lr){3-8}\cmidrule(lr){9-9}
20 & MOSEG & 14 & 31 & \bf{50} & 5 & 0 & 0 & 0.3016 \\
 & MOSEG.MS & 8 & 32 & 48 & 12 & 0 & 0 & 0.2726 \\
 & VPWBS & 44 & 13 & 18 & 13 & 9 & 3 & \bf{0.2302} \\
 & DPDU & 49 & 3 & 45 & 3 & 0 & 0 & 0.5300 \\
\cmidrule(lr){1-2}\cmidrule(lr){3-8}\cmidrule(lr){9-9}
30 & MOSEG & 14 & 25 & {50} & 11 & 0 & 0 & 0.2765 \\
 & MOSEG.MS & 11 & 30 & 41 & 18 & 0 & 0 & 0.2848 \\
 & VPWBS & 24 & 20 & 33 & 9 & 9 & 5 & \bf{0.1843}\\
 & DPDU & 26 & 9 & \bf{51} & 14 & 0 & 0 & 0.3294 \\
 \bottomrule
 \end{tabular}
}
\end{table}

\begin{table}[h!t!b!]
\caption{\ref{sim:three} Performance of MOSEG, MOSEG.MS, VPWBS and {DPDU} over $100$ realisations.
The best performer in each setting is denoted in bold.}
\label{tab:sim:three}
\centering
{\footnotesize
\begin{tabular}{cc cccccc c}
\toprule
& & \multicolumn{6}{c}{$\wh q-q$}& \\
$\sqrt{10}\delta$ & Method & $-2$ & $-1$ & 0 & 1 & 2 & $\geq$ 3 & $d_H$ \\
\cmidrule(lr){1-2}\cmidrule(lr){3-8}\cmidrule(lr){9-9} 
0.2 & MOSEG & 12 & 19 & 63 & 6 & 0 & 0 & 0.2367 \\
 & MOSEG.MS & 4 & 16 & \bf{64} & 15 & 1 & 0 & \bf{0.1609} \\
 & VPWBS & 77 & 9 & 5 & 6 & 1 & 2 & 0.3025 \\
 & DPDU & 4 & 4 & 36 & 24 & 17 & 15 & 0.1779 \\
 \cmidrule(lr){1-2}\cmidrule(lr){3-8}\cmidrule(lr){9-9} 
0.4 & MOSEG & 7 & 9 & \bf{81} & 3 & 0 & 0 & \bf{0.1401} \\
 & MOSEG.MS & 3 & 21 & 71 & 5 & 0 & 0 & 0.1488 \\
 & VPWBS & 53 & 20 & 12 & 8 & 4 & 3 & 0.2681 \\
 & DPDU & 0 & 0 & 21 & 23 & 34 & 22 & 0.1498 \\
 \cmidrule(lr){1-2}\cmidrule(lr){3-8}\cmidrule(lr){9-9} 
0.8 & MOSEG & 6 & 10 & \bf{82} & 2 & 0 & 0 & 0.1242 \\
 & MOSEG.MS & 2 & 14 & 77 & 6 & 1 & 0 & 0.1099 \\
 & VPWBS & 13 & 7 & 58 & 14 & 6 & 2 & \bf{0.1061} \\
 & DPDU & 0 & 0 & 11 & 29 & 19 & 41 & 0.1644 \\
 \cmidrule(lr){1-2}\cmidrule(lr){3-8}\cmidrule(lr){9-9} 
1.6 & MOSEG & 3 & 5 & \bf{91} & 1 & 0 & 0 & 0.0732 \\
 & MOSEG.MS & 0 & 10 & 88 & 2 & 0 & 0 & 0.0737 \\
 & VPWBS & 1 & 1 & 84 & 10 & 4 & 0 & \bf{0.0404} \\
 & DPDU & 0 & 0 & 14 & 12 & 25 & 49 & 0.1682 \\
 \bottomrule
 \end{tabular}
}
\end{table}

\begin{table}[h!t!b!]
\caption{\ref{sim:four} Performance of MOSEG, MOSEG.MS, VPWBS and {DPDU} over $100$ realisations.
The best performer in each setting is denoted in bold.}
\label{tab:sim:four}
\centering
{\footnotesize
\begin{tabular}{cc ccccccc c}
\toprule
& & \multicolumn{7}{c}{$\wh q-q$}&  \\
$\sqrt{10} \delta$ & Method & $-3$ & $-2$ & $-1$ & 0 & 1 & 2 & $\geq$ 3 & $d_H$ \\
\cmidrule(lr){1-2}\cmidrule(lr){3-9}\cmidrule(lr){10-10} 
0.2 & MOSEG & 45 & 6 & 4 & 5 & 21 & 9 & 10 & 0.4518 \\
 & MOSEG.MS & 17 & 9 & 19 & \bf{12} & 15 & 8 & 20 & \bf{0.2263} \\
 & VPWBS & 95 & 1 & 2 & 1 & 1 & 0 & 0 & 0.4073 \\
 & DPDU & 99 & 0 & 0 & 0 & 1 & 0 & 0 & 0.9578 \\
\cmidrule(lr){1-2}\cmidrule(lr){3-9}\cmidrule(lr){10-10} 
0.4 & MOSEG & 44 & 7 & 10 & 10 & 8 & 5 & 16 & 0.4347 \\
 & MOSEG.MS & 15 & 12 & 16 & \bf{17} & 7 & 16 & 17 & \bf{0.2015} \\
 & VPWBS & 73 & 2 & 7 & 6 & 7 & 5 & 0 & 0.3247 \\
 & DPDU & 80 & 5 & 3 & 5 & 5 & 2 & 2 & 0.7317 \\
\cmidrule(lr){1-2}\cmidrule(lr){3-9}\cmidrule(lr){10-10} 
0.8 & MOSEG & 4 & 23 & 31 & 27 & 9 & 5 & 1 & 0.1978 \\
 & MOSEG.MS & 0 & 3 & 34 & {38} & 15 & 9 & 1 & {0.0834} \\
 & VPWBS & 13 & 40 & 29 & 11 & 4 & 2 & 1 & 0.1165 \\
 & DPDU & 0 & 0 & 0 & \bf{43} & 36 & 21 & 8 & \bf{0.0629} \\
\cmidrule(lr){1-2}\cmidrule(lr){3-9}\cmidrule(lr){10-10} 
1.6 & MOSEG & 0 & 7 & 45 & 43 & 3 & 2 & 0 & 0.0970 \\
 & MOSEG.MS & 0 & 1 & 32 & \bf{59} & 7 & 1 & 0 & \bf{0.0387} \\
 & VPWBS & 3 & 35 & 38 & 19 & 1 & 3 & 1 & 0.0900\\
 & DPDU & 0 & 0 & 0 & 54 & 32 & 14 & 5 & 0.0444 \\
 \bottomrule
 \end{tabular}
}
\end{table}
  
\begin{table}[h!t!b!]
\caption{\ref{sim:zero} Proportions of detecting false positives when $q = 0$ for MOSEG and MOSEG.MS over $100$ realisations.
}
\label{tab:sim:zero}
\centering
{\footnotesize
\begin{tabular}{c cccc}
\toprule
& \multicolumn{4}{c}{$\delta$} \\
Method & 1 & 1.2 & 1.4 & 1.6 \\
\cmidrule(lr){1-1} \cmidrule(lr){2-5}
MOSEG & 0.05 & 0.01 & 0.01 & 0.02 \\
MOSEG.MS& 0.04 & 0.01 & 0.01 & 0.02 \\
\bottomrule
\end{tabular}}
\end{table}

\section{Real data application}
\label{sec:real}

There exists an extensive literature on the prediction of the equity premium,
which is defined as the difference between the compounded return on the S\&P 500 index and the three month Treasury bill rate.
Using $14$ macroeconomic and financial variables (see Table~\ref{tab:eqprem_data} for full descriptions),
\cite{welch2008comprehensive} demonstrate the difficulty of this prediction problem, in part due to the time-varying nature of the data.
\cite{koo2016high} note that the majority of the variables are highly persistent with strong, positive autocorrelations, and develop an $\ell_1$-penalised regression method that identifies co-integration relationships among the variables.
Accordingly, we transform the data by taking the first difference of any variable labelled as being persistent by \cite{koo2016high}, and scale each covariate series to have unit standard deviation.
With the thus-transformed variables, we propose to model the monthly equity premium observed from $1927$ to $2005$ as $Y_t$, with the $14$ variables at lags $1, 2, 3$ and $12$ as regressors $\mbf x_t$ via piecewise stationary linear regression; in total, we have $n = 936$ and $p = 57$ including the intercept.

\begin{figure}[h!t!]
\centering	
\includegraphics[width = .8\textwidth]{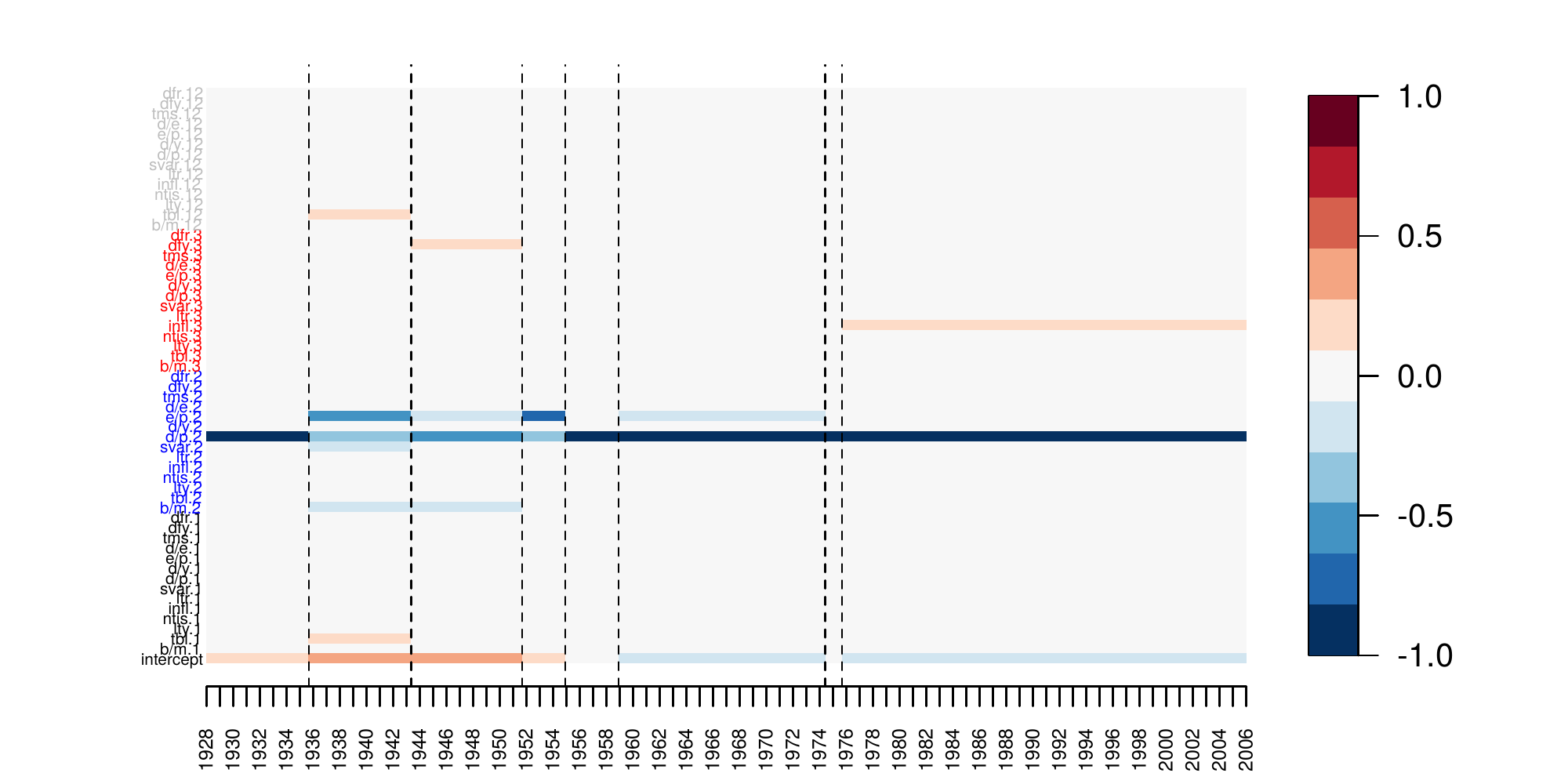}
\caption{Equity premium data: Parameter estimates from each estimated segment obtained by MOSEG.MS. Variables at different lags are coloured differently in the $y$-axis.}
\label{fig:equity}
\end{figure}

We apply MOSEG.MS with $\mc G = \{72, 96, 120\}$ in line with the choice described in Section~\ref{sec:tuning} but we select $G_h$ to be multiples of $12$ for interpretability as the observation frequency is monthly.
MOSEG.MS returns $\wh q = 7$ change point estimators reported in Table~\ref{tab:eq}, and takes $45$ seconds in total (including CV).
When applied to the same dataset, DPDU takes $25$ minutes and VPWBS takes $15$ minutes, and neither detects any change point. 
In Figure~\ref{fig:equity}, we plot the local parameter estimates obtained from each of the seven estimated segments.
We can relate the change detected in $1954$ to the findings reported in \cite{rapach2010out}, where they attribute the instability in the pairwise relationships between the equity premium and each of the $14$ variables to the Treasury-Federal Reserve Accord and the transition from the wartime economy.
Dividend price ratio (\verb+d/p+, at lag two) is active throughout the observation period which agrees with the observations made in \cite{welch2008comprehensive}.
They also remark that the recession from $1973$ to $1975$ due to the Oil Shock drives the good predictive performance of many models proposed for equity premium forecasting, and most perform poorly over the $30$ year period ($1975$--$2005$) following the Oil Shock. 
The two last segments defined by the change point estimators reported in Table~\ref{tab:eq} are closely located with these important periods, which supports the validity of the segmentation returned by MOSEG.MS.
We note that regardless of the choice of bandwidths, both of the two estimators in $1974$ and $1975$ defining the two periods are detected separately.

\begin{table}[h!t!]
\caption{Equity premium data: Change point estimators detected by MOSEG.MS.}
\label{tab:eq}
\centering
{\footnotesize
\begin{tabular}{c ccccccc} 
\toprule
Estimator & $\wh{\cp}_1$ & $\wh{\cp}_2$ &$\wh{\cp}_3$ &$\wh{\cp}_4$ &$\wh{\cp}_5$ &$\wh{\cp}_6$ &$\wh{\cp}_7$  \\ 
\cmidrule(lr){1-1} \cmidrule(lr){2-8}
Date & Oct 1935  & Apr 1943 &  Aug 1951  & Nov 1954  & Nov 1958  &
May 1974 &  Aug 1975 \\
\bottomrule
\end{tabular}
}
\end{table}

\section{Conclusions}
\label{sec:conc}

In this paper, we propose MOSEG, a high-dimensional data segmentation methodology for detecting multiple changes in the parameters under a linear regression model.
It proceeds in two steps, first grid-based scanning of the data for large changes in local parameter estimators over a moving window,
followed by a computationally efficient location refinement step.
We further propose its multiscale extension, MOSEG.MS, which alleviates the necessity to select a single bandwidth.
Both numerically and theoretically, we demonstrate the efficiency of  the proposed methodologies.
Computationally, they are highly competitive thanks to the careful design of the algorithms that limit the required number of Lasso estimators.
Theoretically, we show the consistency of MOSEG and MOSEG.MS in a general setting permitting serial dependence and heavy tails and establish their (near-)minimax optimality under Gaussianity. 
In particular, the consistency of MOSEG.MS is derived for a parameter space that simultaneously permits large changes over short intervals and small changes over long stretches of stationarity, which is much broader than that typically adopted in the literature.
Comparative simulation studies and findings from the application of MOSEG.MS to equity premium data support its efficacy.
The R software implementing MOSEG and MOSEG.MS is available from \url{https://github.com/Dom-Owens-UoB/moseg}.




\bibliographystyle{apalike}
\bibliography{ref}

\clearpage

\appendix 

\numberwithin{equation}{section}
\numberwithin{figure}{section}
\numberwithin{table}{section}
\numberwithin{thm}{section}

\section{Literature review and comparison with the existing methods}
\label{sec:comparison}

\subsection{{Under (sub-)Gaussianity}}

Table~\ref{tab:comparison} provides an overview of the theoretical properties of MOSEG and MOSEG.MS in comparison with the methods proposed in \cite{wang2021statistically}, \cite{kaul2019detection} and \cite{xu2022change} for the change point problem in~\eqref{eq:model} under Gaussianity, as well as their computational complexity.

Specifically, Table~\ref{tab:comparison} reports the {\it separation} and {\it localisation} rates associated with each method, which are defined as below.
For a given methodology, let $\wh{\mc K}$ denote the set of estimated change points.
Then, when the magnitude of change $\Delta$, measured by either 
\begin{align*}
\Delta^{(1)} = \min_{1 \le j \le q} \delta_j^2 \cdot \min_{0 \le j \le q} (\cp_{j + 1} - \cp_j) \text{ \ or \ } 
\Delta^{(2)} = \min_{1 \le j \le q} \delta_j^2 \min(\cp_j - \cp_{j - 1}, \cp_{j + 1} - \cp_j),
\end{align*}
diverges faster than the separation rate $s_{n, p}$ associated with the method, all $q$ changes are detected by $\wh{\mc K}$ with asymptotic power one and their locations are consistently estimated with the localisation rate $\ell_{n, p}$, such that $\min_{1 \le j \le q} \min_{\wh k \in \wh{\mc K}} w_j \vert \wh k - \cp_j \vert = O_P(\ell_{n, p})$.
Here, $w_j$ refers to the relative difficulty in locating $\cp_j$ which is related to the jump size $\delta_j$.

\begin{table}[!h!t!b]
\caption{Comparison of data segmentation methods developed for the model~\eqref{eq:model}
in their theoretical properties under Gaussianity and computational complexity (for given tuning parameters). 
Here, $\mathfrak{s} = \max_{0 \le j \le q} \vert \mc S_j \vert$ and
$\mathfrak{S} = \vert \cup_{j = 0}^{q} \mc S_j \vert$.
Refer to the text for the definitions of $s_{n, p}$, $\ell_{n, p}$, $\Delta$ and $w_j$.
{For \cite{xu2022change}, we report the rates associated with their preliminary estimators; see the text for further explanation.}
}
\label{tab:comparison}
\centering
\resizebox{\columnwidth}{!}{\footnotesize
\begin{tabular}{l cc cc c}
\toprule
& \multicolumn{2}{c}{Separation} & \multicolumn{2}{c}{Localisation} & Computational \\
& $s_{n, p}$ & $\Delta$ & $\ell_{n, p}$ & $w_j$ &  complexity 
\\
\cmidrule(lr){1-1} \cmidrule(lr){2-3} \cmidrule(lr){4-5} \cmidrule(lr){6-6}
MOSEG  & $\mathfrak{s} \log(p \vee n)$ & $\Delta^{(1)}$ & $\mathfrak{s} \log(p \vee n)$ & $\delta_j$ & $O(\frac{n}{rG} \cdot \text{Lasso}(p))$  
\\ 
MOSEG.MS  & $\mathfrak{s} \log(p \vee n)$ & $\Delta^{(2)}$ & $\mathfrak{s} \log(p \vee n)$
& $\delta_j$ & $O(\frac{n}{rG_1} \cdot \text{Lasso}(p))$    \\ 
\cmidrule(lr){1-1} \cmidrule(lr){2-3} \cmidrule(lr){4-5} \cmidrule(lr){6-6}
\cite{wang2021statistically}  & $\mathfrak{s} \log(p \vee n)$ & $\Delta^{(1)}$ & $\mathfrak{s} \log(n)$ & $\delta_j$ & 
$O(n \log^2(n) \cdot \text{GroupLasso}(p))$ 
\\ 
\cite{kaul2019detection}  & $\mathfrak{S} \log(p \vee n)$ & $\Delta^{(1)}$ & $\mathfrak{S} \log(p)$ & $\delta$ &
$O( \tilde{q} \cdot \text{Lasso}(p) + \text{SA}(\tilde{q}))$
\\
\cite{xu2022change} & $\mathfrak{s} \log(p \vee n)$ & $\Delta^{(1)}$ & $\mathfrak{s} \log(p \vee n)$ & $\delta_j$ & $O(n^2p^2 + n^2 \cdot \text{Lasso}(p))$ 
\\
\bottomrule
\end{tabular}}
\end{table}

\cite{wang2021statistically} propose a method which learns the projection that is well-suited to reveal a change over each local segment and combines it with the wild binary segmentation algorithm \citep{fryzlewicz2014wild} for multiple change point detection.
\cite{kaul2019detection} propose to minimise an $\ell_0$-penalised cost function given a set of candidate estimators of size $\tilde{q}$.
Their theoretical analysis implicitly assumes that $\min_j (\cp_{j + 1} - \cp_j)$ scales linearly in $n$, and the simulated annealing adopted for minimising the penalised cost, denoted by SA($\tilde{q}$) in Table~\ref{tab:comparison},
has complexity ranging from $O(\tilde{q}^4)$ on average to being exponential in the worst case. 
In specifying the properties of \cite{kaul2019detection}, the global sparsity $\mathfrak{S} = \vert \cup_{j = 0}^{q} \mc S_j \vert$ can be much greater than the segment-wise sparsity $\mathfrak{s}$, particularly when the number of change points $q$ is large.

{\cite{xu2022change} investigate the dynamic programming algorithm of \cite{rinaldo2020localizing} for minimising an $\ell_0$-penalised cost function in a general setting permitting non-Gaussianity and temporal dependence, as is the case in the current paper. 
In Table~\ref{tab:comparison}, we report the separation and localisation rates derived in \cite{xu2022change} for their {\it preliminary} estimators from the dynamic programming algorithm.
These estimators are further refined by a procedure similar to Stage~2 of MOSEG, which are shown to attain the rate $O_P(\delta_j^{-2})$ under a stronger condition on the size of changes, namely that $\Delta^{(1)}/(\mathfrak{s}^2 \log^3(pn)) \to \infty$.
We note that the refined rate is derived for the individual change points, rather than when the estimation of multiple change points is simultaneously considered as is the case for the rates reported in Table~\ref{tab:comparison}.}

We also mention \cite{zhang2015change} where the data segmentation problem is treated as a high-dimensional regression problem with a group Lasso penalty, which only provides that the estimation bias is of $o_P(n)$.
\cite{leonardi2016computationally} consider both dynamic programming and binary segmentation algorithms are considered for change point estimation, and we refer to \cite{rinaldo2020localizing} for a detailed discussion on their results.

From Table~\ref{tab:comparison}, we conclude that MOSEG.MS is highly competitive both computationally and statistically.
We investigate the theoretical properties of MOSEG.MS in the broadest parameter space possible which is formulated with $\Delta^{(2)}$ instead of $\Delta^{(1)}$ as in all the other papers; recall that from the discussion following~\eqref{eq:sep:rate:two}, we always have $\Delta^{(1)} \le \Delta^{(2)}$ and the former can be much smaller than the latter when large shifts over short intervals and small changes over long stretches of stationarity are simultaneously present in the signal.

\subsection{{Beyond (sub-)Gaussianity and independence}}

The theoretical properties of MOSEG.MS reported in Table~\ref{tab:comparison} do not require independence or Gaussianity, unlike other works with the exception of \cite{xu2022change}.
In the presence of serial dependence and sub-Weibull tails (through having $\gamma > 1$ as in Condition~\ref{assum:lin}~\ref{cond:exp}), \cite{xu2022change} require that $\Delta^{(1)} \gtrsim (\mathfrak{s}\log(np))^{4\gamma + 2\gamma^\prime -1}$ for the detection of all change points, where a smaller value of $\gamma^\prime \in (0, \infty)$ imposes a faster decay of the serial dependence.
This is comparable to the detection boundary of MOSEG, $\Delta^{(1)} \gtrsim (\mathfrak{s}\log(np))^{4\gamma + 3}$, see Proposition~\ref{prop:xe}~\ref{prop:xe:one} and Corollary~\ref{cor:one}~\ref{cor:one:one}.
In characterising temporal dependence, Condition~\ref{assum:lin} considers linear processes with algebraically decaying dependence whereas $\gamma^\prime$ of \cite{xu2022change} governs the rate of exponentially decaying, possibly non-linear dependence.
The localisation rate in Corollary~\ref{cor:one}~\ref{cor:one:one} is also comparable to that attained by the preliminary estimators of \cite{xu2022change} produced by a dynamic programming algorithm; as noted above, under a stronger condition on $\Delta^{(1)}$, they derive a further refined rate.

\clearpage

\section{Additional simulations}
\label{sec:add:sim}

\subsection{Choice of the grid}
\label{sec:sim:grid}

\begin{figure}[h!t!b!]
\centering
\includegraphics[width = 0.6\textwidth]{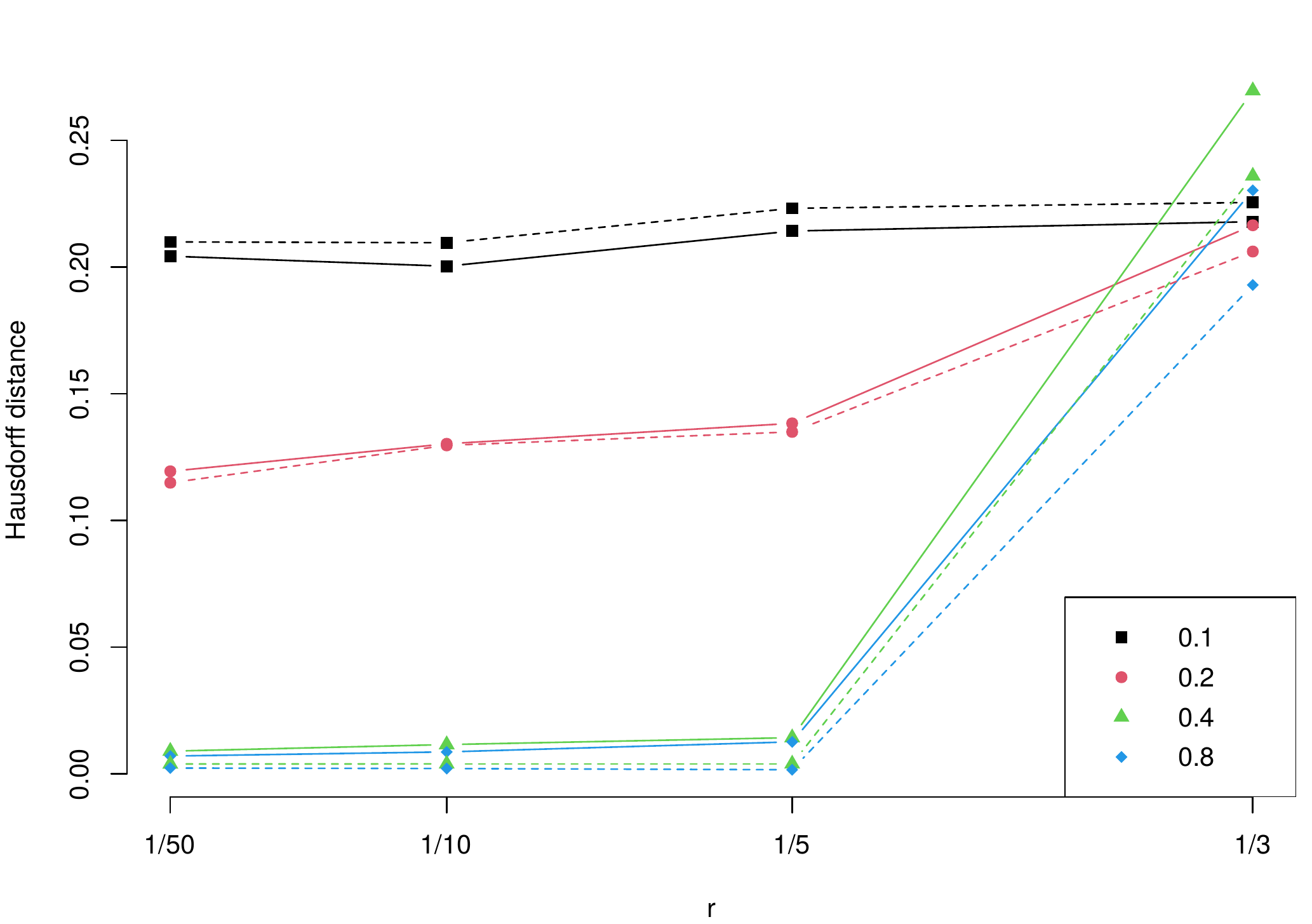}
\caption{Hausdorff distance $d_H$ against $r$ for Stage~1 (solid line) and Stage~2 (dashed line) estimators from MOSEG, as the size of changes varies.}
\label{fig:r_v_haus}
\end{figure}

{We investigate the performance of MOSEG as the coarseness of the grid varies with $r \in \{1/3, 1/5, 1/10, 1/G\}$. 
Recall that when $r = 1/G$, we use the full grid $\mc T = \{G, \ldots, n - G\}$ in Stage~1 of MOSEG, see~\eqref{eq:grid}.}
For this, we set $n = 300$, $p = 100$, $\mathfrak{s} = 2$ and $q = 1$, and generate the data under~\eqref{eq:model} with $\mbf x_t \sim_{\iid} \mc N_p(\mbf 0, \mbf I)$ and $\vep_t \sim_{\iid} \mc N(0, 1)$.
For each realisation, the change point $\cp_1$ is randomly sampled from $\{51, \dots, 250 \}$. 
Varying $\delta \in \{0.1, 0.2, 0.4, 0.8\}$, we generate $\bm\beta_0 = (\beta_{0, 1}, \ldots, \beta_{0, p})^\top$ with $\beta_{0, i} = \delta \cdot (- 1)^{i - 1}$ for $i \in \{1, \ldots, \mathfrak{s}\}$ and have $\bm\beta_1 = - \bm\beta_0$.
Setting $G = 50$, we select the maximiser of the MOSUM statistic as the pre-estimator $\wt\cp_1$ in Stage~1 of MOSEG, which then is refined as in~\eqref{eq:refined} in Stage~2.
Table~\ref{tab:sim:grid} reports the average and the standard error of $n^{-1}\vert \wt\cp_1 - \cp_1 \vert$ and $n^{-1}\vert \wh\cp_1 - \cp_1 \vert$ over $100$ realisations when different grids are used. 
{See also Figure~\ref{fig:r_v_haus} which plots the Hausdorff distance~$d_H$ (see~\eqref{eq:hausdorff}) against~$r$.
When the size of change is very small, estimators from both Stages~1 and~2 perform equally poorly regardless of the choice of $r$. 
However, as $\delta$ increases, we quickly observe that the estimation error becomes close to zero for the estimators from both stages provided that $r$ is not too large.
Also, for $\delta \ge 0.2$, we observe that Stage~2 brings in small improvement in the localisation performance.
From this, we conclude that the performance of MOSEG is robust to the choice of $r$ provided that it is chosen reasonably small, say $r \le 1/5$.}

\begin{table}[htb]
\caption{Comparison of Hausdorff distance $d_H$ for Stage~1 and Stage~2 estimators from MOSEG when different grids are used. The average and the standard error of estimation errors over $100$ realisations are reported.}
\label{tab:sim:grid}
\centering
{\small
\begin{tabular}{c cccc cccc}
\toprule
 & \multicolumn{4}{c}{$r = G^{-1}$} & \multicolumn{4}{c}{$r = 1/10$}  \\
\cmidrule(lr){2-5} \cmidrule(lr){6-9}
 & \multicolumn{2}{c}{Stage 1} &  \multicolumn{2}{c}{Stage 2} & \multicolumn{2}{c}{Stage 1} &  \multicolumn{2}{c}{Stage 2} \\
$\delta$ & Mean & SD  & Mean & SD & Mean & SD & Mean & SD \\
\cmidrule(lr){1-1} \cmidrule(lr){2-5} \cmidrule(lr){6-9}
0.1 & 0.2043 & 0.1561 & 0.2099 & 0.1670 & 0.2003 & 0.1525 & 0.2096 & 0.1683 \\
0.2 & 0.1194 & 0.1367 & 0.1149 & 0.1442 & 0.1302 & 0.1402 & 0.1296 & 0.1549 \\
0.4 & 0.0089 & 0.0104 & 0.0038 & 0.0053 & 0.0115 & 0.0179 & 0.0039 & 0.0050 \\
0.8 & 0.0070 & 0.0089 & 0.0022 & 0.0052 & 0.0086 & 0.0096 & 0.0020 & 0.0049\\
\midrule
 & \multicolumn{4}{c}{$r = 1/5$} & \multicolumn{4}{c}{$r = 1/3$}  \\
\cmidrule(lr){2-5} \cmidrule(lr){6-9}
 & \multicolumn{2}{c}{Stage 1} &  \multicolumn{2}{c}{Stage 2} & \multicolumn{2}{c}{Stage 1} &  \multicolumn{2}{c}{Stage 2} \\
$\delta$ & Mean & SD  & Mean & SD & Mean & SD & Mean & SD \\
 \cmidrule(lr){1-1} \cmidrule(lr){2-5} \cmidrule(lr){6-9}
0.1 & 0.2142 & 0.1525 & 0.2232 & 0.1683 & 0.2179 & 0.1525 & 0.2255 & 0.1683 \\
0.2 & 0.1383 & 0.1402 & 0.1349 & 0.1549 & 0.2165 & 0.1402 & 0.2061 & 0.1549 \\
0.4 & 0.0142 & 0.0179 & 0.0039 & 0.0050 & 0.2696 & 0.0179 & 0.2359 & 0.0050 \\
0.8 & 0.0126 & 0.0096 & 0.0016 & 0.0049 & 0.2303 & 0.0096 & 0.1929 & 0.0049\\
\bottomrule
\end{tabular}}
\end{table}

\subsection{Heavy-tailedness and temporal dependence}
\label{sec:sim:heavy}

We examine the performance of MOSEG.MS, VPWBS \citep{wang2021statistically} {and DPDU \citep{xu2022change}} in the presence of heavy-tailed noise and temporal dependence.
For this, we generate datasets with $n = 300$, $p = 100$, $\mathfrak{s} = 10$ and $q = 2$ where the two change points are located at $\cp_j = jn/3$, $j = 1, 2$.
We use $\bm\beta_0$ generated as in Appendix~\ref{sec:sim:grid} with $\delta \in \{0.2, 0.4, 0.8, 1.6\}$ and set $\bm\beta_j = (- 1)^j \cdot \bm\beta_0$.

We consider the following three settings for the generation of $\mbf x_t$ and $\vep_t$.
\begin{enumerate}[label = (E\arabic*)]
\item \label{e:gauss} $\mbf x_t \sim_{\iid} \mc N_p(\mbf 0, \mbf I)$ and $\vep_t \sim_{\iid} \mc N(0, 1)$ for all $t$. 

\item \label{e:heavy} $X_{it} \sim_{\iid} \sqrt{3/5} \cdot t_5$ for all $i$ and $t$
and $\vep_t \sim_{\iid} \sqrt{3/5} \cdot t_5$ for all $t$.

\item \label{e:timeseries} $\{(\mbf x_t, \vep_t)\}_{t = 1}^n$ is generated as in~\eqref{eq:wold} where $\mbf D_1$ is a diagonal matrix with $0.3$ on its diagonals, $\mbf D_\ell = \mbf O$ for $\ell \ge 2$ and $\bm\zeta_t \sim_{\iid} \mc N_{p + 1}(\mbf 0, \sqrt{1-0.3^2}\mbf I)$ for all $t$. 
\end{enumerate}

Under~\ref{e:heavy}--\ref{e:timeseries}, the data are permitted to be heavy-tailed and serially correlated, respectively; \ref{e:gauss} serves as a benchmark.
Table~\ref{tab:heavy-dep} reports the average and standard error of the Hausdorff distance in~\eqref{eq:hausdorff} and $\wh q - q$ over $100$ realisations.
{It shows that generally, the performance of all three methods is less sensitive to heavy-tailedness or temporal dependence, compared to their sensitivity to the size of changes.}
VPWBS shows good localisation performance, while MOSEG.MS tends to achieve better detection accuracy when the size of change is small.
{DPDU performs very well in the more challenging setting with small $\delta$. However, as noted in Section~\ref{sec:sim}, it is more prone to over-estimate the number of change points as $\delta$ increases with which the localisation performance also deteriorates.}

\begin{table}[h!t!b!]
\caption{Performance of MOSEG.MS and VPWBS under~\ref{e:gauss}--\ref{e:timeseries} over $100$ realisations.
The best performer in each setting is denoted in bold.}
\label{tab:heavy-dep}
\centering
{\footnotesize
\begin{tabular}{ccc cccccc c}
\toprule
& & & \multicolumn{6}{c}{$\wh q-q$}&   \\
$\delta$ & Setting & Method & $-2$ & $-1$ & 0 & 1 & 2 & $\geq$ 3 & $d_H$ \\
\cmidrule(lr){1-3}\cmidrule(lr){4-9}\cmidrule(lr){10-10} 
0.2 & \ref{e:gauss} & MOSEG.MS & 10 & 35 & {46} & 9 & 0 & 0 & 0.2905 \\
 &  & VPWBS & 56 & 10 & 9 & 14 & 9 & 2 & {0.2583} \\
 &  & DPDU & 0 & 0 & \bf{86} & 12 & 1 & 1 & \bf{0.0239} \\
\cmidrule(lr){2-3}\cmidrule(lr){4-9}\cmidrule(lr){10-10} 
 & \ref{e:heavy} & MOSEG.MS & 6 & 53 & {34} & 6 & 1 & 0 & 0.2779 \\
 &  & VPWBS & 60 & 10 & 7 & 13 & 10 & 0 & {0.2663} \\
 &  & DPDU & 0 & 0 & \bf{89} & 11 & 0 & 0 & \bf{0.0168} \\
\cmidrule(lr){2-3}\cmidrule(lr){4-9}\cmidrule(lr){10-10} 
 & \ref{e:timeseries} & MOSEG.MS & 8 & 30 & {44} & 17 & 1 & 0 & 0.2644 \\
 &  & VPWBS & 10 & 15 & 18 & 13 & 28 & 16 & {0.1671} \\
 &  & DPDU & 0 & 0 & \bf{85} & 13 & 2 & 0 & \bf{0.0211} \\
\cmidrule(lr){1-3}\cmidrule(lr){4-9}\cmidrule(lr){10-10} 
0.4 & \ref{e:gauss} & MOSEG.MS & 0 & 9 & 86 & 4 & 1 & 0 & 0.0766 \\
 &  & VPWBS & 1 & 3 & \bf{87} & 7 & 2 & 0 & {0.0361} \\
 &  & DPDU & 0 & 0 & 82 & 16 & 1 & 1 & \bf{0.0240} \\
\cmidrule(lr){2-3}\cmidrule(lr){4-9}\cmidrule(lr){10-10} 
 & \ref{e:heavy} & MOSEG.MS & 1 & 11 & \bf{83} & 5 & 0 & 0 & 0.0830 \\
 &  & VPWBS & 1 & 7 & \bf{83} & 4 & 5 & 0 & {0.0567} \\
 &  & DPDU & 0 & 0 & 82 & 16 & 2 & 0 & \bf{0.0248} \\
\cmidrule(lr){2-3}\cmidrule(lr){4-9}\cmidrule(lr){10-10} 
 & \ref{e:timeseries} & MOSEG.MS & 1 & 9 & \bf{81} & 9 & 0 & 0 & 0.0678 \\
 &  & VPWBS & 0 & 1 & 80 & 14 & 4 & 1 & {0.0397} \\
 &  & DPDU & 0 & 0 & 71 & 24 & 5 & 0 & \bf{0.0359} \\
\cmidrule(lr){1-3}\cmidrule(lr){4-9}\cmidrule(lr){10-10} 
 0.8 & \ref{e:gauss} & MOSEG.MS & 0 & 0 & \bf{99} & 1 & 0 & 0 & 0.0119 \\
 &  & VPWBS & 0 & 0 & 97 & 3 & 0 & 0 & \bf{0.0095} \\
 &  & DPDU & 0 & 0 & 81 & 18 & 1 & 0 & 0.0227 \\
\cmidrule(lr){2-3}\cmidrule(lr){4-9}\cmidrule(lr){10-10} 
 & \ref{e:heavy} & MOSEG.MS & 0 & 0 & \bf{98} & 2 & 0 & 0 & \bf{0.0100} \\
 &  & VPWBS & 0 & 0 & \bf{98} & 2 & 0 & 0 & 0.0104 \\
 &  & DPDU & 0 & 0 & 80 & 17 & 1 & 2 & 0.0209 \\
\cmidrule(lr){2-3}\cmidrule(lr){4-9}\cmidrule(lr){10-10} 
 & \ref{e:timeseries} & MOSEG.MS & 0 & 1 & 96 & 3 & 0 & 0 & 0.0153 \\
 &  & VPWBS & 0 & 0 & \bf{98} & 2 & 0 & 0 & \bf{0.0103} \\
 &  & DPDU & 0 & 0 & 73 & 24 & 3 & 0 & 0.0285 \\
\cmidrule(lr){1-3}\cmidrule(lr){4-9}\cmidrule(lr){10-10} 
1.6 & \ref{e:gauss} & MOSEG.MS & 0 & 0 & 97 & 3 & 0 & 0 & 0.0097 \\
 &  & VPWBS & 0 & 0 & \bf{100} & 0 & 0 & 0 & \bf{0.0037} \\
 &  & DPDU & 0 & 0 & 75 & 22 & 1 & 2 & 0.0309 \\
\cmidrule(lr){2-3}\cmidrule(lr){4-9}\cmidrule(lr){10-10} 
 & \ref{e:heavy} & MOSEG.MS & 0 & 0 & \bf{100} & 0 & 0 & 0 & 0.0036 \\
 &  & VPWBS & 0 & 0 & \bf{100} & 0 & 0 & 0 & \bf{0.0033} \\
 &  & DPDU & 0 & 0 & 78 & 18 & 1 & 3 & 0.0249 \\
\cmidrule(lr){2-3}\cmidrule(lr){4-9}\cmidrule(lr){10-10} 
 & \ref{e:timeseries} & MOSEG.MS & 0 & 1 & 96 & 3 & 0 & 0 & 0.0076 \\
 &  & VPWBS & 0 & 0 & \bf{99} & 1 & 0 & 0 & \bf{0.0045}\\
 &  & DPDU & 0 & 0 & 69 & 24 & 6 & 1 & 0.0307 \\
 \bottomrule
 \end{tabular}
}
\end{table}

\subsection{When the dimensionality is large}
\label{sec:sim:hd}

We additionally examine the case where $p = 1000$, adopting the simulation setting~\ref{e:gauss} from Appendix~\ref{sec:sim:heavy}.
We exclude VPWBS \citep{wang2021statistically} and DPDU \citep{xu2022change} which, as shown in Section~\ref{sec:runtime}, tends to take considerably longer time to run compared to MOSEG.MS. 
Table~\ref{tab:p1000} shows that, in comparison to the the results under \ref{e:gauss} in Table~\ref{tab:heavy-dep} obtained when $p = 100$, the greater sample size is required to detect smaller changes. 
Also, the localisation performance worsens as $p$ increases.
Nonetheless, MOSEG.MS demonstrates itself to be scalable as the dimensionality increases when the size of change is sufficiently large, which is in line with the theoretical requirements.

\begin{table}[htb]
\caption{Performance of MOSEG.MS under~\ref{e:gauss} when $\mathbf{p = 1000}$ over $100$ realisations.}
\label{tab:p1000}
\centering
{\footnotesize
\begin{tabular}{c cccccc c}
\toprule
 & \multicolumn{6}{c}{$\wh q-q$}&  \\
$\delta$  & $-2$ & $-1$ & 0 & 1 & 2 & $\geq$ 3 & $d_H$  \\
\cmidrule(lr){1-1}\cmidrule(lr){2-7}\cmidrule(lr){8-8} 
0.2 & 6 & 47 & 29 & 18 & 0 & 0 & 0.3051 \\
0.4 & 10 & 34 & 44 & 11 & 1 & 0 & 0.2972 \\
0.8 & 1 & 22 & 65 & 11 & 1 & 0 & 0.1391 \\
1.6 & 4 & 3 & 92 & 1 & 0 & 0 & 0.0673 \\
 \bottomrule
 \end{tabular}
}
\end{table}

\clearpage

\section{Proofs}

In what follows, for any vector $\mbf a \in \R^p$ and
a set $\mc A \subset \{1, \ldots, p\}$,
we denote by $\mbf a(\mc A) = (a_i, \, i \in \mc A)^\top$ 
the sub-vector of $\mbf a$ supported on $\mc A$.
We write the population counterpart of $T_k(G)$ with $\bm\beta_{s, e}^*$
defined in~\eqref{eq:mixture} as
\begin{align*}
T^*_k(G) = \sqrt{\frac{G}{2}} 
\l\vert \bm{\beta}^*_{k, k+G} - \bm{\beta}^*_{k - G, k} \r\vert_2.
\end{align*}  
Further, we write $\mc S_{s, e} = \text{supp}(\bm\beta^*_{s, e})$.

\subsection{Proof of Theorem~\ref{thm:one}}

\subsubsection{Supporting lemmas}

\begin{lem} 
\label{lem:w:norm}
We have
\begin{align*}
T^*_k(G) =
\l\{\begin{array}{ll}
\frac{1}{\sqrt{2G}} (G - \vert k - \cp_j \vert ) \delta_j 
& \text{if } \{k - G + 1, \ldots, k + G\} \cap \Cp = \{\cp_j\}, \\
0& \text{if } \{k - G + 1, \ldots, k + G\} \cap \Cp = \emptyset \\
\end{array}\r. 
\end{align*}
\end{lem}

\begin{lem} 
\label{lem:lasso}
Define $\bm\Delta_{s, e} = \wh{\bm{\beta}}_{s, e} - \bm{\beta}^{*}_{s, e}$.
With $\lambda \ge 4 \cdev \rho_{n, p}$, 
we have $\p(\mc B) \ge 1 - \p(\mc R^{(1)} \cap \mc D^{(2)})$ where
\begin{align*}
\mc B &= \l\{
\l\vert \bm\Delta_{s, e} \r\vert_2
\le \frac{12 \sqrt{2\mathfrak{s}} \lambda}{\omega \sqrt{e - s}}
\text{ and } 
\l\vert \bm\Delta_{s, e}(\mc S_{s, e}^c) \r\vert_1 \le 
3 \l\vert \bm\Delta_{s, e}(\mc S_{s, e}) \r\vert_1
\text{ for all } 0 \le s < e \le n
\r.
\\
& \l. \text{ with } \vert \{s + 1, \ldots, e\} \cap \Cp \vert \le 1 \text{ and } 
e - s \ge C_0 \max\l[ (\omega^{-1} \mathfrak{s} \log(p))^{\frac{1}{1 - \tau}},
\rho^2_{n, p} \r] \r\}.
\end{align*}
\end{lem}

\begin{proof}
For given $0 \le s < e \le n$, 
we have
\begin{align*}
\sum_{t = s + 1}^e \l( Y_t - \mbf x_t^\top \wh{\bm\beta}_{s, e} \r)^2 + 
\lambda \sqrt{e - s} \vert \wh{\bm\beta}_{s, e} \vert_1 \le 
\sum_{t = s + 1}^e \l( Y_t - \mbf x_t^\top \bm\beta^*_{s, e} \r)^2 + 
\lambda \sqrt{e - s} \vert \bm\beta^*_{s, e} \vert_1,
\end{align*}
from which it follows that
\begin{align*}
\lambda \sqrt{e - s} \l( \vert \bm{\beta}^*_{s, e} \vert_1 - 
\vert \wh{\bm\beta}_{s, e} \vert_1 \r)
& \ge 
\sum_{t = s + 1}^e \l[ 
(\mbf{x}_t^\top\wh{\bm\beta}_{s, e})^2 -  (\mbf{x}_t^\top\bm{\beta}^*_{s, e})^2 - 
2 Y_t \mbf{x}_t^\top (\wh{\bm\beta}_{s, e} - \bm{\beta}^*_{s, e}) \r]
\\
& =
\sum_{t = s + 1}^e \l[\bm\Delta_{s, e}^\top \mbf x_t \mbf x_t^\top \bm\Delta_{s, e}
- 2 (Y_t - \mbf x_t^\top \bm{\beta}^*_{s, e}) \mbf x_t^\top \bm\Delta_{s, e} \r].
\end{align*}
Then, noting that $\bm\beta^*_{s, e}(\mc S_{s, e}^c) = \mbf 0$,
\begin{align} 
\frac{1}{\sqrt{e - s}} \sum_{t = s + 1}^e \l[\bm\Delta_{s, e}^\top \mbf x_t \mbf x_t^\top \bm\Delta_{s, e}
- 2 (Y_t - \mbf x_t^\top \bm{\beta}^*_{s, e}) \mbf x_t^\top \bm\Delta_{s, e} \r]
+ \lambda \l\vert \wh{\bm\beta}_{s, e} (\mc S_{s, e}^c) \r\vert_1 
\nn \\
\le \lambda \l( \l\vert \bm{\beta}^*_{s, e}(\mc S_{s, e}) \r\vert_1 
- \l\vert \wh{\bm\beta}_{s, e}(\mc S_{s, e}) \r\vert_1 \r)
\le \lambda \l\vert \bm{\Delta}_{s, e}(\mc S_{s, e}) \r\vert_1.
\label{eq:lem:lasso:one}
\end{align}
Since $\lambda \ge 4 \cdev \rho_{n, p}$, 
it follows from~\eqref{eq:lem:lasso:one} that on $\mc D^{(2)}$,
\begin{align*}
& \frac{1}{\sqrt{e - s}} \sum_{t = s + 1}^e \bm\Delta_{s, e}^\top \mbf x_t \mbf x_t^\top \bm\Delta_{s, e} - \frac{\lambda}{2} \l\vert \bm\Delta_{s, e} \r\vert_1
+ \lambda \l\vert \bm\Delta_{s, e}(\mc S_{s, e}^c) \r\vert_1
\le \lambda \l\vert \bm{\Delta}_{s, e}(\mc S_{s, e}) \r\vert_1,
\\
& \therefore \, 0 \le \frac{1}{\sqrt{e - s}} \sum_{t = s + 1}^e \bm\Delta_{s, e}^\top \mbf x_t \mbf x_t^\top \bm\Delta_{s, e}
\le \frac{\lambda}{2} \l(3 \l\vert \bm{\Delta}_{s, e}(\mc S_{s, e}) \r\vert_1 -
\l\vert \bm{\Delta}_{s, e}(\mc S_{s, e}^c) \r\vert_1 \r),
\end{align*}
such that 
\begin{align}
\label{eq:lem:lasso:two}
\l\vert \bm{\Delta}_{s, e}(\mc S_{s, e}^c) \r\vert_1 \le 
3 \l\vert \bm{\Delta}_{s, e}(\mc S_{s, e}) \r\vert_1.
\end{align}
This in particular leads to
\begin{align}
\nn
\l\vert \bm{\Delta}_{s, e} \r\vert_1 \le 
4 \l\vert \bm{\Delta}_{s, e}(\mc S_{s, e}) \r\vert_1
\le 4\sqrt{2\mathfrak{s}} \l\vert \bm{\Delta}_{s, e} \r\vert_2
\end{align}
from the definition of $\mathfrak{s}$. Then on $\mc R^{(1)}$, we have
\begin{align*}
6 \sqrt{2\mathfrak{s}} \lambda \l\vert \bm{\Delta}_{s, e} \r\vert_2 
&\ge
\frac{1}{\sqrt{e - s}} \sum_{t = s + 1}^e \bm\Delta_{s, e}^\top \mbf x_t \mbf x_t^\top \bm\Delta_{s, e} 
\\
&\ge
\omega \sqrt{e - s} \l\vert \bm{\Delta}_{s, e} \r\vert_2^2 -
\frac{32 \crsc \mathfrak{s} \log(p)(e - s)^\tau }{\sqrt{e - s}} \l\vert \bm{\Delta}_{s, e} \r\vert_2^2
\ge \frac{\omega}{2} \sqrt{e - s} \l\vert \bm{\Delta}_{s, e} \r\vert_2^2,
\end{align*}
where the last inequality follows for
$(e - s)^{1 - \tau} \ge 64 \crsc \omega^{-1} \mathfrak{s} \log(p)$.
In summary,
\begin{align}
\label{eq:lem:lasso:three}
\l\vert \bm{\Delta}_{s, e} \r\vert_2 \le \frac{12 \sqrt{2\mathfrak{s}} \lambda}{\omega \sqrt{e - s}}.
\end{align}
Combining~\eqref{eq:lem:lasso:two} and~\eqref{eq:lem:lasso:three},
the proof is complete.
\end{proof}

\subsubsection{Proof of Theorem~\ref{thm:one}~\ref{thm:one:one}}

Let $\mc T_j = \{\cp_j - \lfloor\eta  G\rfloor + 1, \ldots, \cp_j + \lfloor\eta  G\rfloor\} \cap \mc T$ for $1 \le j \le q$.
Under Assumptions~\ref{assum:bounded} and~\ref{assum:bandwidth},
we have $G \ge C_\delta^{-2} C_1 \max\{\omega^{-2} \mathfrak{s} \rho_{n, p}^2,
(\omega^{-1} \mathfrak{s} \log(p))^{1/(1 - \tau)} \}$ 
such that the lower bound on $(e - s)$ made in 
$\mc B$ (see Lemma~\ref{lem:lasso}) is met by all $s = k$ and $e = k + G$, $k = 0, \ldots, n - G$.
By Lemma~\ref{lem:lasso}, 
\begin{align}
& \max_{G \le k \le n - G} \l\vert T_k(G) - T^*_k(G) \r\vert \le 
\nn \\
& \max_{G \le k \le n - G} 
\sqrt{\frac{G}{2}} \l( \l\vert \wh{\bm\beta}_{k - G, k} - \bm\beta^*_{k - G, k} \r\vert_2
+ \l\vert \wh{\bm\beta}_{k, k + G} - \bm\beta^*_{k, k + G} \r\vert_2 \r)
\le \frac{24 \sqrt{\mathfrak{s}} \lambda}{\omega}.
\label{eq:thm:one:a:one}
\end{align}

First, consider some $k$ for which 
$\{ k - G + 2, \ldots, k + G - 1 \} \cap \Cp = \emptyset$.
Then, we have $T^*_k(G) = 0$ from Lemma~\ref{lem:w:norm} such that
by~\eqref{eq:thm:one:a:one},
\begin{align}
\label{eq:thm:one:a:zero}
\max_{k: \, \min_{1 \le j \le q} \vert k - \cp_j \vert \ge G}
T_k(G) \le \max_{G \le \ell \le n - G} \l\vert T_\ell(G) - T^*_\ell(G) \r\vert 
\le \frac{24 \sqrt{\mathfrak{s}} \lambda}{\omega} \le D.
\end{align}
This ensures that any $\wt\cp \in \wt{\Cp}$ satisfies
$\min_{1 \le j \le q} \vert \wt\cp - \cp_j \vert < G$.
Next, let $\cp^{\lft}_j$ and $\cp^{\rgt}_j$ 
denote two points within $\mc T_j$ which are the closest to $\cp_j$
from the left and and the right of $\cp_j$, respectively,
with $\cp^{\lft}_j = \cp^{\rgt}_j$ when $r = 1/G$.
Then by construction of $\mc T$,
\begin{align}
\label{eq:thm:one:a:two}
\max(k_j - \cp^{\lft}_j, \cp^{\rgt}_j - \cp_j) \le \lfloor rG \rfloor 
\quad \text{and} \quad
\min(\cp_j - \cp^{\lft}_j, \cp^{\rgt}_j - \cp_j) \le \frac{\lfloor rG \rfloor}{2},
\end{align}
such that from Lemma~\ref{lem:w:norm},
\begin{align*}
\max\l( T^*_{\cp^{\lft}_j}(G), T^*_{\cp^{\rgt}_j}(G) \r) 
\ge \frac{\delta_j (G - \lfloor rG \rfloor/2)}{\sqrt{2G}} 
\ge \sqrt{\frac{G}{2}} \delta_j (1 - r/2).
\end{align*}
From this and by~\eqref{eq:thm:one:a:one},
at $\wt\cp_j = \arg\max_{k \in \mc T_j} T_k(G)$, we have
\begin{align*}
T_{\wt\cp_j}(G) \ge 
\max\l( T_{\cp^{\lft}_j}(G), T_{\cp^{\rgt}_j}(G) \r) 
\ge \sqrt{\frac{G}{2}} \delta_j \l( 1 - \frac{r}{2} \r) - 
\frac{24 \sqrt{\mathfrak{s}} \lambda}{\omega}
> \frac{1 - r/2}{2} \sqrt{\frac{G}{2}} \delta_j > D, 
\end{align*}  
where the second last inequality follows from
Assumption~\ref{assum:bandwidth}~\ref{assum:bandwidth:b},
and the last one from~\eqref{eq:threshold}.
When $\eta = 1$, this and~\eqref{eq:thm:one:a:zero} indicates that
such $\wt\cp_j$ satisfies~\eqref{eq:local:max}.
When $\eta < 1$, note that
\begin{align*}
& \max\l( T_{\cp^{\lft}_j}(G), T_{\cp^{\rgt}_j}(G) \r) - 
\max\l\{ T_k(G): \, \vert k - \cp_j \vert > (1 - \eta) G, \, k \in \mc T \r\}
\\
& \ge \sqrt{\frac{G}{2}} \delta_j \l( \eta - \frac{3r}{2} \r) - \frac{48 \sqrt{\mathfrak{s}} \lambda}{\omega} 
\ge
\frac{5\eta}{8\sqrt{2}} \min_{1 \le j \le q} \delta_j \sqrt{G} - \frac{48 \sqrt{\mathfrak{s}} \lambda}{\omega} > 0
\end{align*}
from~\eqref{eq:threshold}.
These arguments ensure that we detect at least one change point in $\mc T_j$ at $t = \wt\cp_j$ for each $j = 1, \ldots, q$.
For such $\wt\cp_j$, suppose that $\cp^\circ_j = {\arg\min}_{k \in \{ \cp^{\lft}_j, \cp^{\rgt}_j \}} \vert \wt\cp_j - k \vert$.
Then,
\begin{align*}
\frac{\delta_j}{\sqrt{2G}} (G - \vert \wt\cp_j - \cp_j \vert) + \frac{24 \sqrt{\mathfrak{s}} \lambda}{\omega}
\ge T_{\wt\cp_j}(G) \ge T_{\cp^\circ_j}(G)
\ge \frac{\delta_j}{\sqrt{2G}} (G - \vert \cp^\circ_j - \cp_j \vert) -
\frac{24 \sqrt{\mathfrak{s}} \lambda}{\omega}
\end{align*}
and re-arranging, we obtain
\begin{align*}
\frac{\delta_j}{\sqrt{2G}} \l(\vert \wt\cp_j - \cp_j \vert - \vert \cp^\circ_j - \cp_j \vert \r) \le \frac{48 \sqrt{\mathfrak{s}} \lambda}{\omega}, 
\text{ such that }
\vert \wt\cp_j - \cp_j \vert \le \frac{48 \sqrt{2\mathfrak{s} G}\lambda}{\omega \delta_j}
+ \lfloor rG \rfloor < \l\lfloor \frac{G}{2} \r\rfloor,
\end{align*}
for large enough $C_1$ in Assumption~\ref{assum:bandwidth}~\ref{assum:bandwidth:b}.

Finally, let $\mc L_{\mc T}(t)$ denote the largest time point $k^\prime \in \mc T$
that satisfies $k^\prime \le t$, and define $\mc R_{\mc T}(t)$ analogously.
Then, we establish that
\begin{align}
T_{\mc L_{\mc T}(\cp_j - \frac{\eta G}{2} m)}(G)
> \max\l\{ T_k(G): \, \frac{\eta G}{2} (m + 1) \le \cp_j - k \le \frac{\eta G}{2} (m + 2), \, k \in \mc T \r\},
\label{eq:thm:one:lft} \\
T_{\mc R_{\mc T}(\cp_j + \frac{\eta G}{2} m)}(G)
> \max\l\{ T_k(G): \, \frac{\eta G}{2} (m + 1) \le k - \cp_j \le \frac{\eta G}{2} (m + 2), \, k \in \mc T \r\},
\label{eq:thm:one:rgt}
\end{align}
for $m = 0, \ldots, \lceil 2/\eta \rceil - 2$.
The inequality in~\eqref{eq:thm:one:lft} follows from noting that
\begin{align*}
& T_{\mc L_{\mc T}(\cp_j - \frac{\eta G}{2} m)}(G)
- \max\l\{ T_k(G): \, \frac{\eta G}{2} (m + 1) \le \cp_j - k \le \frac{\eta G}{2} (m + 2), \, k \in \mc T \r\} \\
& \ge \sqrt{\frac{G}{2}} \delta_j \l(\frac{\eta}{2} - r \r) - \frac{48\sqrt{\mathfrak{s}} \lambda}{\omega} \ge
\frac{\eta}{4\sqrt{2}} \min_{1 \le j \le q} \delta_j \sqrt{G} - \frac{48 \sqrt{\mathfrak{s}} \lambda}{\omega} > 0
\end{align*}
under~\eqref{eq:threshold}, and the inequality in~\eqref{eq:thm:one:rgt} follows analogously. This ensures that $\wt\cp_j$ by its construction is the unique local maximiser 
of $T_k(G)$ within the interval $\{\cp_j - G + 1, \ldots, \cp_j + G\} \cap \mc T$ satisfying~\eqref{eq:local:max} for each $j = 1, \ldots, q$, which completes the proof.


\subsubsection{Proof of Theorem~\ref{thm:one}~\ref{thm:one:two}}

Recalling~\eqref{eq:q}, we write
\begin{align*}
Q_j(k)
= \sum_{t = \wt\cp_j - G + 1}^k (Y_t - \mbf x_t^\top \wh{\bm\beta}^{\lft}_j)^2
+ 
\sum_{t = k + 1}^{\wt\cp_j + G} (Y_t - \mbf x_t^\top \wh{\bm\beta}^{\rgt}_j)^2.
\end{align*}
Theorem~\ref{thm:one}~\ref{thm:one:one} establishes that
for each $j = 1, \ldots, q$, we have $\wt\cp_j \in \wt{\Cp}$ that satisfies
$\vert \wt\cp_j - \cp_j \vert < G/2$, and $\wt{\Cp}$ contains no other estimator.
Then under Assumption~\ref{assum:bandwidth}~\ref{assum:bandwidth:a}, 
we have the following statements satisfied for all $j$.
\begin{enumerate}[label = (\roman*)]
\item Defining $\mc I(\wt\cp_j) = \{\wt\cp_j - G + 1, \ldots, \wt\cp_j + G\}$, it fulfils
$\mc I(\wt\cp_j) \cap \Cp = \{ \cp_j \}$.

\item $\{\wt \cp^{\lft}_j - G + 1, \ldots, \wt \cp^{\lft}_j\} \subset \{\cp_{j - 1} + 1, \ldots, \cp_j\}$ 
and 
$\{\wt \cp^{\rgt}_j + 1, \ldots, \wt \cp^{\rgt}_j + G\} \subset \{\cp_j + 1, \ldots, \cp_{j + 1}\}$,
such that
denoting by $\bm\Delta^{\lft}_j = \wh{\bm\beta}^{\lft}_j - \bm\beta_{j - 1}$
and $\bm\Delta^{\rgt}_j = \wh{\bm\beta}^{\rgt}_j - \bm\beta_j$,
we have
\begin{align}
& \max\l(\l\vert \bm\Delta^{\lft}_j \r\vert_2, \l\vert \bm\Delta^{\rgt}_j \r\vert_2\r)
\le \frac{12\sqrt{2\mathfrak{s}}\lambda}{\omega\sqrt{G}}, 
\nn \\
& \l\vert \bm\Delta^{\lft}_j(\mc S_{j - 1}^c) \r\vert_1 \le 
3 \l\vert \bm\Delta^{\lft}_j(\mc S_{j - 1}) \r\vert_1
\text{ and }
\l\vert \bm\Delta^{\rgt}_j(\mc S_j^c) \r\vert_1 \le 
3 \l\vert \bm\Delta^{\rgt}_j(\mc S_j) \r\vert_1
\label{eq:thm:one:b:one}
\end{align}
in $\mc B$, see Lemma~\ref{lem:lasso}.
\end{enumerate}
Then we show that for all $k \in \mc I(\wt\cp_j)$ satisfying $\delta_j^2 \vert k - \cp_j \vert > v_{n, p}$
with
\begin{align}
\label{eq:vn}
v_{n, p} = \max\l( \mathfrak{s} \rho_{n, p}^2, 
\l(\mathfrak{s} \log(p)\r)^{\frac{1}{1 - \tau}} \r) \cdot 
\max\l\{
C_\delta^2 \max\l[ \frac{9\crsc}{2\omega},
\frac{32 \crsc}{\bar{\omega}}
\r]^{\frac{1}{1 - \tau}},
\l(\frac{96 \cdev}{\omega}\r)^2\r\},
\end{align}
we have $Q_j(k) - Q_j(\cp_j) > 0$,
which completes the proof. 

First, suppose that $k \ge \cp_j + 1$. Then,
\begin{align} 
& Q_j(k) - Q_j(\cp_j) =  
\sum_{t = \cp_j + 1}^k \l[ (Y_t - \mbf{x}^\top_t \wh{\bm\beta}^{\lft}_j)^2   
- (Y_t - \mbf{x}^\top_t \wh{\bm\beta}^{\rgt}_j)^2 \r] 
\nn \\
=&  \sum_{t = \cp_j + 1}^k (\bm{\beta}_j  - \wh{\bm\beta}^{\lft}_j)^\top \mbf{x}_t\mbf{x}_t^\top (\bm{\beta}_j  - \wh{\bm\beta}^{\lft}_j)
- \sum_{t = \cp_j + 1}^k (\wh{\bm\beta}^{\rgt}_j - \bm{\beta}_j)^\top \mbf{x}_t\mbf{x}_t^\top (\wh{\bm\beta}^{\rgt}_j - \bm{\beta}_j)
\nn \\
& +  2 \sum_{t = \cp_j + 1}^k \vep_t \mbf{x}_t^\top \l[ 
(\bm{\beta}_j  - \bm{\beta}_{j - 1}) + (\wh{\bm\beta}^{\rgt}_j - \bm{\beta}_j)
- (\wh{\bm\beta}^{\lft}_j - \bm{\beta}_{j - 1}) \r]
= I_1 + I_2 + I_3.
\nn
\end{align}
From the definition of $\mathfrak{s}$ and the Cauchy-Schwarz inequality,
\begin{align}
\label{eq:thm:one:b:three}
\vert \bm{\beta}_j  - \bm{\beta}_{j - 1} \vert_1
\le \sqrt{2\mathfrak{s}} \vert \bm{\beta}_j  - \bm{\beta}_{j - 1} \vert_2
\end{align}
and from~\eqref{eq:thm:one:b:one}, we have
\begin{align}
\label{eq:thm:one:b:two}
\vert \bm\Delta^{\lft}_j \vert_1 \le 4 \vert \bm\Delta^{\lft}_j(\mc S_j) \vert_1
\le 4 \sqrt{2\mathfrak{s}} \vert \bm\Delta^{\lft}_j \vert_2 
\text{ and analogously, }
\vert \bm\Delta^{\rgt}_j \vert_1 \le 
4 \sqrt{2\mathfrak{s}} \vert \bm\Delta^{\rgt}_j \vert_2.
\end{align}
From~\eqref{eq:thm:one:b:three}--\eqref{eq:thm:one:b:two}, we derive 
\begin{align*}
& \l\vert \wh{\bm\beta}^{\lft}_j  - \bm{\beta}_j \r\vert_2
\le \delta_j \l(1 + \frac{12 \sqrt{2\mathfrak{s}} \lambda}{\omega \delta_j \sqrt{G}} \r)
\le \frac{3\delta_j}{2} \text{ and similarly, }
\l\vert \wh{\bm\beta}^{\lft}_j  - \bm{\beta}_j \r\vert_2 \ge \frac{\delta_j}{2},
\\
& \l\vert \wh{\bm\beta}^{\lft}_j  - \bm{\beta}_j \r\vert_1 \le 
\sqrt{\mathfrak{s}} \delta_j \l(1 + \frac{96 \sqrt{\mathfrak{s}} \lambda}{\omega \delta_j \sqrt{G}} \r) \le \frac{3 \sqrt{\mathfrak{s}} \delta_j}{2},
\end{align*}
for a large enough $C_1$ in Assumption~\ref{assum:bandwidth}~\ref{assum:bandwidth:b}.
Then on $\mc R^{(1)}$, we have
\begin{align}
I_1 \ge \vert k - \cp_j \vert \omega \delta_j^2 \l(\frac{1}{4} - \frac{9 \crsc \mathfrak{s} \log(p)}{4 \vert k - \cp_j \vert^{1 - \tau} \omega} \r)  \ge \frac{\omega}{8} \delta_j^2 \vert k - \cp_j \vert
\label{eq:t1}
\end{align}
from that $\vert k - \cp_j \vert > \delta_j^{-2} v_{n, p} \ge C_\delta^{-2} v_{n, p}$ 
(from Assumption~\ref{assum:bounded}) and~\eqref{eq:vn}.
As for $I_2$, from Lemma~\ref{lem:lasso}, 
\eqref{eq:vn} and~\eqref{eq:thm:one:b:two}
we have on $\mc R^{(2)}$,
\begin{align}
\label{eq:t2}
\l\vert I_2 \r\vert \le \l\vert \bm\Delta^{\rgt}_j \r\vert_2^2 
\l[\vert k - \cp_j \vert \bar{\omega} + 32 \crsc \mathfrak{s} \log(p) \vert k - \cp_j \vert^\tau \r]
\le 2 \vert k - \cp_j \vert \bar{\omega} \l\vert \bm\Delta^{\rgt}_j \r\vert_2^2 
\le \frac{576 \bar{\omega} \mathfrak{s} \vert k - \cp_j \vert \lambda^2}{\omega^2 G}.
\end{align}
Turning our attention to $I_3$, 
from~\eqref{eq:thm:one:b:three}--\eqref{eq:thm:one:b:two},
\begin{align*}
& \l\vert (\bm{\beta}_j  - \bm{\beta}_{j - 1}) + (\wh{\bm\beta}^{\rgt}_j - \bm{\beta}_j)
- (\wh{\bm\beta}^{\lft}_j - \bm{\beta}_{j - 1}) \r\vert_1 
\le \l\vert \bm{\beta}_j  - \bm{\beta}_{j - 1} \r\vert_1  + 
\l\vert \wh{\bm\beta}^{\rgt}_j - \bm{\beta}_j \r\vert_1 +
\l\vert \wh{\bm\beta}^{\lft}_j - \bm{\beta}_{j - 1} \r\vert_1 
\\
& \le \sqrt{\mathfrak{s}} \delta_j \l(1 + \frac{192 \sqrt{\mathfrak{s}} \lambda}{\omega \delta_j \sqrt{G}}\r) \le 2 \sqrt{\mathfrak{s}} \delta_j,
\end{align*}
where the last inequality follows from Assumption~\ref{assum:bandwidth}~\ref{assum:bandwidth:b}.
Then on $\mc D^{(1)}$,
\begin{align}
\frac{1}{2} \l\vert I_3 \r\vert &\le 
\l\vert \sum_{t = \cp_j + 1}^k \vep_t \mbf{x}_t^\top \r\vert_\infty \;
\l\vert (\bm{\beta}_j  - \bm{\beta}_{j - 1}) + (\wh{\bm\beta}^{\rgt}_j - \bm{\beta}_j)
- (\wh{\bm\beta}^{\lft}_j - \bm{\beta}_{j - 1}) \r\vert_1
\nn \\
&\le 2\cdev \delta_j \sqrt{\mathfrak{s} (k - \cp_j)} \rho_{n, p}.
\label{eq:t3}
\end{align}
Then from~\eqref{eq:t1}, \eqref{eq:t2} and~\eqref{eq:t3}, we derive 
\begin{align*}
\frac{\vert I_2 \vert}{I_1} = \frac{4608 \bar{\omega} \mathfrak{s} \lambda^2}{\omega^3 \delta_j^2 G} \le \frac{1}{3} \quad \text{and} \quad 
\frac{\vert I_3 \vert}{I_1} = \frac{32 \cdev\sqrt{\mathfrak{s}} \rho_{n, p}}{\omega \delta_j \sqrt{k - \cp_j}} \le \frac{1}{3}
\end{align*}
under Assumption~\ref{assum:bandwidth}~\ref{assum:bandwidth:b},
for all $k \in \mc I_j$ satisfying $\delta_j^2 \vert k - \cp_j \vert > v_{n, p}$
from~\eqref{eq:vn}.
Analogous arguments apply when $k \le \cp_j$, 
and the above arguments are deterministic on $\mc M$.
In summary, we have
\begin{align*}
\min_{1 \le j \le q} \min_{\substack{k \in \mc I_j \\ \delta_j^2 \vert k - \cp_j \vert > v_{n, p}}}
\l( Q_j(k) - Q_j(\cp_j) \r) > \frac{\omega}{24} v_{n, p} > 0,
\end{align*}
which concludes the proof.

\subsection{Proof of Proposition~\ref{prop:xe}}

\subsubsection{Supporting lemmas}

Define $\mathbb{K}(b) = \mathbb{B}_0(b) \cap \mathbb{B}_2(1)$
with some $b \ge 1$,
where $\mathbb{B}_d(r) = \{\mbf a: \, \vert \mbf a \vert_d \le r\}$
with the dimension of $\mbf a$ determined within the context.
Let $\mbf e_i$ denote a vector that contains zeros
except for its $i$th component set to be one.
We denote the time-varying vector of parameters under~\eqref{eq:model} by
$\bm\beta(t) = \sum_{j = 1}^{q + 1} \bm\beta_j \mathbb{I}_{\{\cp_{j - 1} + 1 \le t \le \cp_j\}}$.

%

Denote by $\mbf Z_t = (\mbf x_t^\top, \vep_t)^\top \in \R^{p + 1}$
which admits $\mbf Z_t = \sum_{\ell = 0}^\infty \mbf D_\ell \bm\xi_{t - \ell}$
under~\eqref{eq:wold}.
For some $\mbf a, \mbf b \in \mathbb{B}_2(1)$, define
$U_t(\mbf a) = \mbf a^\top \mbf Z_t$ and
$W_t(\mbf a, \mbf b) = \mbf a^\top \mbf Z_t \mbf Z_t^\top \mbf b$.
Let $\bm\xi_t^\prime$ denote an independent copy of $\bm\xi_t$,
and define $\mbf Z_{t, \{0\}} = \sum_{\ell = 0, \, \ell \ne t}^\infty \mbf D_\ell \bm\xi_{t - \ell}
+ \mbf D_t \bm\xi^\prime_0$.
We denote the functional dependence measure and 
the dependence-adjusted norm
for $U_t(\mbf a)$ as defined in \cite{zhang2017gaussian},  by
\begin{align*}
\delta_{t, \nu}(\mbf a) = \l\Vert \mbf a^\top \mbf Z_t - \mbf a^\top \mbf Z_{t, \{0\}} \r\Vert_\nu \quad \text{and} \quad
\vertiii{U_{\cdot}(\mbf a)}_{\nu}
= \sum_{t = 0}^\infty \delta_{t, \nu}(\mbf a),
\end{align*}
respectively. Analogously, we define 
\begin{align*}
\delta_{t, \nu}(\mbf a, \mbf b) = \l\Vert \mbf a^\top \mbf Z_t \mbf Z_t^\top \mbf b - \mbf a^\top \mbf Z_{t, \{0\}} \mbf Z_{t, \{0\}}^\top \mbf b \r\Vert_\nu \quad \text{and} \quad
\vertiii{ W_{\cdot}(\mbf a, \mbf b) }_{\nu}
= \sum_{t = 0}^\infty \delta_{t, \nu}(\mbf a, \mbf b)
\end{align*}
for $W_t(\mbf a, \mbf b)$.
Finally, for some $\kappa \ge 0$, 
we denote the dependence adjusted sub-exponential norm
of $W_t(\mbf a, \mbf b)$ by
$\Vert W_{\cdot}(\mbf a, \mbf b) \Vert_{\psi_\kappa}
= \sup_{\nu \ge 2} \nu^{-\kappa} \vertiii{ W_{\cdot}(\mbf a, \mbf b) }_{\nu}$.
In what follows, we denote by $C_\Pi$ with $\Pi \subset \{\gamma, \nu, \Xi, \varsigma\}$ a constant that depends on the parameters included in $\Pi$
which may vary from one occasion to another.

\begin{lem}
\label{lem:func:dep}
Suppose that Condition~\ref{assum:lin} holds.
\begin{enumerate}[label = (\roman*)]
\item \label{lem:func:dep:one} Under Condition~\ref{assum:lin}~\ref{cond:exp}, we have
$\sup_{\mbf a, \mbf b \in \mathbb{B}_2(1)} 
\Vert W_{\cdot}(\mbf a, \mbf b) \Vert_{\psi_\kappa} 
\le C_{\gamma, \Xi, \varsigma} C_\xi^2 < \infty$
with $\kappa = 2\gamma + 1$.

\item \label{lem:func:dep:two} Under Condition~\ref{assum:lin}~\ref{cond:gauss}, we have 
$\sup_{\mbf a \in \mathbb{B}_2(1)} \vertiii{ U_{\cdot}(\mbf a) }_2 \le C_{\Xi, \varsigma}$.
\end{enumerate}
\end{lem}

\begin{proof}
In what follows, we denote by $\mu_\nu = \Vert \xi_{it} \Vert_\nu$.
For given $\nu > 1$, we have
\begin{align}
\label{eq:lem:func:dep:one}
\sup_{\mbf a \in \mathbb{B}_2(1)} \delta_{t, \nu}(\mbf a) = \l\Vert \mbf a^\top \mbf D_t (\bm\xi_0 - \bm\xi^\prime_0) \r\Vert_\nu
\le C_\nu \mu_\nu \sqrt{ 2 \sup_{\mbf a \in \mathbb{B}_2(1)} \l\vert \mbf a^\top \mbf D_t \r\vert_2^2 }
\le C_\nu \mu_\nu \Xi (1 + t)^{-\varsigma}
\end{align}
with $C_\nu = \max(1/(\nu - 1), \sqrt{\nu - 1})$,
where the inequality follows from
Lemma~2 of \cite{chen2021inference} (Burkholder's inequality) and Minkowski inequality,
and the second from Condition~\ref{assum:lin}
and from that $\Vert \mbf D_t \Vert_2 \le \sqrt{\Vert \mbf D_t \Vert_1 \Vert \mbf D_t \Vert_\infty}$ (with $\Vert \cdot \Vert_a$ denoting the induced matrix norms).
Therefore, under Condition~\ref{assum:lin}~\ref{cond:gauss}, 
\begin{align*}
\sup_{\mbf a \in \mathbb{B}_2(1)} \vertiii{ U_{\cdot}(\mbf a) }_2 \le \Xi \sum_{t = 0}^\infty (1 + t)^{-\varsigma} 
\le C_{\Xi, \varsigma},
\end{align*}
which proves~\ref{lem:func:dep:two}.
Note that by H\"{o}lder and Minkowski's inequalities,
\begin{align*}
\delta_{t, \nu}(\mbf a, \mbf b) \le &
\l\Vert \sum_{\ell = 0}^\infty \mbf a^\top \mbf D_\ell \bm\xi_{t - \ell} \r\Vert_{2\nu} 
\l\Vert \mbf b^\top \mbf D_t (\bm\xi_0 - \bm\xi^\prime_0) \r\Vert_{2\nu}
\\
& + \l\Vert \sum_{\ell = 0, \, \ell \ne t}^\infty \mbf b^\top \mbf D_\ell \bm\xi_{t - \ell} + \mbf b^\top \mbf D_t \bm\xi^\prime_0 \r\Vert_{2\nu}
\l\Vert \mbf a^\top \mbf D_t (\bm\xi_0 - \bm\xi^\prime_0) \r\Vert_{2\nu}.
\end{align*}
For given $\nu > 2$, similarly as in~\eqref{eq:lem:func:dep:one}, we can show that 
\begin{align}
&\sup_{\mbf a \in \mathbb{B}_2(1)} 
\l\Vert \sum_{\ell = 0}^\infty \mbf a^\top \mbf D_\ell \bm\xi_{t - \ell} \r\Vert_{2\nu}
\le \sum_{\ell = 0}^\infty \sup_{\mbf a \in \mathbb{B}_2(1)}  \l\Vert \mbf a^\top \mbf D_\ell \bm\xi_{t - \ell} \r\Vert_{2\nu}
\nn \\
& \le C_{2\nu} \mu_{2\nu} 
\sum_{\ell = 0}^\infty
\sqrt{\sup_{\mbf a \in \mathbb{B}_2(1)} \l\vert \mbf a^\top \mbf D_\ell \r\vert_2^2 }
\le  C_{2\nu} \mu_{2\nu} \sum_{\ell = 0}^\infty \Xi (1 + \ell)^{-\varsigma}
\le C_{\gamma, \Xi, \varsigma} C_\xi \nu^{\gamma + 1/2}
\label{eq:lem:func:dep:two}
\end{align}
under Condition~\ref{assum:lin}~\ref{cond:exp}. Then, 
\eqref{eq:lem:func:dep:one}--\eqref{eq:lem:func:dep:two} lead to
\begin{align*}
& \sup_{\mbf a, \mbf b \in \mathbb{B}_2(1)} \delta_{t, \nu}(\mbf a, \mbf b) \le 
C_{\gamma, \Xi, \varsigma} C_\xi^2 \nu^{2\gamma + 1} (1 + t)^{-\varsigma}, \quad \text{and}
\\
& \sup_{\mbf a, \mbf b \in \mathbb{B}_2(1)} \vertiii{ W_{\cdot}(\mbf a, \mbf b) }_{\nu} \le 
C_{\gamma, \Xi, \varsigma} C_\xi^2 \nu^{2\gamma + 1} \sum_{t = 0}^\infty (1 + t)^{-\varsigma}
\le C_{\gamma, \Xi, \varsigma} C_\xi^2 \nu^{2\gamma + 1},
\end{align*}
such that we have $\sup_{\mbf a, \mbf b \in \mathbb{B}_2(1)} 
\Vert W_{\cdot}(\mbf a, \mbf b) \Vert_{\psi_\kappa} \le C_{\gamma, \Xi, \varsigma} C_\xi^2$ with $\kappa = 2\gamma + 1$,
which proves~\ref{lem:func:dep:one}.
\end{proof}

\begin{lem}
\label{lem:exp:nagaev}
Under Condition~\ref{assum:lin}~\ref{cond:exp},
there exist fixed constants $C^\prime, C^{\prime\prime} > 0$ such that
for all $0 \le s < e \le n$ and $z > 0$, we have
\begin{align*}
\sup_{\mbf a, \mbf b \in \mathbb{B}_2(1)} \p\l(
\frac{1}{\sqrt{e - s}} \l\vert \sum_{t = s + 1}^e \mbf a^\top \mbf Z_t \mbf Z_t^\top \mbf b - \E\l(\sum_{t = s + 1}^e \mbf a^\top \mbf Z_t \mbf Z_t^\top \mbf b\r) \r\vert \ge z
\r) \le C^\prime \exp\l( - C^{\prime\prime} z^{\frac{2}{4\gamma + 3}} \r).
\end{align*}
\end{lem}
\begin{proof}
By Lemma~\ref{lem:func:dep}~\ref{lem:func:dep:one} and
Lemma~C.4 of \cite{zhang2017gaussian},
there exist constants $C^\prime, C^{\prime\prime} > 0$ that depend on $\gamma, \Xi, \varsigma$ and $C_\xi$, such that for all $z > 0$,
\begin{align*}
& \sup_{\mbf a, \mbf b \in \mathbb{B}_2(1)} \p\l(
\frac{1}{\sqrt{e - s}} \l\vert \sum_{t = s + 1}^e \mbf a^\top \mbf Z_t \mbf Z_t^\top \mbf b - \E\l(\sum_{t = s + 1}^e \mbf a^\top \mbf Z_t \mbf Z_t^\top \mbf b\r) \r\vert \ge z
\r) 
\\
& \le C^\prime \exp\l(- \frac{(4\gamma + 3) z^{\frac{2}{4\gamma + 3}}}{4 e (C_{\gamma, \Xi, \varsigma} C_\xi^2)^{\frac{2}{4\gamma + 3}} }\r)
\le C^\prime \exp\l(-C^{\prime\prime} z^{\frac{2}{4\gamma + 3}} \r).
\end{align*}
\end{proof}

\begin{lem}
\label{lem:gauss:hw}
Under Condition~\ref{assum:lin}~\ref{cond:gauss},
there exists a fixed constants $C^{\prime\prime\prime} > 0$ such that
for all $0 \le s < e \le n$ and $0 < z < C_{\Xi, \varsigma}^2 \sqrt{e - s}$, we have
\begin{align*}
& \sup_{\mbf a, \mbf b \in \mathbb{B}_2(1)} \p\l(
\frac{1}{\sqrt{e - s}} \l\vert \sum_{t = s + 1}^e \mbf a^\top \mbf Z_t \mbf Z_t^\top \mbf b - \E\l(\sum_{t = s + 1}^e \mbf a^\top \mbf Z_t \mbf Z_t^\top \mbf b\r) \r\vert \ge z
\r) \le 6 \exp(- C^{\prime\prime\prime} z^2).
\end{align*}
\end{lem}
\begin{proof}
By Lemma~\ref{lem:func:dep}~\ref{lem:func:dep:two} and
Theorem~6.6 of \cite{zhang2021}, there exists an absolute constant $C > 0$
such that for all $0 < z < C_{\Xi, \varsigma}^2 \sqrt{e - s}$,
\begin{align}
& \sup_{\mbf a \in \mathbb{B}_2(1)} \p\l(
\frac{1}{\sqrt{e - s}} \l\vert \sum_{t = s + 1}^e \mbf a^\top \mbf Z_t \mbf Z_t^\top \mbf a - \E\l(\sum_{t = s + 1}^e \mbf a^\top \mbf Z_t \mbf Z_t^\top \mbf a\r) \r\vert \ge z
\r) \nn
\\
& \le 2 \exp\l[ - C \min\l(\frac{z^2}{C_{\Xi, \varsigma}^4}, \frac{z \sqrt{e - s}}{C_{\Xi, \varsigma}^2}\r) \r] \le 2 \exp(- C C^{-4}_{\Xi, \varsigma} z^2).
\nn 
\end{align}
Then noting that
\begin{align*}
& \sup_{\mbf a, \mbf b \in \mathbb{B}_2(1)} \p\l(
\frac{2}{\sqrt{e - s}} \l\vert \sum_{t = s + 1}^e \mbf a^\top \mbf Z_t \mbf Z_t^\top \mbf b - \E\l(\sum_{t = s + 1}^e \mbf a^\top \mbf Z_t \mbf Z_t^\top \mbf b\r) \r\vert \ge z
\r) \le 
\\
& \sup_{\mbf a, \mbf b \in \mathbb{B}_2(1)} \p\l(
\frac{1}{\sqrt{e - s}} \l\vert \sum_{t = s + 1}^e (\mbf a + \mbf b)^\top \mbf Z_t \mbf Z_t^\top (\mbf a + \mbf b) - \E\l(\sum_{t = s + 1}^e (\mbf a + \mbf b)^\top \mbf Z_t \mbf Z_t^\top (\mbf a + \mbf b)\r) \r\vert \ge \frac{z}{3}
\r)
\\
& + 2 \sup_{\mbf a \in \mathbb{B}_2(1)} \p\l(
\frac{1}{\sqrt{e - s}} \l\vert \sum_{t = s + 1}^e \mbf a^\top \mbf Z_t \mbf Z_t^\top \mbf a - \E\l(\sum_{t = s + 1}^e \mbf a^\top \mbf Z_t \mbf Z_t^\top \mbf a\r) \r\vert \ge \frac{z}{3}
\r) 
\le 6 \exp\l(- \frac{C  z^2}{9C^4_{\Xi, \varsigma}} \r),
\end{align*}
we can find $C^{\prime\prime\prime}$ that depends on $\Xi$ and $\varsigma$.
\end{proof}

\subsubsection{Proof of Proposition~\ref{prop:xe}~\ref{prop:xe:one}}

Recalling $C^\prime$ from Lemma~\ref{lem:exp:nagaev},
we set $c_1 = 3C^\prime$.

\noindent\textit{Verification of Assumption~\ref{assum:dev}:}

By assumption, we have $\E(\mbf x_t \vep_t) = \mbf 0$.
Then setting $\mbf a = \mbf e_i, \, i = 1, \ldots, p$, $\mbf b = \mbf e_{p + 1}$
and $z = \cdev \log^{2\gamma + 3/2}(p \vee n)$ in Lemma~\ref{lem:exp:nagaev},
\begin{align}
\label{eq:prop:ex:one:one}
\p(\mc D^{(1)}) \ge 1 - C^\prime p n^2 \exp\l(-C^{\prime\prime} \cdev^{\frac{2}{4\gamma + 3}} \log(p \vee n) \r).
\end{align}
Next, by construction,
\begin{align}
\label{eq:beta:one}
\sum_{t = s + 1}^e (\bm\beta(t) - \bm\beta^*_{s, e}) = \mbf 0 
\quad \text{and} \quad
\max_{\substack{0 \le s < e \le n \\ 
\vert \{s + 1, \ldots, e\} \cap \Cp \vert \le 1}} \max_{s < t \le e} 
\l\vert \bm\beta(t) - \bm\beta^*_{s, e} \r\vert_2 \le C_\delta
\end{align}
under Assumption~\ref{assum:bounded}, and
\begin{align}
\label{eq:beta:two}
\E\l[ \sum_{t = s + 1}^e \mbf x_t \mbf x_t^\top (\bm\beta(t) - \bm\beta^*_{s, e}) \r]
= \bm\Sigma_x \sum_{t = s + 1}^e  (\bm\beta(t) - \bm\beta^*_{s, e}) = \mbf 0
\end{align}
under Assumption~\ref{assum:xe}.
Then setting $\mbf a = \mbf e_i, \, i = 1, \ldots, p$, 
$\mbf b = \bm\beta(t) - \bm\beta^*_{s, e}$ for given $s, e$ and $t \in \{s + 1, \ldots, e\}$
and $z = \cdev C_\delta \log^{2\gamma + 3/2}(p \vee n)$ in Lemma~\ref{lem:exp:nagaev},
\begin{align}
\label{eq:prop:ex:one:two}
\p(\mc D^{(2)}) \ge 1 - C^\prime p n^3 \exp\l(- C^{\prime\prime} (\cdev C_\delta)^{\frac{2}{4\gamma + 3}} \log(p \vee n) \r),
\end{align}
from~\eqref{eq:beta:one} and~\eqref{eq:beta:two}.
Combining~\eqref{eq:prop:ex:one:one} and~\eqref{eq:prop:ex:one:two},
we can find large enough $\cdev$ that depends only on $C^{\prime\prime}$, $\gamma$, $C_\delta$ and $c_2$ such that
$\p(\mc D^{(1)} \cap \mc D^{(2)}) \ge 1 - 2c_1(p \vee n)^{-c_2}/3$.

\noindent\textit{Verification of Assumption~\ref{assum:rsc}:}

Let $b_{s, e}$ denote an integer that depends on $(e - s)$ for some $0 \le s < e \le n$, and define
\begin{align*}
\mc R &= \l\{ \sup_{\mbf a \in \mathbb{K}(2b_{s, e})} \frac{1}{e - s} \l\vert \sum_{t = s + 1}^e \mbf a^\top\l(\mbf x_t\mbf x_t^\top - \bm\Sigma_x\r) \mbf a \r\vert \ge \frac{\Lambda_{\min}(\bm\Sigma_x)}{54} \text{ for all }
0 \le s < e \le n \r. 
\nn \\  
& \qquad \l. \text{with } e - s \ge C_0 \log^{4\gamma + 3}(p \vee n) \r\}.
\end{align*}
By Lemma~\ref{lem:exp:nagaev} and Lemma~F.2 of \cite{basu2015regularized},
we have
\begin{align}
\p\l( \mc R^c \r) &\le 
\sum_{\substack{0 \le s < e \le n \\ e - s \ge C_0 \log^{4\gamma + 3}(p \vee n) 
}}
C^\prime \exp\l[ - C^{\prime\prime} \l(\frac{\sqrt{e - s}\Lambda_{\min}(\bm\Sigma_x)}{54}\r)^{\frac{2}{4\gamma + 3}} + 2 b_{s, e} \log(p) \r]
\nn \\
&\le C^\prime n^2 \exp\l[ -\frac{C^{\prime\prime}}{2} \l(\frac{C_0^{1/2} \Lambda_{\min}(\bm\Sigma_x)}{54}\r)^{\frac{2}{4\gamma + 3}} \log(p \vee n) \r],
\nn
\end{align}
where the last inequality follows with 
\begin{align*}
b_{s, e} = \l\lfloor \frac{C^{\prime\prime}}{4\log(p)}\l(\frac{ \sqrt{e - s} \Lambda_{\min}(\bm\Sigma_x)}{54}\r)^{\frac{2}{4\gamma + 3}} \r\rfloor,
\end{align*}
which satisfies $b_{s, e}  \ge 1$ for large enough $C_0$.
Further, we can find $C_0$ that depends only on $C^{\prime\prime}$, $\Lambda_{\min}(\bm\Sigma_x)$, $\gamma$ and $c_2$ which leads to
$\p(\mc R) \ge 1 - c_1 (p \vee n)^{-c_2}/3$.
Then, by Lemma~12 of \cite{loh2012high}, on $\mc R$, we have
\begin{align*}
\sum_{t = s + 1}^e \mbf a^\top \mbf x_t \mbf x_t^\top \mbf a \ge & \Lambda_{\min}(\bm\Sigma_x) (e - s) \vert \mbf a \vert_2^2
\\
& - \frac{\Lambda_{\min}(\bm\Sigma_x)}{2} (e - s) \l( \vert \mbf a \vert_2^2 + \frac{4\log(p)}{C^{\prime\prime}} \l(\frac{54}{ \sqrt{e - s} \Lambda_{\min}(\bm\Sigma_x)}\r)^{\frac{2}{4\gamma + 3}} \vert \mbf a \vert_1^2 \r)
\\
\ge & \omega (e - s) \vert \mbf a \vert_2^2 - \crsc \log(p) (e - s)^{\frac{4\gamma + 2}{4\gamma + 3}} \vert \mbf a \vert_1^2
\end{align*}
for all $\mbf a \in \R^p$,
with $\omega = \Lambda_{\min}(\bm\Sigma_x)/2$ and
$\crsc$ depending only on $C^{\prime\prime}$, $\gamma$ and $\Lambda_{\min}(\bm\Sigma_x)$.
Analogously we have on $\mc R$, 
\begin{align*}
\sum_{t = s + 1}^e \mbf a^\top \mbf x_t \mbf x_t^\top \mbf a \le 
\bar{\omega} (e - s) \vert \mbf a \vert_2^2 + \crsc \log(p) (e - s)^{\frac{4\gamma + 2}{4\gamma + 3}} \vert \mbf a \vert_1^2
\end{align*}
for all $\mbf a \in \R^p$, with $\bar{\omega} = 3\Lambda_{\max}(\bm\Sigma_x)/2$.

Combining the arguments above, we have
$\p(\mc D^{(1)} \cap \mc D^{(2)} \cap \mc R^{(1)} \cap \mc R^{(2)})
\ge 1 - c_1 (p \vee n)^{-c_2}$,
with $\tau = (4\gamma + 2)/(4\gamma + 3)$ and 
$\rho_{n, p} = \log^{2\gamma + 3/2}(p \vee n)$.

\subsubsection{Proof of Proposition~\ref{prop:xe}~\ref{prop:xe:two}}

We set $c_1 = 18$.

\noindent\textit{Verification of Assumption~\ref{assum:dev}:}

By assumption, we have $\E(\mbf x_t \vep_t) = \mbf 0$.
Then setting $\mbf a = \mbf e_i, \, i = 1, \ldots, p$, $\mbf b = \mbf e_{p + 1}$
and $z = \cdev \sqrt{\log(p \vee n)}$ in Lemma~\ref{lem:gauss:hw},
\begin{align}
\label{eq:prop:ex:two:one}
\p(\mc D^{(1)}) \ge 1 - 6 p n^2 \exp\l(-C^{\prime\prime\prime} \cdev^2 \log(p \vee n) \r),
\end{align}
provided that $C_0 > C^{-4}_{\Xi, \varsigma} \cdev^2$.
Also, setting $\mbf a = \mbf e_i, \, i = 1, \ldots, p$, 
$\mbf b = \bm\beta(t) - \bm\beta^*_{s, e}$ for given $s, e$ and $t \in \{s + 1, \ldots, e\}$
and $z = \cdev C_\delta \sqrt{\log(p \vee n)}$ in Lemma~\ref{lem:gauss:hw},
\begin{align}
\label{eq:prop:ex:two:two}
\p(\mc D^{(2)}) \ge 1 - 6 p n^3 \exp\l(- C^{\prime\prime\prime} \cdev^2 C_\delta^2 \log(p \vee n) \r),
\end{align}
from~\eqref{eq:beta:one} and~\eqref{eq:beta:two}.
Combining~\eqref{eq:prop:ex:two:one} and~\eqref{eq:prop:ex:two:two},
we can find large enough $\cdev$ that depends only on $C^{\prime\prime\prime}$, $C_\delta$ and $c_2$ such that
$\p(\mc D^{(1)} \cap \mc D^{(2)}) \ge 1 - 2c_1(p \vee n)^{-c_2}/3$.

\noindent\textit{Verification of Assumption~\ref{assum:rsc}:}

Let $b_{s, e}$ denote an integer that depends on $(e - s)$ for some $0 \le s < e \le n$, and define
\begin{align*}
\mc R &= \l\{ \sup_{\mbf a \in \mathbb{K}(2b_{s, e})} \frac{1}{e - s} \l\vert \sum_{t = s + 1}^e \mbf a^\top\l(\mbf x_t\mbf x_t^\top - \bm\Sigma_x\r) \mbf a \r\vert \ge \frac{\Lambda_{\min}(\bm\Sigma_x)}{54} \text{ for all }
0 \le s < e \le n \r. 
\nn \\  
& \qquad \l. \text{with } e - s \ge C_0 \log(p \vee n)\r\}.
\end{align*}
Then by Lemma~\ref{lem:gauss:hw} and Lemma~F.2 of \cite{basu2015regularized},
we have
\begin{align}
\p\l( \mc R^c \r) &\le 
\sum_{\substack{0 \le s < e \le n \\ e - s \ge C_0 \log(p \vee n)}} 
6 \exp\l[ - C^{\prime\prime\prime} (e - s) \l(\frac{\Lambda_{\min}(\bm\Sigma_x)}{54}\r)^2 + 2 b_{s, e} \log(p) \r]
\nn \\
&\le 6 n^2 \exp\l[ -\frac{C^{\prime\prime\prime} C_0}{2} \l(\frac{\Lambda_{\min}(\bm\Sigma_x)}{54}\r)^2 \log(p \vee n) \r],
\nn
\end{align}
where the last inequality follows with 
\begin{align*}
b_{s, e} = \l\lfloor \frac{C^{\prime\prime\prime} (e - s)}{4\log(p)}\l(\frac{\Lambda_{\min}(\bm\Sigma_x)}{54}\r)^2 \r\rfloor,
\end{align*}
which satisfies $b_{s, e}  \ge 1$ for large enough $C_0$.
Further, we can find $C_0$ that depends only on $C^{\prime\prime\prime}$, $\Lambda_{\min}(\bm\Sigma_x)$ and $c_2$ which leads to
$\p(\mc R) \ge 1 - c_1 (p \vee n)^{-c_2}/3$.
Then, by Lemma~12 of \cite{loh2012high}, on $\mc R$, we have
\begin{align*}
\sum_{t = s + 1}^e \mbf a^\top \mbf x_t \mbf x_t^\top \mbf a \ge & \Lambda_{\min}(\bm\Sigma_x) (e - s) \vert \mbf a \vert_2^2 \\
- &\frac{\Lambda_{\min}(\bm\Sigma_x)}{2} (e - s) \l( \vert \mbf a \vert_2^2 + \frac{4\log(p)}{C^{\prime\prime\prime} (e - s)} \l(\frac{54}{ \Lambda_{\min}(\bm\Sigma_x)}\r)^2 \vert \mbf a \vert_1^2 \r)
\\
\ge & \omega (e - s) \vert \mbf a \vert_2^2 - \crsc \log(p) \vert \mbf a \vert_1^2
\end{align*}
for all $\mbf a \in \R^p$,
with $\omega = \Lambda_{\min}(\bm\Sigma_x)/2$ and
$\crsc$ depending only on $C^{\prime\prime\prime}$ and $\Lambda_{\min}(\bm\Sigma_x)$.
Analogously we have on $\mc R$, 
\begin{align*}
\sum_{t = s + 1}^e \mbf a^\top \mbf x_t \mbf x_t^\top \mbf a \le 
\bar{\omega} (e - s) \vert \mbf a \vert_2^2 + \crsc \log(p) \vert \mbf a \vert_1^2
\end{align*}
for all $\mbf a \in \R^p$, with $\bar{\omega} = 3\Lambda_{\max}(\bm\Sigma_x)/2$.

Combining the arguments above, we have
$\p(\mc D^{(1)} \cap \mc D^{(2)} \cap \mc R^{(1)} \cap \mc R^{(2)})
\ge 1 - c_1 (p \vee n)^{-c_2}$,
with $\tau = 0$ and 
$\rho_{n, p} = \sqrt{\log(p \vee n)}$.

\subsection{Proof of Theorem~\ref{thm:multiscale}}
\label{sec:pf:thm:multiscale}

In what follows, we operate on 
$\mc M = \mc D^{(1)} \cap \mc D^{(2)} \cap \mc R^{(1)} \cap \mc R^{(2)} \cap \mc B$.
Under Assumption~\ref{assum:multiscale},
we have all $G \in \mc G$ satisfy $G \ge C_0 \max\{\rho_{n, p}^2,
(\omega^{-1} \mathfrak{s} \log(p))^{1/(1 - \tau)} \}$ 
such that the lower bound on $(e - s)$ made in 
$\mc B$ (see Lemma~\ref{lem:lasso}) is met by all
$s = k$ and $e = k + G$, $k = 0, \ldots, n - G$.
{Additionally, thanks to the construction of the bandwidths described in Remark~\ref{rem:bandwidths}, the ordered bandwidths $G_1 < G_2 < \ldots < G_H$, satisfy $G_h < G_{h + 1} \le 2 G_h$ for all $1 \le h \le H - 1$.
This ensures that for each $j$, there exists some $G_h$ which satisfies
\begin{align*}
4 G_h \le \min(\cp_j - \cp_{j - 1}, \cp_{j + 1} - \cp_j) < 4 G_{h + 1} \le 8 G_h.
\end{align*}
Then, under Assumption~\ref{assum:multiscale}, it follows that
\begin{align*}
\delta_j^2 G_h \ge 4C_1 \max\l\{
\omega^{-2} \mathfrak{s} \rho_{n, p}^2,
\l(\omega^{-1} \mathfrak{s} \log(p)\r)^{1/(1 - \tau)} \r\}.
\end{align*}
Referring to such $G_h$ by $G_{(j)}$, we summarise these observations in the following: For each change point $\cp_j, \,  j = 1, \ldots, q$, there exists a bandwidth $G_{(j)} \in \mc G$ such that
\begin{enumerate}[label = (\alph*)]
\item \label{assum:multiscale:a}
$4 G_{(j)} \le \min(\cp_{j + 1} - \cp_j, \cp_j - \cp_{j - 1})$, and
\item \label{assum:multiscale:b}
$\delta_j^2 G_{(j)} \ge 4 C_1 \max\l\{
\omega^{-2} \mathfrak{s} \rho_{n, p}^2,
\l(\omega^{-1} \mathfrak{s} \log(p)\r)^{1/(1 - \tau)} \r\}$.
\end{enumerate}
Also, from~\eqref{eq:multiscale:threshold:one}, we have
\begin{align}
\label{eq:multiscale:threshold}
\frac{48\sqrt{\mathfrak{s}}\lambda}{\omega} < D < \frac{\eta}{4\sqrt{2}} \min_{1 \le j \le q} \delta_j \sqrt{G_{(j)}}.
\end{align}
}

For some $k$ and $G \in \mc G$, we write $\mc I(k, G) = \{k - G + 1, \ldots, k + G\}$.
Recall that for each pre-estimator $\wt\cp \in \wt{\Cp}(G)$, we denote by
$\mc I(\wt\cp) = \mc I(\wt\cp, G)$ its detection interval.
By the same arguments adopted in~\eqref{eq:thm:one:a:one}
and Lemmas~\ref{lem:w:norm} and~\ref{lem:lasso},
we have
\begin{align}
\label{eq:thm:multiscale:one}
\max_{G \in \mc G} \max_{\substack{G \le k \le n - G \\ \vert \mc I(k, G) \cap \Cp \vert \le 1}}
\l\vert T_k(G) - T^*_k(G) \r\vert \le \frac{24\sqrt{\mathfrak{s}} \lambda}{\omega}
\quad \text{and} \quad
T^*_k(G) = 0 \text{ \ if \ } \mc I(k, G) \cap \Cp = \emptyset.
\end{align}
Then, we make the following observations. 
\begin{enumerate}[label = (\roman*)]
\item \label{eq:thm:multiscale:i} From~\eqref{eq:thm:multiscale:one}
and the requirement on $D$ in~\eqref{eq:multiscale:threshold},
we have $\mc I(\wt\cp) \cap \Cp \ne \emptyset$
for all $\wt\cp \in \wt{\Cp}(\mc G)$,
i.e.\ each pre-estimator in $\wt{\Cp}(\mc G)$ has (at least) one change point
in its detection interval.

\item \label{eq:thm:multiscale:ii} Under Assumption~\ref{assum:multiscale},
for each $\cp_j, \, j = 1, \ldots, q$, there exists one pre-estimator
$\wt\cp \in \wt{\Cp}(G_{(j)})$ such that $\mc I(\wt\cp) \cap \Cp = \{\cp_j\}$
and $\vert \wt\cp - \cp_j \vert < \lfloor G_{(j)}/2 \rfloor$,
by the arguments used in the proof of Theorem~\ref{thm:one}~\ref{thm:one:one}.
\end{enumerate}

Thanks to~\ref{eq:thm:multiscale:ii}, 
there exists an anchor estimator $\wt\cp^A \in \wt{\Cp}^A$ for each $\cp_j$, 
in the sense that $\cp_j \in \mc I(\wt\cp^A)$ 
and further, this anchor estimator $\wt\cp^A$ is detected with some bandwidth $G \le G_{(j)}$. 
At the same time, there is at most a single anchor estimator $\wt\cp^A$ 
fulfilling $\cp_j \in \mc I(\wt\cp^A)$ by its construction in~\eqref{eq:alg:multiscale:anchor},
and~\ref{eq:thm:multiscale:i} ensures that all anchor estimators
contain one change point in its detection interval.
Therefore, we have $\wh q = \vert \wt{\Cp}^A \vert = q$
and we may write $\wt{\Cp}^A = \{\wt\cp_j^A, \, 1 \le j \le q: \,
\wt\cp_1^A < \ldots < \wt\cp_q^A\}$.

Next, by~\ref{eq:thm:multiscale:ii}, there exists some $\wt\cp \in \wt{\Cp}(G_{(j)})$
fulfilling~\eqref{eq:alg:multiscale:clustering} for each $j = 1, \ldots, q$. 
To see this, note that
if $\wt\cp \in \wt{\Cp}(G_{(j)})$ detects $\cp_j$ in the sense that $\cp_j \in \mc I(\wt\cp)$,
\begin{align*}
& \l\{\wt\cp -  G_{(j)} - \l\lfloor \frac{G_{(j)}}{2} \r\rfloor + 1, \ldots, \wt\cp + G_{(j)} + \l\lfloor \frac{G_{(j)}}{2} \r\rfloor \r\}
\subset \l\{ \cp_j - 2 G_{(j)} + 1, \cp_j + 2 G_{(j)} \r\}, \text{ while} 
\\
& \mc I(\wt\cp^A_{j - 1}) \subset \l\{ \cp_{j - 1} - 2 G_{(j - 1)} + 1, \ldots, \cp_{j - 1} + 2 G_{(j - 1)} \r\} \text{ and }
\\
& \mc I(\wt\cp^A_{j + 1}) \subset \l\{ \cp_{j + 1} - 2 G_{(j + 1)} + 1, \ldots, \cp_{j + 1} + 2 G_{(j + 1)} \r\},
\end{align*}
and the sets on RHS do not overlap under~\ref{assum:multiscale:a}.
This in turn implies that we have $\vert \mc C_j \vert \ge 1$.
Also for $\wt\cp^M_j \in \mc C_j$, we have that its detection bandwidth
$G^M_j$ satisfies
\begin{align*}
\frac{3}{2} G^M_j \le \min(\cp_{j + 1} - \cp_j, \cp_j - \cp_{j - 1}) 
\quad \text{and} \quad
G^M_j \ge G_{(j)}
\end{align*}
by the construction of $\mc C_j$.
Also, the bandwidths generated as in Remark~\ref{rem:bandwidths} satisfy
\begin{align*}
G_{\ell - 1} + \frac{1}{2} G_{\ell - 1} \le G_{\ell - 1} + G_{\ell - 2} = G_\ell \le 2 G_{\ell - 1}, \quad
\text{such that} \quad
\frac{1}{2} G_\ell \le G_{\ell - 1} \le \frac{2}{3} G_\ell \text{ for } \ell \ge 2,
\end{align*}
and therefore
\begin{align}
\label{eq:thm:multiscale:two}
\frac{1}{4} G_{(j)} \le G^*_j
\quad \text{and} \quad 
G^*_j \le \l(\frac{3}{4} \cdot \frac{2}{3} + \frac{1}{4}\r) G^M_j \le \frac{1}{2} \min(\cp_{j + 1} - \cp_j, \cp_j - \cp_{j - 1}).
\end{align}
Further,  by that $\vert \wt\cp^m_j - \cp_j \vert < G^m_j$ (see~\ref{eq:thm:multiscale:i}) 
and
\begin{align*}
2 G^m_j + G^*_j = \frac{11}{4} G^m_j + \frac{1}{4} G^M_j 
\le \frac{11}{4} G_{(j)} + \frac{1}{4} G^M_j
\le 
\frac{41}{48} \min(\cp_{j + 1} - \cp_j, \cp_j - \cp_{j - 1}),
\end{align*}
we have
\begin{align}
\label{eq:thm:multiscale:three}
\{\wt\cp^m_j - G^m_j - G^*_j + 1, \ldots, \wt\cp^m_j  - G^m_j \} \cap 
\{\wt\cp^m_j + G^m_j + 1, \ldots, \wt\cp^m_j  + G^m_j + G^*_j \} \cap \Cp = \emptyset.
\end{align}
From~\eqref{eq:thm:multiscale:two} and~\ref{assum:multiscale:b}, we have
\begin{align*}
\delta_j^2 G_j^* \ge C_1 \max\l\{
\omega^{-2} \mathfrak{s} \rho_{n, p}^2,
\l(\omega^{-1} \mathfrak{s} \log(p)\r)^{1/(1 - \tau)} \r\}
\end{align*}
and from~\eqref{eq:thm:multiscale:three} and Lemma~\ref{lem:lasso}, we have
$\bm\Delta^{\lft}_j = \wh{\bm\beta}^{\lft}_j - \bm\beta_{j - 1}$
and $\bm\Delta^{\rgt}_j = \wh{\bm\beta}^{\rgt}_j - \bm\beta_j$ satisfy
\begin{align*}
& \max\l(\l\vert \bm\Delta^{\lft}_j \r\vert_2, \l\vert \bm\Delta^{\rgt}_j \r\vert_2\r)
\le \frac{12\sqrt{2\mathfrak{s}}\lambda}{\omega\sqrt{G_j^*}}
\le \frac{24\sqrt{2\mathfrak{s}}\lambda}{\omega\sqrt{G_{(j)}}}, 
\nn \\
& \l\vert \bm\Delta^{\lft}_j(\mc S_{j - 1}^c) \r\vert_1 \le 
3 \l\vert \bm\Delta^{\lft}_j(\mc S_{j - 1}) \r\vert_1
\text{ and }
\l\vert \bm\Delta^{\rgt}_j(\mc S_j^c) \r\vert_1 \le 
3 \l\vert \bm\Delta^{\rgt}_j(\mc S_j) \r\vert_1,
\end{align*}
such that the arguments analogous to those employed in the proof of Theorem~\ref{thm:one}~\ref{thm:one:two} 
are applicable to establish the localisation rate of $\check\cp_j$,
which completes the proof.

\clearpage

\section{Algorithms}
\label{app:alg}

\begin{algorithm}[h!t!b!]
\caption{MOSEG: Single-bandwidth two-stage data segmentation methodology under a regression model.}
\label{algo:mosumsr}
\SetAlgoLined
\DontPrintSemicolon
\SetKwInOut{Initialise}{initialise}
\SetKwInOut{Input}{input}
\Input{Bandwidth $G$, grid resolution $r$, penalty $\lambda$, threshold $D$, $\eta \in (0, 1]$}
\BlankLine

\Initialise{$\wt{\Cp} = \emptyset$, $\wh{\Cp} = \emptyset$}
\BlankLine

// Stage~1
\BlankLine

Compute $T_k(G)$ in~\eqref{eq:wald} for all $k \in \mc T = \mc T(r, G)$
\BlankLine

Add all $\wt\cp$ satisfying
$T_{\wt\cp}(G) > D$ and 
$\wt\cp = {\arg\min}_{k \in \{\wt\cp -  \lfloor \eta G \rfloor + 1, \ldots, \wt\cp +  \lfloor \eta G \rfloor\} \cap \mc T} T_k(G)$
to $\wt{\Cp}$, and
set $\wt{\Cp} = \{\wt\cp_j, \, 1 \le j \le \wh q \}$
\BlankLine

// Stage~2
\BlankLine
	
\For{$j  = 1, \ldots, \wh q$}{
Identify $\wh \cp_j = {\arg\min}_{\wt\cp_j - G + 1 \le j \le \wt\cp_j + G} 
Q(k; \wt\cp_j - G, \wt\cp_j + G, \wh{\bm\beta}^{\lft}_j, \wh{\bm\beta}^{\rgt}_j)$
with $\wh{\bm\beta}^{\lft}_j$ and $\wh{\bm\beta}^{\rgt}_j$ 
computed as in~\eqref{eq:beta:lr},
and add it to $\wh{\Cp}$
}
  
\Return{$\wh{\Cp}$ }
\end{algorithm}

\begin{algorithm}[h!t!b!]
\caption{MOSEG.MS: Multiscale extension of MOSEG.}
\label{algo:multiscale}
\SetAlgoLined
\DontPrintSemicolon
\SetKwInOut{Initialise}{initialise}
\SetKwInOut{Input}{input} 
\Input{A set of bandwidths $\mc G$, grid resolution $r$, penalty $\lambda$, threshold $D$, $\eta \in (0, 1]$}
\BlankLine

\Initialise{$\wt{\Cp}^A = \emptyset$, $\check{\Cp} = \emptyset$, $\mc C_j = \emptyset$ for all $j$} 
\BlankLine

// Pre-estimator generation
\BlankLine
\For{$h = 1, \ldots, H$}{
	Initialise $\wt{\Cp}(G_h) = \emptyset$
	\BlankLine 
	
	Compute $T_k(G_h)$ in~\eqref{eq:wald} for all $k \in \mc T_h = \mc T(r, G_h)$
	\BlankLine

	Add all $\wt\cp$ satisfying $T_{\wt\cp}(G_h) > D$ and
	$\wt\cp = {\arg\min}_{k \in \mc I_\eta(\wt\cp) \cap \mc T_h} T_k(G_h)$, 
	to $\wt{\Cp}(G_h)$
}
\BlankLine

// Anchor change point estimator identification
\BlankLine
Identify all $\wt\cp(G) \in \cup_{h = 1}^H \wt{\Cp}(G_h)$ satisfying~\eqref{eq:alg:multiscale:anchor},
and add all such estimators to $\wt{\Cp}^A$,
which is denoted by
$\wt{\Cp}^A = \{\wt\cp^A_j, \, 1 \le j \le \wh q: \, 
\wt\cp^A_1 < \ldots < \wt\cp^A_{\wh q}\}$
\BlankLine

\BlankLine
\For{$j = 1, \ldots, \wh q$}{
	\BlankLine
	// Pre-estimator clustering
	\BlankLine
	
	Identify all $\wt\cp \in \cup_{h = 1}^H \wt{\Cp}(G_h)$ satisfying~\eqref{eq:alg:multiscale:clustering}
	and add it to $\mc C_j$
	\BlankLine
	
	// Location refinement
	\BlankLine
	Add $\check \cp_j$ obtained as in~\eqref{eq:alg:multiscale:final} to $\wh{\Cp}$
}

\Return{$\check{\Cp}$}
\end{algorithm} 

\clearpage

\section{Further information on the real dataset}
\label{sec:real:info}

Table~\ref{tab:eqprem_data} lists the covariates included in the dataset analysed in Section~\ref{sec:real}. 

\begin{table}[h!t!b!]
\caption{Covariates contained in the equity premium dataset analysed in Section~\ref{sec:real} (cf.\ \cite{koo2016high}, Table~3)}
    \label{tab:eqprem_data}
\resizebox{\columnwidth}{!}{
    \begin{tabular}{ll}
    \toprule
    Name & Description \\ 
          \cmidrule(lr){1-1}\cmidrule(lr){2-2}
    d/p &
Dividend price ratio: difference between the log of dividends and the log of prices \\
d/y &
Dividend yield: difference between the log of dividends and the log of lagged prices \\
e/p &
Earnings price ratio: difference between the log of earnings and the log of prices \\
d/e &
Dividend payout ratio: difference between the log of dividends and the log of earnings \\
b/m &
Book-to-market ratio: ratio of book value to market value for the Dow Jones Industrial Average \\
ntis &
Net equity expansion: ratio of 12-month moving sums of net issues by NYSE listed stocks over \\& the total end-of-year market capitalization of NYSE stocks \\
tbl &
Treasury bill rates: 3-month Treasury bill rates \\
lty &
Long-term yield: long-term government bond yield \\
tms &
Term spread: difference between the long term bond yield and the Treasury bill rate \\
dfy &
Default yield spread: difference between Moody’s BAA and AAA-rated corporate bond yields \\
dfr &
Default return spread: difference between the returns of long-term corporate and government bonds \\
svar &
Log of stock varianceobtained as the sum of squared daily returns on S\&P500 index \\
infl &
Inflation: CPI inflation for all urban consumers \\
ltr & Long-term return: return of long term government bonds \\
        \bottomrule
    \end{tabular}
    }
\end{table} 

\end{document}